\documentclass[11pt,a4paper,final]{iopart}
\usepackage[utf8]{inputenc}
\usepackage{color}
\usepackage{graphicx}
\usepackage{bm}
\usepackage{rotating}
\usepackage{multirow}

\usepackage{hyperref}
\usepackage[normalem]{ulem}
\usepackage{comment}
\usepackage{iopams}
\usepackage{titlesec}

\expandafter\let\csname equation*\endcsname\relax

\expandafter\let\csname endequation*\endcsname\relax

\usepackage{amsmath}
\usepackage{amssymb}

\def\orcid#1{\kern .08em\href{https://orcid.org/#1}{\includegraphics[keepaspectratio,width=0.7em]{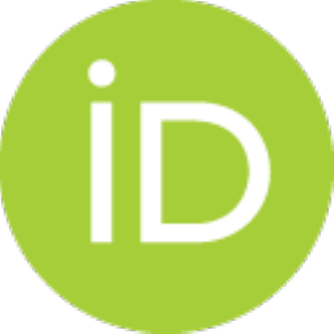}}}

\usepackage[belowskip=-5pt]{caption}
\usepackage{pifont}

\begin{document}
\title{Optical potentials for the rare-isotope beam era}

\author{C. Hebborn$\,\,\!\!$\orcid{0000-0002-0084-2561}$^{1,2,*}$, F. M. Nunes\,\,\!\!\orcid{0000-0001-8765-3693}$^{1,3}$, G. Potel\,\,\!\!\orcid{0000-0002-4887-7499}$^2$, W.~H.~ Dickhoff\,\,\!\!\orcid{0000-0003-1738-3979}$^4$, J. W. Holt\,\,\!\!\orcid{0000-0003-4373-3856}$^5$, M.~C.~Atkinson\,\,\!\!\orcid{0000-0002-5423-1432}$^{2,6}$, R.~B.~Baker\,\,\!\!\orcid{0000-0002-3909-4425}$^7$, C. Barbieri\,\,\!\!\orcid{0000-0001-8658-6927}$^{8,9}$, G. Blanchon$^{10,11}$, M.~Burrows\,\,\!\!\orcid{0000-0002-4574-4711}$^{12}$, R. Capote\,\,\!\!\orcid{0000-0002-1799-3438}$^{13}$, P.~Danielewicz\,\,\!\!\orcid{0000-0002-1989-5241}$^{1,3}$, M.~Dupuis$^{10,11}$, Ch. Elster\,\,\!\!\orcid{0000-0002-2459-1226}$^{7}$, J. E. Escher\,\,\!\!\orcid{0000-0002-0829-9153}$^2$, L. Hlophe$^2$, A.~Idini\,\,\!\!\orcid{0000-0001-7624-8975}$^{14}$, H.~Jayatissa\,\,\!\!\orcid{0000-0001-8746-0234}$^{15}$, B.~P.~Kay\,\,\!\!\orcid{0000-0001-8675-0731}$^{15}$, K. Kravvaris\,\,\!\!\orcid{0000-0002-1715-0967}$^2$, J.~J.~Manfredi\,\,\!\!\orcid{0000-0003-3703-7424}$^{16}$, A. Mercenne$^{17}$, B. Morillon$^{10,11}$, G.~Perdikakis\,\,\!\!\orcid{0000-0002-8539-8737}$^{18}$, C. D.~Pruitt\,\,\!\!\orcid{0000-0003-0607-9461}$^2$, G. H. Sargsyan\,\,\!\!\orcid{0000-0002-3589-2315}$^2$, I.~J.~Thompson\,\,\!\!\orcid{0000-0002-8058-5963}$^2$, M. Vorabbi\,\,\!\!\orcid{0000-0002-1012-7238}$^{19}$ and T. R. Whitehead\,\,\!\!\orcid{0000-0001-8909-2033}$^1$ }
\address{$^{*}$E-mail: hebborn@frib.msu.edu}
\address{$^1$Facility for Rare Isotope Beams, East Lansing, MI 48824, USA.}
\address{$^2$Lawrence Livermore National Laboratory, P.O. Box 808, L-414, Livermore, CA 94551, USA.}
\address{$^3$Department of Physics and Astronomy, Michigan State University, East Lansing, MI 48824-1321, USA.}
\address{$^4$Department of Physics, Washington University, St. Louis, MO 63130, USA.}
\address{$^5$Department of Physics and Astronomy and Cyclotron Institute, Texas A\&M University, College Station, TX 77843, USA.}
\address{$^6$Theory Group, TRIUMF, Vancouver, BC V6T 2A3, Canada.}
\address{$^7$Institute of Nuclear and Particle Physics, and Department of Physics and Astronomy, Ohio University, Athens, OH 45701, USA.}
\address{$^8$Dipartimento di Fisica, Università degli Studi di Milano, Via Celoria 16, I-20133 Milano, Italy.}
\address{$^9$INFN, Sezione di Milano, Via Celoria 16, I-20133 Milano, Italy.}
\address{$^{10}$CEA, DAM, DIF, 91297 Arpajon, France.}
\address{$^{11}$Université Paris-Saclay, CEA, Laboratoire Matière sous Conditions Extrêmes, 91680 Bruyères-Le-Châtel, France.}
\address{$^{12}$Department of Physics and Astronomy, Louisiana State University, Baton Rouge, LA 70803, USA.}
\address{$^{13}$NAPC--Nuclear Data Section, International Atomic Energy Agency, A-1400, Vienna, Austria.}
\address{$^{14}$Department of Physics, LTH, Lund University, P.O. Box 118, S-22100 Lund, Sweden.}
\address{$^{15}$Physics Division, Argonne National Laboratory, Lemont, IL, 60439, USA.}
\address{$^{16}$Department of Engineering Physics, Air Force Institute of Technology, Wright-Patterson AFB, OH 45433, USA.}
\address{$^{17}$Center for Theoretical Physics, Sloane Physics Laboratory, Yale University, New Haven, Connecticut 06520, USA.}
\address{$^{18}$Department of Physics, Central Michigan University, Mount Pleasant, MI 48859, USA.}
\address{$^{19}$National Nuclear Data Center, Brookhaven National Laboratory, Upton, NY 11973-5000, USA.}

\date{\today}
\newcommand*{\email}[1]{\href{mailto:#1}{\nolinkurl{#1}} } 
\newcommand{\cmark}{\ding{51}}%
\newcommand{\xmark}{\ding{55}}%
\newcommand {\nc} {\newcommand}
\newcommand {\bit} {\begin{itemize}}
\newcommand {\eit} {\end{itemize}}

\nc {\IR} [1]{\textcolor{red}{#1}}
\nc {\IB} [1]{\textcolor{blue}{#1}}
\nc {\IE} [1]{\textcolor{Plum}{#1}}

\newcommand{\red}[1]{{ \color{red} #1 }}
\newcommand{\green}[1]{{ \color{green} #1 }}
\newcommand{\magenta}[1]{{ \color{magenta} #1 }}

\newcommand{\ket}[1]{| #1 \rangle}
\newcommand{\bra}[1]{\langle #1 |}

\newcommand{\ie}{\emph{i.e.}}
\newcommand{\eg}{\emph{e.g.}}
\newcommand{\etc}{\emph{etc.}}

\maketitle
\begin{abstract} We review recent progress and motivate the need for further developments in nuclear optical potentials that are widely used in the theoretical analysis of nucleon elastic scattering and reaction cross sections. In regions of the nuclear chart away from stability, which represent a frontier in nuclear science over the coming decade and which will be probed at new rare-isotope beam facilities worldwide, there is a targeted need to quantify and reduce theoretical reaction model uncertainties, especially with respect to nuclear optical potentials. We first describe the primary physics motivations for an improved description of nuclear reactions involving short-lived isotopes, focusing on  its benefits for fundamental science discoveries and 
applications to   medicine, energy, and security. We then outline the various methods in use today to build   optical potentials starting from phenomenological, microscopic, and \textit{ab initio} methods, highlighting in particular the strengths and weaknesses of each approach. We then discuss publicly-available tools and resources  facilitating the propagation of recent progresses in the field to  practitioners. Finally, we provide a set of open challenges and recommendations for the field to advance the fundamental science goals of nuclear reaction studies in the rare-isotope beam era.\\

    This paper is the outcome of the Facility for Rare Isotope Beams Theory Alliance (FRIB - TA) topical program “Optical Potentials in Nuclear Physics” held in March 2022 at FRIB. Its content is non-exhaustive,  was chosen by the participants and  reflects  their efforts related to  optical potentials.

\end{abstract}

\tableofcontents

\section{Introduction}

Nuclear reactions drive the chemical evolution of the universe, enable a wide range of societally beneficial technologies on Earth, and provide a tool to study the structure of atomic nuclei and properties of the nuclear force in the laboratory. In particular, nuclear reactions involving exotic short-lived isotopes are fundamental to address numerous open questions in contemporary nuclear science research to be investigated at  frontier radioactive ion beam facilities, such as the Facility for Rare Isotope Beams (FRIB). Nuclear reaction theories with quantified uncertainties will be crucial to maximize the scientific impact of rare isotope facilities worldwide. However, even the simplest reactions that involve light nuclei at low center-of-mass energies are challenging to understand from fundamental \textit{ab initio} nuclear theory \cite{navratil16}. The continued development of microscopic models that capture salient features of the quantum many-body problem, such as antisymmetry and multiple scattering, is therefore needed to advance nuclear reaction science in the rare-isotope beam era.

The nuclear optical model is the primary tool to reduce the complexity of quantum many-body scattering into a form that is tractable across a large range of energies, target isotopes, and reaction channels. The idea is to replace fundamental two-body and many-body forces between projectile and target with a complex and energy-dependent two-body local potential $U(r,E) = V(r,E) + i W(r,E)$, in analogy with the scattering and absorption of light in a dielectric medium  with complex index of refraction~\cite{Hodg63,Hodg71,Hodg97} . The imaginary part, $W(r,E)$, of the nuclear optical model potential (OMP) accounts for the loss of flux in the elastic scattering channel due to open reaction channels (e.g., inelastic, pick-up, break-up, and similar reactions) dictated by the projectile energy $E$. The energy dependence of the optical potential  accounts for temporal nonlocalities due to virtual excitations of the nucleus and implicitly accounts for spatial nonlocalities that arise from exchange scattering on indistinguishable nucleons in the target. Formally, spatial nonlocality gives rise to a momentum-dependent potential, but for elastic scattering this can be approximated  \cite{perey62} in terms of an equivalent energy-dependent local mean field. During the last 60 years the nuclear optical model has been broadly applied to analyze the elastic scattering of pions, nucleons and heavier ions by nuclei, over a wide range of energies \cite{Hodg63,Hodg71,Hodg97}. Inelastic scattering was included by the coupled-channels formalism~\cite{bu63,tamura65}, and consideration of dispersion effects from the requirement of causality~\cite{mangsa86} allows for the description of both bound and scattering states by the same complex nuclear mean field \cite{masa91,masa91rev,Dickhoff17}.

Following decades of refinements \cite{becchetti69,varner91,koning03,ripl3,PhysRevC.80.034608}, the phenomenological optical model has achieved an impressive description of nucleon-nucleus scattering on stable target isotopes up to projectile energies $E \leq 200$\,MeV. The functional form of modern phenomenological optical potentials includes complex-valued volume, surface, and spin-orbit terms, together with a central Coulomb interaction:
\begin{equation}
U(r,E) = V_V(r,E) + V_D(r,E) + i W_V(r,E) + i W_D(r,E) + V_{so}(r,E) \vec l \cdot \vec s + i W_{so}(r,E) \vec l \cdot \vec s + V_C(r).
\label{kdpar}
\end{equation}
The energy dependence in Eq.\ \eqref{kdpar} is typically decoupled from the radial dependence, e.g.,
\begin{equation}
V_V(r,E) = v_V(E) \, f(r,R_V,a_V),
\label{kdpar2}
\end{equation}
  where the form factor $f(r)$ is usually defined as  a Woods-Saxon shape characterized by a radius $R$ and diffuseness $a$, i.e.
 \begin{equation}    
  f(r,R,a)= \frac{1}{1+e^{(r-R)/a}}.
 \end{equation}
Note that other radial dependencies can be used, such as Gaussian and squares of Woods-Saxon form factors. 
 Whereas the real and imaginary volume terms ($V_V$, $W_V$) are proportional to form factors, the real and imaginary surface contribution ($V_D$, $W_D$) and the spin-orbit contributions  ($V_{so}$, $W_{so}$) are proportional to the radial derivatives of these form factors.   In general, the fitted parameters that describe the energy and geometry dependence also vary with the mass number $A$ and isospin asymmetry  $\delta=(N-Z)/A$ . In total, approximately 40 fitted parameters are used in the construction of the widely used non-dispersive phenomenological global optical potential of Koning and Delaroche~\cite{koning03}. Enforcing the appropriate dispersion relations  
 \begin{equation}
     {\rm Re} \{U(E)\}=-\frac{1}{\pi} \mathcal{P}\int dE'  \frac{{\rm Im} \{U(E')\}}{E-E'}\label{eq:full_dispersion}
 \end{equation}
  allows the number of parameters to be significantly reduced.   Moreover, the description of both elastic scattering at incident energies below 5~MeV and the bound states is improved as compared to non-dispersive approaches . For nuclear reactions involving unstable isotopes, for which scattering data are scarce, the quality of phenomenological optical potential extrapolations is uncertain. For this purpose, microscopic or semi-microscopic   optical potentials derived from fundamental nucleon-nucleon and many-nucleon forces may provide useful starting points. 

Since the 1950s, several approaches have been developed to derive nucleon-nucleus optical potentials starting from microscopic many-body theory. These include the Watson multiple scattering theory, Feshbach projection operator formalism, and Green's function theory.   Before briefly presenting each approach, let us  define the many-body Hamiltonian for the $(A+1)$-body system
\begin{equation}
{\cal H} = H_A(\vec r_1,\dots,\vec r_A) + h_0 + V(\vec r,\vec r_1,\dots,\vec r_A),
\end{equation}
where $H_A$ is the Hamiltonian for the $A$-particles of the target nucleus with eigenstates satisfying $H_A\Phi_k = \epsilon_k \Phi_k$, $h_0$ is the kinetic energy of the projectile, and $V$ is the many-body  interaction potential between projectile and nucleons in the target. The exact eigenvalues and eigenstates of the $(A+1)$-particle system satisfy ${\cal H}\Psi = {\cal E}\Psi$.    

The aim of the multiple scattering theory of Watson \cite{watson53} and its later extension by Kerman, McManus, and Thaler~\cite{kerman59} is to derive an equation for the optical potential in terms of free-space nucleon-nucleon scattering amplitudes. This can be formally justified only under certain assumptions, including the so-called ``impulse approximation'', whereby it is assumed that the projectile nucleon is traveling at sufficiently high energy that it interacts strongly with only one or few nucleons of the target at some instant in time.   
The theory is therefore expected to be valid at energies sufficiently above the excitation energies of the nucleus, i.e. larger than roughly 60 to 100~MeV, where impacts of Pauli-blocking and three-body forces are diminished.  In the impulse approximation and at leading order, the nucleon-nucleus optical potential with kinetic energy $E$  can be represented as~\cite{kerman59}
\begin{equation}
\hat{U}\left({\vec  k^{\prime}},{\vec  k};E\right)=
\sum_{\alpha=p,n} \int d^{3}{\vec  p^{\prime}}
d^{3}{\vec  p} \left\langle {{\vec  k^{\prime}}{\vec  p^{\prime}}}
\mid \hat{\tau}_{\alpha} (E)
\mid {{\vec  k  \vec p }} \right\rangle \rho_{\alpha} \left({{\vec  p^{\prime}}
+\frac{{\vec  k'}}{A}}, {\vec  p}+\frac{{\vec  k}}{A}\right)
\delta^{3} ({\vec  k^{\prime}} + {\vec  p^{\prime}} -
{\vec  k} - {\vec  p}), \label{eq:MS6}
\end{equation}
where the momenta ${\vec  k'}$ and ${\vec  k}$ are the final and initial momenta
of the projectile in the frame of zero total nucleon-nucleus
momentum, $\alpha$ sums over target neutrons and protons. The quantity $\hat{\tau}_{\alpha} (E)$ represents a nucleon-nucleon amplitude, and $\rho_{\alpha}$ the ground-state density of the target for protons ($p$) and neutrons ($n$). Note that extensions of this approach going beyond the impulse approximation and including three-body forces are possible and will be discussed further in the text.

At lower projectile energies, where Pauli blocking and three-body forces play an enhanced role, optical model potentials may be constructed using either the Feshbach projection formalism or Green's function theory.  Neglecting first the antisymmetrization between the projectile and the constituent nucleons in the target, 
Feshbach derived \cite{feshbach1958unified} an equation for the projection $P\Psi(\vec r, \vec r_1,\dots, \vec r_A) = \phi_0(\vec r)\Phi_0(\vec r_1,\dots,\vec r_A)$ of the total wavefunction onto the elastic scattering channel of the form 
\begin{equation}
\left ( h_0 + {\cal V}_{00} + \sum_{j,k\neq 0}{\cal V}_{0j}G_{jk} {\cal V}_{k0}\right ) \phi_0 = E \, \phi_0,  \label{eq:7i}
\end{equation}
 where  $\Phi_0$ is  the ground state of the target and $E={\cal E}-\epsilon_0$ is  the nucleon-nucleus relative energy defined from the ground-state target energy $\epsilon_0$. We define in Eq.~\eqref{eq:7i}; the  Green's function matrix element~$G_{jk}$ 
\begin{equation}
    G_{jk}=\lim_{  \eta {\to} 0^+}\frac{1}{E -H_{jk} + i \eta }\label{eq:8i}
\end{equation}
from the Hamiltonian matrix element $H_{jk}$  and coupling potentials between the elastic scattering channel and the inelastic ones ${\cal V}_{jk}$,
\begin{equation}
H_{jk}=h_0 \delta_{jk}+{\cal V}_{jk} +\epsilon_k \delta_{jk} \quad \text{and} \quad
{\cal V}_{jk} = \langle \Phi_j | V | \Phi_k \rangle.\end{equation} The optical potential is then identified as
\begin{equation}
V_{\rm opt} = {\cal V}_{00} + \sum_{j,k\neq 0}{\cal V}_{0j}\frac{1}{E -H_{jk} + i \eta } {\cal V}_{k0}
\label{vefff}
\end{equation} 
with  $\eta {\to} 0^+$. The potential is complex, energy dependent, and nonlocal.  This formulation was later extended \cite{feshbach62} to treat inelastic scattering processes and to properly account for antisymmetrization between projectile and target.

An alternative derivation of the optical potential makes use of the language of second-quantized many-body theory and Green's functions. Here the propagation of particles and holes in a quantum many-body system is characterized by the one-body Green's function~\cite{Dickhoff08} 
\begin{equation}
G(\vec r, t; \vec r^{\, \prime}, t^{\, \prime}) = -i \langle \Phi_0  | \hat T [ \hat a_H(\vec r,t) \, \hat a^\dagger_H(\vec r^{\, \prime},t^\prime) ] | \Phi_0  \rangle,
\end{equation}
where $\hat T$ is the time ordering operator, $a^\dagger_H$ and $a_H$ are creation/annihilation operators in the Heisenberg representation. The time Fourier transform $G(\vec r, \vec r^{\,\prime}; E)$ of the one-body Green's function can be expressed in terms of the free Green's function  $ G_0(\vec r, \vec r^{\,\prime}; E)$,  defined from the free Hamiltonian $H_0=h_0+H_A$,  
and  the nucleon irreducible self--energy $\Sigma^\star(\vec r, \vec r^{\, \prime}; E)$ through the Dyson equation~\cite{Dickhoff08} 
\begin{equation}
G(\vec r, \vec r^{\,\prime}; E) = G_0(\vec r, \vec r^{\,\prime}; E)+ \int d^3 y \int d^3 y^\prime \, G_0(\vec r, \vec y; E) \, \Sigma^\star(\vec y, \vec y^{\, \prime}; E) \, G(\vec y^{\, \prime}, \vec r^{\, \prime}; E) \, .\label{eq9}
\end{equation}
The self energy can then be shown \cite{bell59, Dickhoff19} to play the role of a nonlocal potential that governs the spectrum of overlap functions $\phi_0(\vec r\,) = \langle \Phi_0 | a(\vec r\,) | \Psi \rangle$  associated with positive-energy elastic scattering states of the $(A+1)$-body system according to the Schr\"odinger equation
\begin{equation}
-\frac{\nabla_r^2}{2\mu}\phi_0(\vec r\,) + \int \Sigma^\star(\vec r, \vec r^{\, \prime}; E) \, \phi_0(\vec r^{\, \prime}) \, d^3 r^\prime = E\phi_0(\vec r\,) , 
\label{wave}
\end{equation}
where $\mu=[A/(A+1)]m_N$ is the reduced mass and $m_N$ is the nucleon mass.   Hence, $\Sigma^\star(\vec r, \vec r^{\, \prime}; E)$ can be identified with the nucleon-nucleus optical potential as well as providing a similar interpretation for the properties of a removed nucleon \cite{Dickhoff19}.

Having briefly reviewed the different formalisms used to derive  optical potentials, the rest of the paper is organized as follows. In Sec.~\ref{Sec2}, we motivate the need for modern optical potentials to address a wide range of applications from fundamental science discoveries to astrophysics to nuclear energy and security. In Sec.~\ref{Sec3}, we review recent advances in constructing optical potentials from phenomenological, semi-microscopic and microscopic approaches with emphasis on the nucleon-nucleus potential. We also comment on the use of microscopic optical potentials to inform phenomenology and discuss limitations and model uncertainties of the two-body approximation used in nearly all optical model applications. In Sec.~\ref{S4}, we present new tools and resources to facilitate the propagation of recent (and future) progress in the field of nuclear optical models to experimentalists and practitioners. In Sec.~\ref{Sec5}, we compare the different approaches presented in this work and discuss their accuracy for different systems.
We end with a summary and outlook.

\section{Applications of optical potentials} \label{Sec2}

\subsection{Direct reactions to probe exotic nuclei}\label{Sec2A}
Direct reactions have played a foundational role in the development of our understanding of nuclear structure and will be an essential tool in the FRIB era~\cite{Bonaccorso18}. In 1950, Butler had already observed that proton angular distributions following the ($d$,$p$)  reaction yield information about the transferred neutron~\cite{Butler50}. The rapid development of reaction theory thereafter, reliant on optical potential models, fostered their use to explore many facets of nuclear structure. More specifically, direct reactions have been instrumental in enriching our knowledge concerning shell evolution, weak binding, pairing, symmetries, deformation and applications to nuclear astrophysics and fundamental symmetries.

In the mid eighties, the development of radioactive-ion beams (RIBs) {and new experimental techniques} enabled the study of nuclei away from stability, revealing exotic structures and challenging the usual description of nuclei. In particular, in regions with extreme neutron-to-proton ratio, nuclei exhibit single-particle levels with a non-standard spin-parity ordering, forming the so-called \textit{islands of inversion} \cite{PhysRevC.41.1147,PhysRevC.12.644,Otsuka20} . {Exploring nuclei far from stability helped elucidate mechanisms for shell evolution, such as the ubiquitous action of the tensor force~\cite{Otsuka05,Otsuka20}.} Even more surprising, halo nuclei can be found close to the drip lines. These nuclei present a strongly clusterized structure, in which one or two loosely-bound nucleons have a high probability of presence far from the rest of the nucleons. With several major facilities around the world now capable of delivering  beams of short-lived nuclei   at Coulomb barrier energies and beyond, in conjunction with specifically tailored instrumentation, a wealth of direct-reaction data is expected. 

{In the absence of a universal, fully} {\it ab initio} description of nuclear reactions {and its connection to nuclear structure}, the theoretical description of the reaction process is often  simplified into a few-body one, where both  projectile and target can be seen as composed of one or more clusters of nucleons.   In this few-body picture, the structure of the nuclei involved in the reaction   is described through an effective interaction reproducing properties of the  low-energy spectrum of the nuclei   while the interactions between the clusters making up the projectile and the target are simulated through optical potentials accounting in an effective fashion for the neglected many-body structure. These cluster-cluster (or nucleus-nucleus)  optical potentials generalize the concept of nucleon-nucleus optical potentials.   The accuracy of the nuclear-structure information inferred from experimental data is therefore strongly influenced by the reliability of both the optical potential  and the few-body reaction model.

Among all nuclear reactions, {elastic scattering}, in which both the projectile and the target nuclei remain in their ground states and the beam is deflected from its incident direction, constitutes one of the simplest probes. The diffraction pattern of the angular distribution can be used to infer the size of the nucleus~\cite{PhysRevLett.105.022701}. {Resonant elastic scattering is also a powerful tool for exploring isobaric analog resonances with RIBs~\cite{Bradt18}.} The difference between the incoming flux and the elastic one defines the reaction cross section, which corresponds to processes in which the colliding species exchange energy and/or mass. As their magnitude grows with the spatial extension of the projectile, reaction cross sections have played a key role in the discovery of halo nuclei~\cite{TANIHATA1985380,PhysRevLett.55.2676}. 

Inelastic scattering is defined as an excitation of the target and/or the projectile during the collision process. The magnitude of inelastic scattering cross sections provide information about the system's response to nuclear and electromagnetic probes~\cite{AN21}, associated with the degree of collectivity of the populated states. A high degree of collectivity is then often associated with the excitation of a rotational band in deformed nuclei, or with vibrations of the nuclear surface{---it is anticipated that inelastic scattering will be a powerful complement to Coulomb excitation studies in the FRIB era, especially in probing octupole deformation~\cite{Butler16}}.

Another key {nuclear-structure} quantity is the Gamow-Teller (GT) strength, which provides an essential connection to nuclear $\beta$ decays. GT strength distributions are typically probed with charge-exchange reactions in which a proton in a target nucleus is exchanged for a neutron in the projectile nucleus, or vice-versa. Charge-exchange cross sections also provide insights on the isovector densities, i.e., the difference between proton and neutron densities inside the nucleus~\cite{Frekers18}, {and play a role in understanding double-$\beta$ decay nuclear matrix elements}~\cite{Cappuzzello18,CAPPUZZELLO2023103999} .

One-nucleon transfer reactions, such as $(d,p)$ and $(p,d)$, {at energies a few MeV per nucleon above the Coulomb barrier in both the entrance and exit channels (typically} between $\sim$5 and 50~MeV/nucleon, depending on the $Q$ value), are 
{highly selective probes} to obtain information about the nuclear response to nucleon addition and removal (single-particle strength)~\cite{Schiffer12,PhysRevLett.104.112701,Wimmer18,TimoJohn2020PPNP} . 
The absolute value of the cross section is proportional to the single-particle content of the populated state, namely {the} spectroscopic factor, while the shape of  the angular partial differential cross section is 
{a strong signature of its} orbital angular momentum. The evolution of the single-particle strength along an isotopic{/isotonic} chain~\cite{Schiffer04,Manfredi21} is an 
{ideal tool to explore} correlations as a function of neutron-proton asymmetry, in particular as one approaches the nucleon drip lines.

Multiple-nucleon transfer reactions, in which several nucleons are transferred from the projectile to the target and vice-versa,  are also commonly used to probe clustering and pairing effects.  In particular, $\alpha$-transfer cross sections for processes such as $(^6{\rm Li},d)$ carry  information about the $\alpha$ clustering inside the nucleus {(see, for example, Ref.~\cite{Aumann20})}, and can be used to infer reaction rates that are of astrophysical interest (see, for example, Refs.~\cite{Tribble_2014,Horowitz19,AstroAnnRev,Tumino:21} and Sec. \ref{Sec2C}). Two-nucleon transfer reactions give insights on pairing modes and pairing correlations~\cite{vonOertzen:01,Potel:13,Frauendorf14,Potel:20}. Quasi-free scattering reactions, mentioned below, which knock out multiple  nucleons \cite{BERTULANI2006155,PhysRevC.83.014605,Nature4N,STEVENS2018374} and or clusters of nucleons, e.g., alpha particles \cite{doi:10.1126/science.abe4688}, can probe similar properties.

The structure of loosely-bound exotic nuclei can also be studied through breakup reactions~\cite{Bonaccorso13,BC12,PhysRevLett.94.072701,OB10,BB21,PhysRevC.89.069901,PhysRevLett.109.232502,PhysRevC.73.024602}, since the counting statistics are high owing to the fragile nature of these nuclei. When performed on heavy targets, the Coulomb repulsion  between the  projectile's clusters and the target dominates, while on light targets, their nuclear interaction is  responsible for most of the breakup cross section.  Consequently, the mechanisms driving the dissociation, and hence the properties of the exotic nuclei probed by the reaction, depend strongly on the nature of the target. For example, electromagnetic strength functions and capture rates of astrophysical interest can be extracted from Coulomb-dominated breakup observables~\cite{AN21,Tribble_2014,AstroAnnRev,Baur_2007,BAUR1986188,PhysRevC.78.069908,PhysRevC.73.024605,MOSCHINI2019367,MORO2020135959}.
 
Inclusive measurements of one-nucleon breakup at intermediate energies ($>$50 MeV/nucleon), often referred to  as   one-nucleon removal, {or heavy-ion knockout,} reactions~\cite{Hansen03,T06,PhysRevC.69.044603,SAUVAN20001}, have even higher counting statistics because only one fragment is measured after the collision. This type of inclusive reaction typically uses a beryllium {or carbon} target and is the favored probe {for nuclei at} the limits of stability for which {only} low-intensity beams are available. The standard associated observable is the parallel-momentum distribution of the remaining core. It acts as a key probe of the single-particle structure of the projectile, as it carries information on the spin, parity and energy of single-particle states~\cite{Hansen03,T06,PhysRevC.66.024607,HC21}. These reactions have therefore been widely used to study the shell structure evolution across regions of the nuclear chart, i.e., from stable to exotic nuclei at the drip lines. 
An open question in the field has been raised by such studies, relating to the nucleon asymmetry dependence of the quenching of cross sections on the difference between proton and neutron separation energies~\cite{Gade08,Tostevin14,Tostevin21} . This issue ties together many of the direct-reaction probes mentioned in this section (see recent review in Ref.~\cite{Aumann21} and references therein).

Quasifree $(p, 2p)$ and $(p, pn)$ scattering reactions are also a key probe of the single-particle degrees of freedom of exotic nuclei~\cite{PhysRevC.88.064610,PhysRevC.92.044605,GOMEZRAMOS2018511}.  Contrary to one-nucleon removal reactions and transfer, quasifree scattering probes the inner part of the single-particle wavefunction. This is a consequence of the high energies  at which these measurements are performed (typically above 200 MeV/nucleon) .  

\begin{figure*}
\centering
\includegraphics[width=\textwidth]{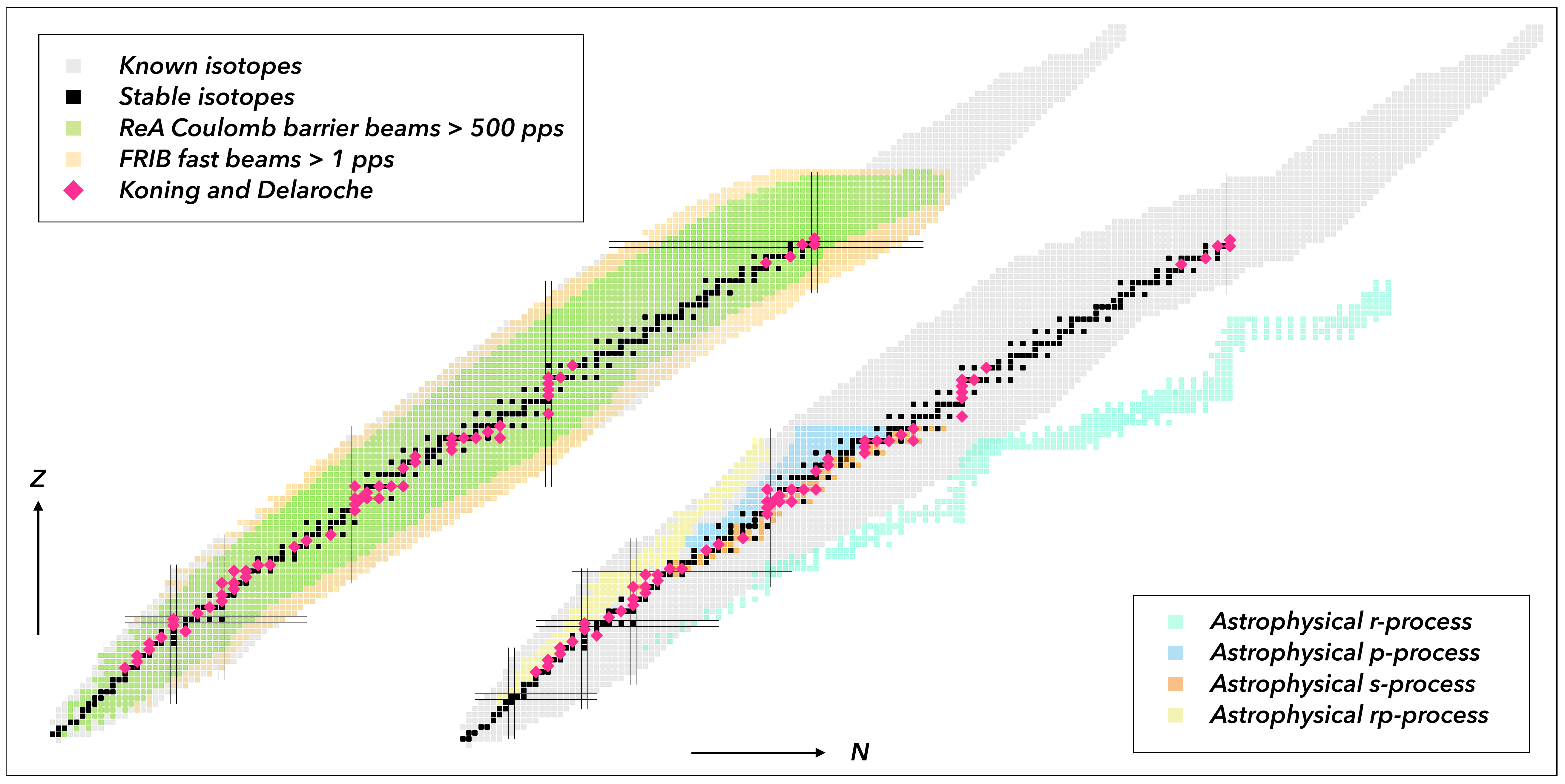}
\caption{\label{figX} The chart of nuclides showing estimates of the reach of reaccelerated beams  at FRIB and `fast' fragmentation beams at FRIB and the well-known astrophysical processes. Also indicated (pink diamonds) are the nuclei whose properties were used to constrain the Koning and Delaroche optical potential~\cite{koning03}, highlighting the dramatic extrapolations made.}
\end{figure*}

To interpret  reaction measurements, to arrive at a more fundamental understanding of the nuclear structure
and reactions, and to plan new experiments, it is crucial to develop accurate reaction models
coupled with realistic interactions between the relevant nuclei.  
In the context of the direct reactions discussed here,  optical potentials are responsible for most of the uncertainties {beyond the description of the bound states}~\cite{Lovell2020}.   Historically, optical potentials have been fitted to elastic-scattering data on  stable targets. Thus,  these interactions are not well constrained for exotic nuclei. 
Fig.~\ref{figX} illustrates the nuclei used to derive the Koning and Delaroche~\cite{koning03} optical model parametrization (pink   diamonds), overlaid on the beams expected to be available at  FRIB,  both for reaccelerated (Coulomb barrier) beams of $>$100~particles per second  (pps) and for `fast' beams of $>$1 pps. Juxtaposed, Fig.~\ref{figX} also contains the regions of the nuclear chart relevant for the various astrophysical processes. Direct reactions involving these nuclei can be used to extract astrophysical rates~\cite{Tribble_2014,AstroAnnRev,Tumino:21}. It is clear that, in the FRIB era, using current phenomenological optical potentials carries tremendous uncertainties when extrapolating to the driplines.

In order to enhance the accuracy of our predictions concerning the reactions to be measured at FRIB, and to provide the associated quantified uncertainties, there is  an urgent need for developing optical potentials across the whole nuclear  chart at energies  ranging from {a few MeV up to} 400 MeV.  These next-generation interactions should include more physics constraints in order to reliably cover the variety of reactions to be studied at FRIB and ultimately provide a comprehensive and accurate account of exotic nuclei.

\subsection{Compound nuclear reactions}
\label{Sec2B}

Compound-nuclear reactions play an important role in nuclear physics and in applications. Their cross sections are required input for astrophysical simulations that describe stellar evolution and nucleosynthesis and for modeling processes that are relevant to medical isotope production, national security applications, and to generating energy.  

\begin{figure}
\centering{\includegraphics[width=0.8\textwidth,angle=0,trim=20 40 20 90,clip=true]{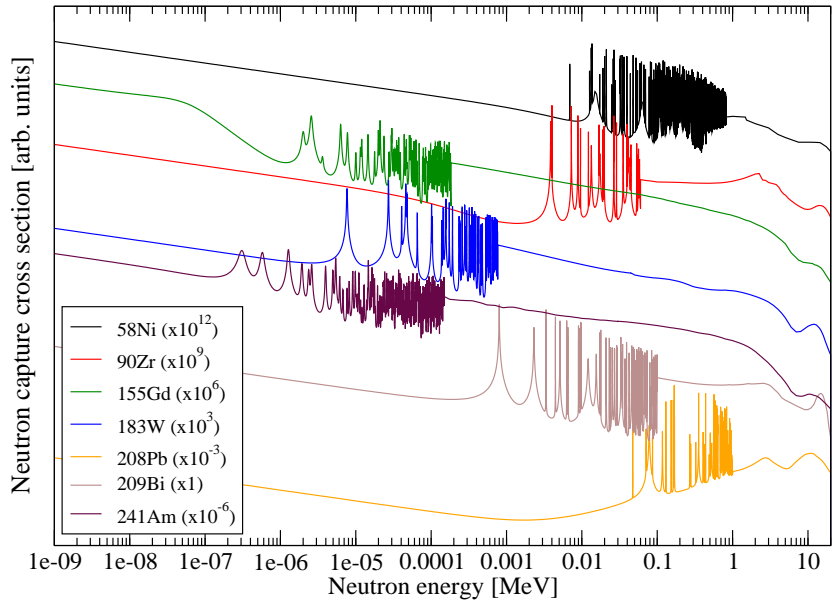}}
\caption{Evaluated neutron capture cross sections for various stable targets~\cite{ENDF_b7:06}.  The depicted evaluations are based on a combination of calculations and measured data (not shown) and illustrate the different energy regimes: resolved resonance region (RRR) at lower energy where individual resonances can be distinguished, unresolved resonance region (URR) at intermediate energy where resonance peaks are still visible but overlapping, Hauser-Feshbach (HF) regime at higher energy where the cross section has a smooth dependence on energy, representing an average over strongly overlapping resonances.   The high-energy behavior of the cross sections can be described in an average way using the HF formalism.}
\label{Fig:CN}
\end{figure}

In a compound-nuclear reaction, a projectile fuses with a target to produce a highly-excited intermediate nuclear system which equilibrates and subsequently decays by particle evaporation, fission, or gamma emission. Compound reactions are very slow; at very low energies they produce narrow, isolated resonances and can be described in the framework of the R-matrix formalism~\cite{Lane:58}.  With increasing projectile energy, the resonances begin to overlap, forming the unresolved-resonance region (URR) and, at even higher energies, one enters a region in which the statistical Hauser-Feshbach (HF) formalism is applicable~\cite{HauserFeshbach:52}. The demarcation between the various regimes depends on the   projectile type and on the structure of the compound nucleus (CN) formed. For reactions involving well-deformed nuclei with large level densities (e.g., n+$^{155}$Gd), the region of strongly overlapping resonances lies at much lower energies than for reactions involving nuclei near closed shell (e.g., n+$^{208}$Pb), see Fig.~\ref{Fig:CN}.

The HF formalism describes the average cross section for forming a CN  at energy $E_{ex}$ by fusing projectile $a$ and target $A$ (channel $\alpha$) and subsequent decay into reaction products $c$ and $C$   (channel $\chi$) 
\begin{equation}
\sigma_{\alpha \chi} (E ) = \sum_{J\pi}    \sigma^{CN}_{\alpha} (E_{ex}, J, \pi) G^{CN}_{\chi} (E_{ex}, J, \pi) W_{\alpha \chi}(E_{ex}, J, \pi).
\label{eq:HF}
\end{equation}
Here $\sigma^{CN}_{\alpha} (E_{ex}, J, \pi)$ is the CN formation cross section for the $\alpha$ channel.  The quantities $G^{CN}_{\chi} (E_{ex}, J, \pi)$ describe the competition between the decay channel of interest ($\chi$) and all other competing channels. Calculating $G^{CN}_{\chi}$ requires nuclear structure information, such as gamma-ray strength functions, fission barriers, and level densities~\cite{ripl3}.  These are traditionally written in terms of products of transmission coefficients (TCs) and level densities (in the residual nuclei).  The TC for gamma emission is related to the gamma-ray strength function, the TC for fission describes tunneling through fission barriers.  
All other TCs describe the probability for particles to be emitted from  the CN    and are obtained from a potential-model or coupled-channels calculation that uses a nucleon-nucleus optical potential: $T_{\beta} = 1 -  |S_{\beta\beta}^{\rm opt}|^2$, where $S_{\beta\beta}^{\rm opt}$   denotes the optical-model S-matrix, that can be obtained from projecting out all non-elastic channels coupled in the calculation.

Eq.~\eqref{eq:HF} above also contains a width fluctuation correction factor, $ W_{\alpha \chi}( E_{ex}, J, \pi )$, which accounts for remnant correlations between the incoming and outgoing channels~\cite{Hilaire:03}. It is a reminder that the reaction cannot be completely separated in two independent processes.  Similarly, in most realistic cases (and in all nuclear reaction evaluations) there are additional, non-compound, reaction processes that have to be accounted for when describing reaction observables.  For that reason, statistical reaction codes have capabilities well beyond the evaluation of the HF expression.  They include descriptions for direct reactions, pre-equilibrium reactions, as well as CN reactions.  As detailed in Sec.~\ref{Sec2A}, the description of   direct-reaction observables, in turn, require OMPs. Statistical reaction codes also typically contain subroutines or auxiliary codes to generate transmission coefficients using OMPs. In addition, they need nuclear structure information – most modern ones can read this from databases, such as the Reference Input Parameter Library (RIPL-3)~\cite{ripl3}. Much work has been devoted to improving statistical reaction calculations.  Nuclear structure inputs have received much attention over the past two decades, with multiple theoretical and experimental efforts aimed at providing more reliable inputs for gamma-ray strength functions and nuclear level densities in particular (see Refs.~\cite{Getal19,ZELEVINSKY2019180,LARSEN201969,SAVRAN2013210,BRACCO2019360} and references therein).

Similarly, a large number of optical potentials has been made available for use with statistical reaction codes~\cite{ripl3}.  This includes phenomenological nucleon-nucleus potentials, such as the spherical potential by Koning and Delaroche~\cite{koning03}, the semi-microscopic nucleon-nucleus potential by Bauge et al~\cite{Bauge:98, Baug2000, Baug2001}, and the dispersive nucleon-nucleus potential by Morillon and Romain~\cite{Morillon:07}.  For deformed nuclei,  the relevant nucleon-nucleus transmission coefficients are generated using coupled-channels calculations.  Multiple efforts have focused on developing appropriate coupling schemes and requisite potentials~\cite{Nobre:15, Soukhovitskii:16, Soukhovitskii:20}.  To describe fusion or emission of composite particles, optical potentials for light ions ($t$, $^3$He, $\alpha$, etc) are required.  These tend  to have larger uncertainties, as there is fewer available data  to place constraints on the shapes and parameters of those potentials.  

As they take  various inputs, statistical reaction calculations are affected by multiple sources of uncertainty:  nuclear-structure information for the various possible decay channels may be lacking,  some reaction mechanisms, e.g., pre-equilibrium contributions, are not sufficiently well known and  optical model uncertainties also affect the predicted cross sections.  Calculations for neutron-induced reactions on actinides are known to be sensitive to the coupled-channels optical potential utilized~\cite{Soukhovitskii:16, Soukhovitskii:20}.

Level densities can be estimated by extrapolating from known discrete levels, and (ideally) by measuring resonances in the interaction of two subsystems of the nucleus. Much better physical accuracy is possible if an optical potential is known for the scattering of those two subsystems. That is because we can simplify R-matrix theory when we can ignore interferences between resonances, and convert an average ratio of widths to a ratio of average widths $\langle \Gamma\rangle$.
Then the ratio of average width reads
\begin{equation}
    \frac{\langle \Gamma\rangle}{D} = \frac{T_\beta} {2\pi} , 
    \label{Eq:AvgWidth}
\end{equation} 
where $D$ is the average level spacing.   The transmission coefficient  $T_\beta$   comes from the S-matrix 
element for optical-model scattering, as explained above. 
Thus, given level densities and optical potentials we can estimate average widths. These estimates are the basis of the HF model of statistical reactions and decays that uses Porter-Thomas \cite{PorterThomas1955} distributions with these statistical averages and also neglects interference between resonances.

These approximations should be tested to gauge the accuracy of the HF model. Much work has gone into improving the average-ratio approximation, and now  width fluctuation corrections (WFC) are standard in HF models \cite{Ernebjerg2005} . The neglect of interference still needs to be tested but should be reasonable for angle-integrated data if not for angular distributions. Improvements have also been given by Simonius \cite{simonius1974} to Eq.~\eqref{Eq:AvgWidth} when the widths are large, since the transmission coefficient has maximum value of one. 

\begin{figure}
\centerline{\includegraphics[clip, trim=0cm 0cm 0cm 0.55cm,width=0.6\textwidth]{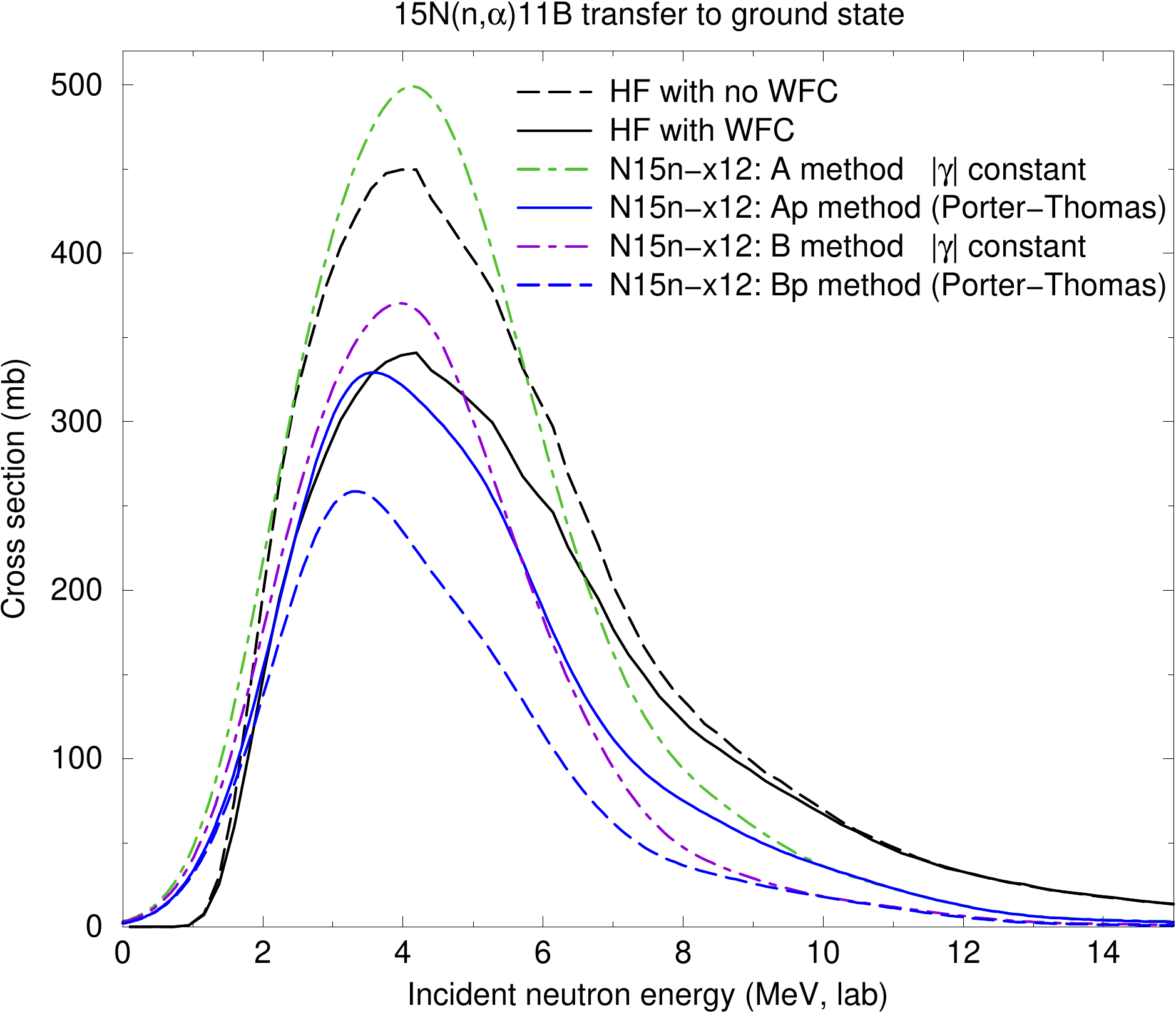}}
\caption{$^{14}$N($n$,$\alpha$) cross sections as a function of neutron energy.
The HF curves are standard Hauser-Feshbach calculations,  without and with the width fluctuation corrections (WFC).
The A curves follow Simonius \cite{simonius1974} and the B curves the linear approximation given in the text.
The Ap and Bp curves use the standard Porter-Thomas statistical distributions \cite{PorterThomas1955} for the R-matrix parameters, while A and B curves have fixed amplitude sizes but random signs.}
\label{Fig:N14n-a}
\end{figure}

One overall test of all these approximations is to derive the resonance parameters for a typical HF model using the common approximations, and then see whether a full R-matrix calculation with those parameters gives the HF predictions.
The results of such a comparison for $n$ + $^{14}$N elastic, inelastic and transfer reactions have been recently calculated.  Fig. \ref{Fig:N14n-a}   shows the $^{14}$N($n$,$\alpha$) cross sections as a function of neutron energy.
We see that, as might be expected,  the best HF calculation (HF with WFC: the solid black line) is closest to the best R-matrix statistical model (Ap: the solid blue line).
The agreement is not perfect and this is only a comparison between models,   but it gives hope that, around the unresolved resonance region, there is a domain where the R-matrix resonance
treatment can match up with the HF statistical treatment. This should help to constrain both the resonance parameters and optical potentials in the two regimes.

It is instructive to briefly focus on specific ways that observables may inform optical potential development and vice versa. The connection between experiment and theory is, of course, facilitated via interpretative or predictive calculations. In the rest of this section, we will spotlight some examples of information exchange between experiment and theory. 
Nuclear reaction yields are related to optical potentials through the transmission coefficients discussed  above. The  shapes of the angular distributions of emitted particles are observables that theoretical calculations can reproduce. Typically, the transmission coefficients for the emitted particles will need to be modified to precisely fit the experimental data. The new experimentally constrained transmission coefficients will correspond to a new set of optical potential parameters that are now phenomenologically constrained and available for future calculations. These new parameters, locally fitted as they are, do not necessarily carry any predictive value away from the vicinity of the target nucleus considered. This shortcoming is becoming quite an issue for nuclear astrophysics applications in which a theoretical prediction is, for most participating nuclei, the only possible way to determine the thermonuclear reaction rates of interest (see e.g., Fig.~\ref{figX}).
A second way to inform nuclear theory from experimental data is by reproducing evaporation spectra from highly excited compound nuclei. While the so-called evaporation technique \cite{Voinov2007} is typically suitable for determining the level densities of the excited compound systems, recently, it has been demonstrated in the literature \cite{Voinov2021} that the evaporated particle spectra can also provide some insight into optical potential  properties.

Beam time at premier experimental facilities such as FRIB is very expensive and the number of available hours at large-scale radioactive ion beam facilities is often limited. This raises the degree of importance placed on high-quality, reliable, and accurate simulations to support the value of the proposed experiment. Optical potentials are necessary ingredients of any such simulation or theoretical prediction. The value of such predicted cross-sections is profound for astrophysics applications involving thousands of species away from stability (for example for r-process or i-process nucleosynthesis). Typically predictions are collected into reaction rate libraries like the widely used JINA REACLIB~\cite{JINAREACLIB}. An improved optical potential thus can influence fields beyond nuclear physics that make use of such libraries. Lastly, theoretical cross-section calculations are typically collected into databases (e.g.\ TENDL~\cite{TENDL}) that are available for use with various simulation tools such as GEANT4~\cite{AGOSTINELLI2003250} and MCNP~\cite{MCNP} that are broadly utilized by experimentalists to optimize their instrumentation and interpret raw data.

\subsection{Astrophysically relevant reactions}
\label{Sec2C}

Another area in which optical potentials are ubiquitous is nucleosynthesis. Astrophysical modelling for a wide range of astrophysical sites require large networks of reactions. These cannot all be measured, and instead models use global optical potentials for the task. In some scenarios, the reaction rates are needed away from stability, and typically the existing parametrizations are extrapolated without an estimate   of uncertainties (as alluded above).
In this section we discuss a few examples to illustrate the ways in which a global optical potential, with quantified uncertainties and valid away from stability, can benefit the field.

First, we consider neutron capture reaction rates with medium mass and heavy nuclei away from stability. These are thought to be responsible for synthesizing the majority of elements heavier than iron in the universe through the so-called rapid neutron capture process (r-process). The nucleosynthetic path of the r-process involves isotopes that are very neutron-rich and are located  closer to the neutron dripline than the valley of stability, where the optical potential has been more thoroughly tested and constrained (see Fig.~\ref{figX}). The standard optical potential parametrization suggested by Bohr and Mottelson takes into account the large neutron-proton asymmetry away from stability through an isospin-dependent isovector term
\begin{equation}
U_{iso}=\frac{1}{2}\,t_z\,\delta\, U_{sym} 
\end{equation} 
where $t_z$ is the nucleon isospin component, $\delta=(N-Z)/{A}$    and $U_{sym}$ is the so-called symmetry potential.

The symmetry potential 
is currently under active research investigation. Phenomenological and semi-microscopic optical model parametrizations adjust the imaginary potential to agree with experimental data \cite{Voinov2021}. It has been already suggested  in 2007 \cite{Goriely2007} and recently corroborated by experimental evidence \cite{Voinov2021} that the isovector component, constrained by data near stability, does not adequately reproduce the effect of the neutron-proton asymmetry that exists in nature. More  experimental data and theoretical investigations are needed to quantify this deficiency of modern theories. Still, the above works suggest a significant effect on neutron captures relevant to the r-process but also on the less neutron-rich nuclei near stability that may be of interest to nucleosynthesis occurring under conditions with lower temperatures and neutron densities compared to the r-process, such as the i-process. 

\begin{figure}
    \centering
    \includegraphics[clip,trim=0.1cm 0.1cm 0.1cm 0.1cm,width=0.8\textwidth]{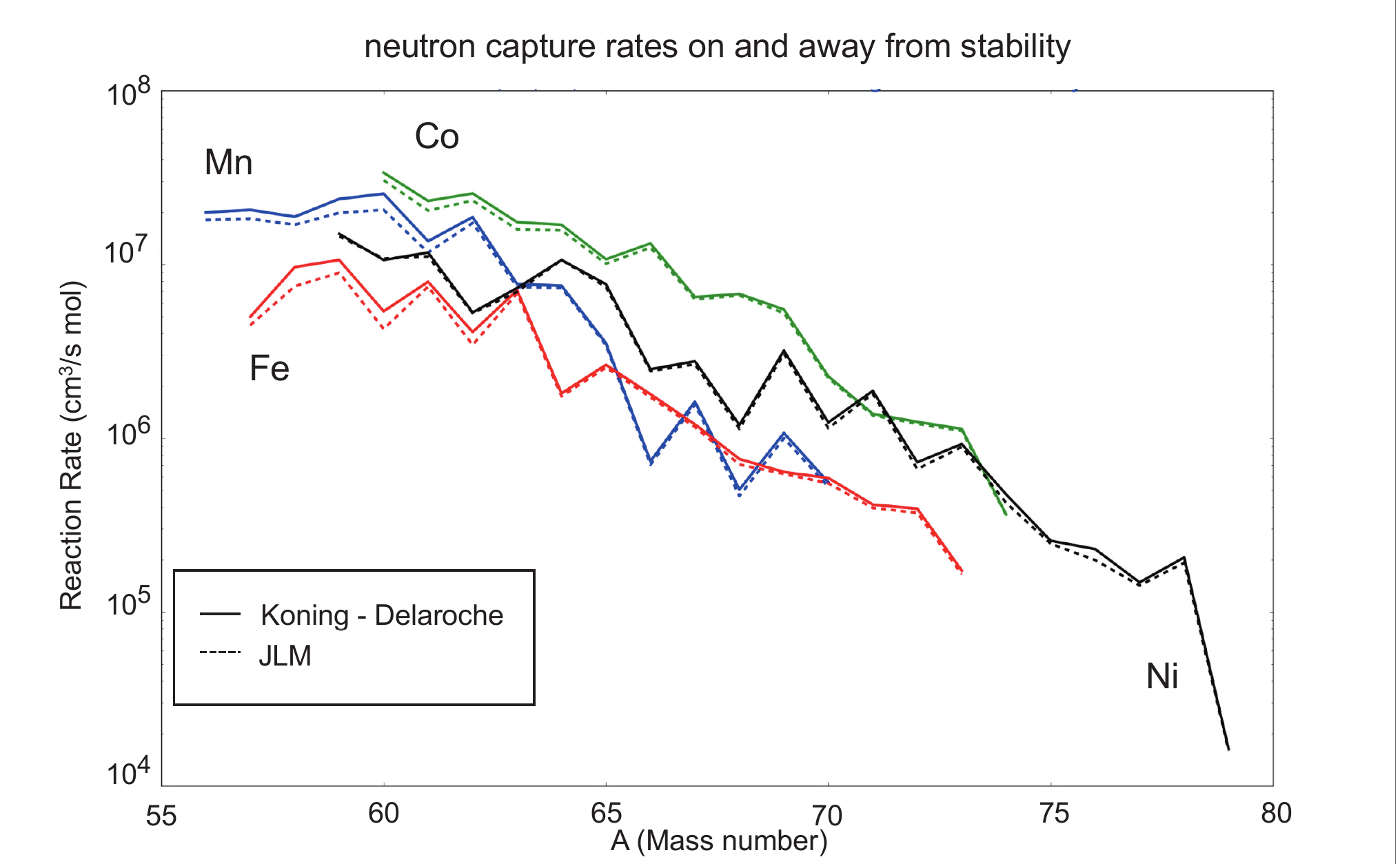}
    \caption{Neutron capture rate calculations using the Koning-Delaroche and JLM optical potentials for Fe, Mn, Co, and Ni isotopes far away from stability. Despite the increasing neutron excess the calculations produce a smooth change with neutron number. Moreover, the resulting reaction rates are in increasingly better agreement with each other as the neutron number increases, suggesting that, if anything, the semi-microscopic JLM potential converges to the phenomenological Koning-Delaroche potential for neutron-rich nuclei.  This suggests that there is no meaningful benefit in using the JLM potential compared to the KD one away from stability.    }
    \label{fig:OMP_comp}
\end{figure}

Significant as the effect of neutron excess may be, current theory does not appear to be sensitive to it. Reaction rate calculations with the typically-used optical potentials developed by Koning and Delaroche (KD) ~\cite{koning03}  and by Jeukenne, Lejeune and Mahaux (JLM)~\cite{Jeukenne77,jeukenne76} produce very similar results even though the latter one is based on a microscopic calculation. In Fig.~ \ref{fig:OMP_comp}, a number of neutron capture rate calculations using the two potentials are shown for Fe, Mn, Co, and Ni isotopes spanning from stability up to several neutrons towards the dripline. Despite the increasing neutron excess, the calculations reproduce a smooth change with neutron number. The difference between the reaction rates computed with KD and JLM decreases for increasing $A$, suggesting  that no difference in the treatment of neutron excess exists between the semi-microscopic and the phenomenological potential away from stability. While we tested only two examples, there is no reason to assume that other potentials based on similar phenomenological approaches would behave differently with increasing neutron excess unless the theoretical description of the effect of neutron-proton asymmetry is fundamentally improved.

Nucleosynthesis in massive stars and in the proton-rich regions predicted to occur in  neutrino-driven winds during core-collapse supernovae involves  rates of $(n,p)$ and ($n$,$\alpha$) reactions at temperatures from approximately 1~GK down to hundreds of MK. In addition, photodisintegration reactions ($\gamma$,$p$) and ($\gamma$,$\alpha$) contribute to the production of the so-called $p$-nuclei, a class of isotopes shielded from the neutron-induced nucleosynthesis processes~\cite{ARNOULD20031,Rauscher_2013} . 
Unfortunately, data in the astrophysically-interesting region below 1~MeV are sparse and of lower precision.  At the same time the focus of most evaluated neutron reaction data libraries is in the broader region of 0-20~MeV, and this energy range translates to very few data at energies below 1 MeV. The fine details of the cross section at these low energies, which can influence strongly nucleosynthesis calculations, are  therefore often not  included.  The accurate  determination of these reactions rates requires  quantitative and predictive nucleon-nucleus optical potentials at energies reaching way below 1 MeV.

\begin{figure}
    \centering
    \includegraphics[width=0.8\textwidth]{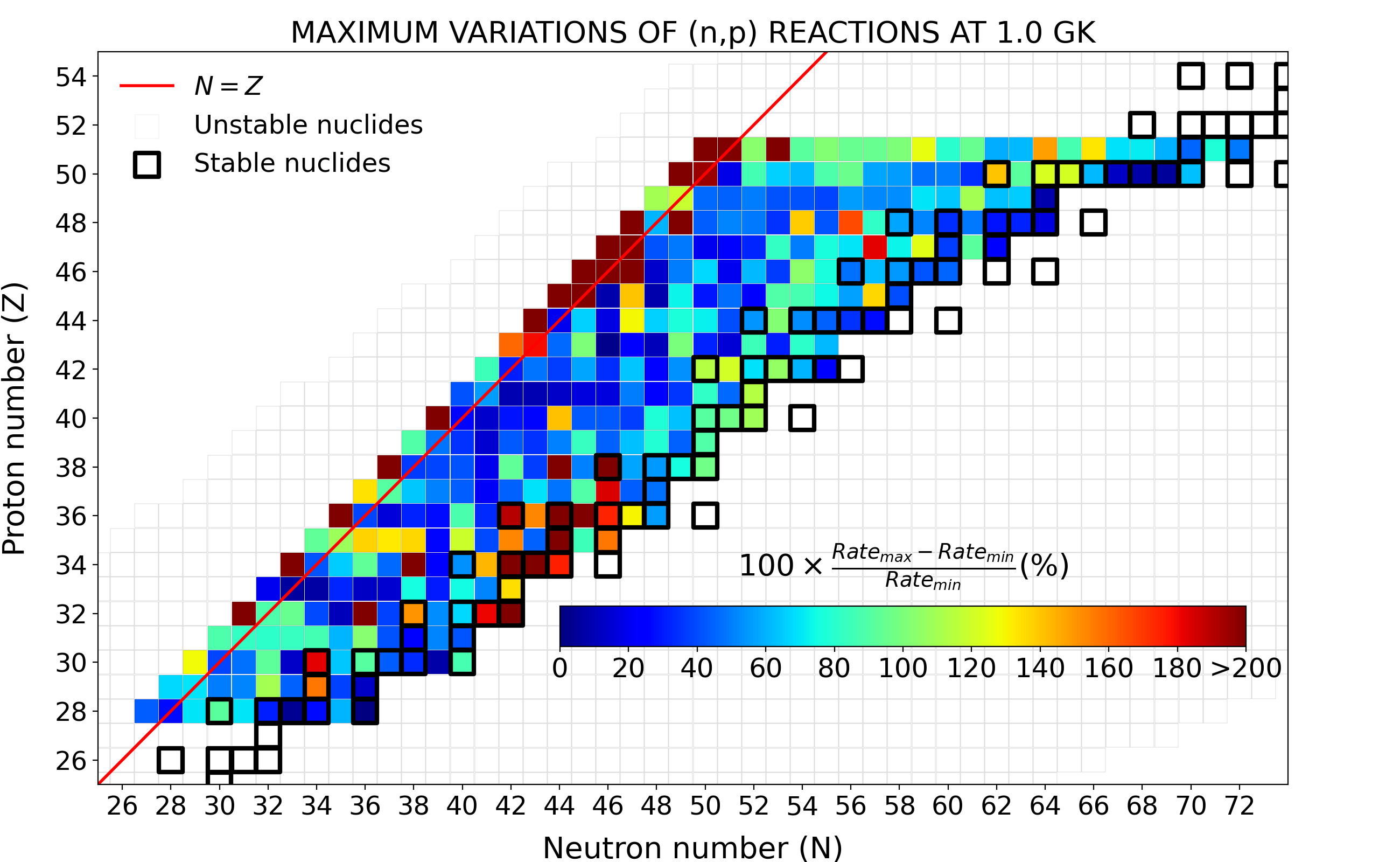}
    \caption{Maximum variation of the reaction rates for $(n,p)$ reactions with isotopes relevant to the neutrino-p process when the JLM scaling parameters $\lambda_x$ are varied within the estimated valid range.} 
    \label{fig:OMP_sens_np}
\end{figure}

We should also note the central role optical potentials play in sensitivity studies, particularly those involving nuclei away from stability. These are essential tools for nuclear astrophysics that guide the experimental and modeling efforts \cite{McKay2020}. Typically, a sensitivity study involves calculating the same astrophysics ensemble multiple times, each time changing the value of a ‘‘parameter" that enters the calculation and recording how the results of the calculation change. In nuclear astrophysics, the most common ‘‘parameters" modified are the values of {\em individual reaction rates} (see for example \cite{Nikas2020}). Ideally, reaction rates should change systematically, connecting with the uncertainties of the underlying theory. 
An example of such a calculation using the currently available optical potentials is shown in Fig. \ref{fig:OMP_sens_np} for proton-rich unstable nuclei relevant for the $\nu$-p process in supernovae. One can note in Fig. \ref{fig:OMP_sens_np} that these uncertainties are not smooth in mass numbers and in isospin asymmetry.

Such sensitivity studies are based on the assumption that the uncertainty of the varied parameters is known to some reasonable degree in order for the result to be valuable. To perform useful sensitivity studies, we need optical potentials that reproduce the changing nuclear structure away from stability and have well-quantified uncertainties. 

Although so far we have focused on nucleon-nucleus optical potentials, astrophysics also has a dire need for global optical potentials on light ions, particularly $\alpha$-nucleus optical potentials.
There are many $\alpha$-induced reactions relevant for nuclear astrophysics due to the heavy abundance of helium in the universe after the Big Bang. 
Chemical elements above $^{56}$Fe are formed either via the slow neutron capture process (s-process) or through the rapid neutron capture process (r-process). The main neutron sources for the s-process occurring in asymptotic giant branch stars are ($\alpha,n$) reactions on heavy nuclei such as $^{13}$C and $^{22}$Ne. In proton-rich explosive stellar environments such as novae and X-ray bursts, the dominant nucleosynthesis ($p,\gamma$) reaction sequence is halted at several waiting point nuclei due to low ($p,\gamma$) reaction $Q$-values resulting in a ($p$,$\gamma$)-($\gamma$,$p$) equilibrium. $\alpha$-capture on these waiting point nuclei allows the nucleosynthesis of heavier nuclei via the ‘‘$\alpha$p-process", a sequence of ($\alpha,p$) reactions followed by ($p,\gamma$) proton captures, which eventually leads to the successive proton captures in the rapid proton capture process ($rp$-process) to synthesize heavier proton-rich nuclei. 

To effectively explain the observed abundances of chemical elements from such stellar environments and nucleosynthesis processes, accurate stellar models are required. These models require nuclear input parameters such as nuclear masses and reaction rates~\cite{Nikas2020,PhysRevC.101.055807,10.1093/mnras/stab772,10.1093/mnras/stz3322} . At present, many of the astrophysically-relevant ($\alpha,\gamma$), ($\alpha,n$), and ($\alpha,p$) reaction rates have not been experimentally constrained within relevant Gamow energies, where measurements are hindered by low counting rates.   The reaction cross sections of  relevant $\alpha$-induced reactions are instead deduced using HF statistical model calculations. 

Statistical HF calculations for $\alpha$-induced reactions require a robust set of $\alpha$-OMPs. Throughout the years, there have been various efforts to determine such OMPs \cite{MCFADDEN66, DEMETRIOU02, Avrigeanu10, Avrigeanu14}. These OMPs are determined by fitting available elastic-scattering angular distribution data. Compared to the amount of scattering data currently available for neutrons, protons and deuterons, the amount of available $\alpha$ scattering data are rather sparse, specially for nuclei further away from stability. This is generally the case for $A=3$ and $A=4$ projectiles compared to   neutrons/protons. Additionally, elastic scattering data are generally obtained for higher energies due to the influence of the Coulomb barrier. The $\alpha$-OMPs by Avrigeanu et al. \cite{Avrigeanu10, Avrigeanu14} has been derived using $\alpha$ scattering data for target atomic masses ranging from 45 to 209. The McFadden \& Satchler $\alpha$-OMPs \cite{MCFADDEN66} have recently been shown to agree with available experimental data within a factor of $\sim$2 for target atomic masses $A = 20 - 50$ \cite{Mohr15}. While these $\alpha$-OMPs can reasonably reproduce experimental reaction cross sections for certain target masses and energies, for experimentally-inaccessible regions of the nuclei chart, they can introduce significant uncertainties for the deduced reaction rates affecting various nucleosynthesis calculations. In order to overcome these difficulties, a cohesive effort by the nuclear physics community is desired to obtain more  scattering data which would aid in significantly improving the nucleon-, deuteron- and $\alpha$-OMPs for more accurate theoretical reaction rate calculations.

\subsection{Nuclear data for energy, security, medical, and other applications}
 \label{Sec2D}
 
A predictive theoretical capability of total, elastic and reaction cross sections is vital for a wide range of applications \cite{Bernstein2019}. Improved optical potential calculations enable sophisticated modeling and guide the experimental efforts necessary to advance the technology readiness level within a given application. Below, we list several specific areas in which more comprehensive and reliable optical potentials would have significant impact.

\subsubsection*{Energy and security applications}~\\

\noindent Most energy and security applications require high-quality evaluated cross sections for neutron-induced reactions up to 20-30 MeV, such as those present in evaluated nuclear data libraries like ENDF/B-VIII.0 \cite{BROWN20181}. The highest-priority targets are actinides and structural/engineering materials. High-quality optical models provide the starting point
for reliable descriptions of reactions needed for modern comprehensive evaluations of neutron induced reactions (e.g., see a recent evaluation of U-238 and U-235 neutron induced reactions \cite{CAPOTE2018254} adopted into the ENDF/B-VIII.0 library \cite{BROWN20181}).
Because of the high societal impact of these applications, many (but not all) relevant cross sections have been thoroughly examined, and sophisticated uncertainty quantification techniques developed and applied \cite{Capote:2010,Koning2015,Capote:2020}. The extensive high-quality experimental data collected to inform the relevant cross sections mean that phenomenological models are in their range of validity, provided sufficient physics -- such as selection of the appropriate reaction model and detailed structural information -- are considered.

There are still needs for OMP development for materials used in next-generation reactor architectures. In some designs such as those involving molten salts, potentials are needed for isotopes beyond typical structural materials and actinides. Experimental data suitable for improving a phenomenological potential on these isotopes are not always available, especially because the bulk of single-nucleon scattering measurements were conducted more than thirty years ago. Also relevant are reactions on light elements which are important both in their own right and because they can provide important constraints on reverse reactions that are difficult to access experimentally but important for neutron economy (e.g., $^{16}$O$(n,\alpha)^{13}$C to improve knowledge of $^{13}$C$(\alpha,n)^{16}$O or vice-versa).

\subsubsection*{Medical applications}~\\

\noindent An important application is the production of  isotopes used  in medical diagnosis and treatment {\cite{Nichols:2014}}.   The two major methods for isotope production are 1) reactor-based fission, capture, and $(n,p)$ reactions and 2) charged-particle-induced reactions using fast charged-particle and/or neutron beams~\cite{tecdoc1211,trs473,PAetal19} .
  For reactor-based isotope production, neutron-induced reaction cross sections similar to those in the energy and security applications are important, so there can be significant overlap in the theoretical and experimental tools needed to constrain these cross sections \cite{trs473}. For charged-particle-induced reactions, much higher energies are required: up to tens of MeV in commercial cyclotrons such as those found in hospital settings, and up to hundreds of MeV for dedicated accelerator facilities such as Los Alamos Neutron Science Center (LANSCE) . 
  
  Isotope production requires not only nuclear physics information which can be partially obtained via an optical model, but also efficient chemical separation techniques that allow the produced   isotope to be extracted. As such, the most important reactions  are those that change the number of protons and thus the chemistry of the target, facilitating chemical separation and increasing the specific activity of the product. For charged-particle beams, the charge-exchange reactions $(p,xn)$, $(d,xn)$, $(\alpha,xn)$ are the most important \cite{Hermanne:2018,Hermanne:2021,Tarkanyi:2019a,Tarkanyi:2019b,Engle:2019}. In a similar vein, the most important neutron-induced reactions in the reactor-based setting are $(n,p)$, fission, or multi-step reactions with subsequent $\beta$- or $\alpha$-decay, yielding a change in element of the reaction products \cite{trs473}. In some cases a direct production of the parent radionuclide is feasible (e.g., $^{100}$Mo$(n,2n)$ reaction is used to produce $^{99}$Mo parent for production of the very important $^{99}$Tc generators) \cite{Tarkanyi:2019a}. 
  In each of these cases, secondary particle reactions often contribute significantly to reaction yields, compounding the
importance of proper reaction modelling to consider all relevant channels.   Because of the inherently complicated nature of these reactions, it should be noted that the OMP provides only the first ingredient required for reliable predictions of the needed cross sections, and that additional information -- such as from a statistical reaction and pre-equilibrium model -- is essential. As currently-available OMPs have been developed  to describe at  $p$, $n$,    $d$, or $\alpha$, several OMPs across a  wide range of energies may be needed simultaneously to describe secondary particle production and follow-on reactions~\cite{ripl3}.
  
  A further medical application is radiotherapy, where cross section information at energies up to 250 MeV/nucleons are important. For example, in a proton radiotherapy procedure, the dose delivered in the beam entrance region (before the Bragg peak) depends heavily on the elastic scattering of protons on tissue, which is one of the easiest quantities to cleanly predict using a suitable OMP and reaction code. However, this information must be combined with atomic and radiological data that are often absent or highly uncertain. As is the case in compound nuclear reactions that combine nuclear structure and OMP information, knowing the relative uncertainty of the OMP versus other reservoirs of uncertainty (e.g., atomic data) can help focus efforts on reducing the most impactful uncertainties first. There are also important quantities that the OMP cannot fully inform, for example, activation data or the production cross section for positron-emitting radionuclides (e.g., $^{11}$C). As uncertainty quantification is still developing for many other types of physics data entering medical and energy/security applications, better OMP uncertainty assessments can potentially provide a methodological guidepost for other fields.

\subsubsection*{Space applications}~\\

\noindent The interaction of radiation with space-based systems is important to understand for both national security and industry. Radiation shielding in space is a balance between maximizing protection and minimizing the amount of shielding material, as there is a strong cost motivation to minimize the overall weight of the system~\cite{Slabaetal16}. In addition, radiation effects on electronic systems are often difficult to directly study in space-like environments~\cite{QUARTEMONT2021165777}. There is sometimes a dearth of experimental data to inform the relevant models in both these cases~\cite{Slabaetal16}. Much like in other applications outlined above, secondary particle production contributes significantly to the relevant doses. Knowledge of neutron production from proton-induced reactions is very important and double-differential cross sections of produced neutrons may be needed. In addition, gamma rays produced by inelastic scattering can potentially have a large impact on overall dose delivered to a space-based system (for example, an electronics circuit) by incoming neutrons~\cite{QUARTEMONT2021165777}. Given that these reaction cross sections are often poorly known, predictive and uncertainty-quantified OMPs, in concert with other theoretical inputs, can have a significant impact. In particular, a recent survey of nuclear data needs for space radiation protection has identified significant shortcomings in experimental data \cite{Norbury2012}. Previous work has been done using optical models to predict nuclear fragmentation processes relevant for space applications \cite{deWet2020, Ramsey1998}. 
 
\newpage
\section{Review of strategies to build nucleon-nucleus optical potentials}\label{Sec3}

\subsection{Standard and dispersive phenomenological approaches}
\label{Sec3A}

\subsubsection{Standard optical potentials}~\
\label{S3.1}

\vspace{0.2cm}
\noindent Early developments of optical potentials sought to describe single-nucleon cross sections phenomenologically, using a local, complex, one-body potential~\eqref{kdpar}--\eqref{kdpar2} with parameters varying smoothly   in $E$ and $A$, analogous to the refractive index for an absorptive medium. By the 1960s, enough cross section data had been collected to contemplate training a ``global'' optical potential suitable for predicting elastic scattering cross sections across a broad range of nuclei and energies, with Becchetti and Greenlees the first to do so \cite{becchetti69}. Increases in computational power and the size of experimental databases led to potentials with a growing number of free parameters and better empirical performance, including the spherical CH89 \cite{varner91} ($40\leq A$, $10\leq E\leq65$~MeV) and the Koning and Delaroche \cite{koning03} ($24\leq A\leq 209$, $1$~keV$\leq E\leq200$~MeV)   potentials, which remain widely used. Most recent phenomenological efforts include additional physics, such as enforcing dispersivity, including non-locality, and/or describing deformation via a coupled-channels approach, each of which can improve the accuracy needed for applications \cite{ripl3}. To facilitate interpretation of experimental data sensitive to nuclear asymmetry, e.g., quasielastic charge-exchange reactions, many single-nucleon potentials adopt a Lane-consistent form wherein the OMP is partitioned into isoscalar and isovector components such that the neutron and proton OMPs differ only by a change in sign of the asymmetry-dependent components \cite{Baug2001, danielewicz_symmetry_2017}. Besides the high-visilibity global efforts, over the decades many hundreds of experimental papers have included isotope- or region-specific optical-potential analyses to assess the impact of their newly collected data (for example, \cite{Guss1989}). However, the basic formula for developing a phenomenological OMP remains essentially unchanged since the 1950s: select a suitable collection of functional forms dependent on $A$, $E$, and/or $\delta$, compile experimental scattering data, predict reaction observables given an OMP, and optimize OMP parameters according to $\chi^{2}$ minimization.

As an example, consider the Koning-Delaroche OMP \cite{koning03}, one of the most widely used OMPs since its introduction in 2003. The nominal range of validity in energy is 1 keV to 200 MeV. Both a global and several local versions are available, spanning near-spherical systems with 24~$\leq$~A~$\leq$~209. To train the OMP, the authors used hundreds of proton and neutron differential elastic scattering and analyzing power data sets from 27~$\leq$~A~$\leq$~209 collected from the 1950s to the 1990s, as well as proton reaction cross section data and neutron total cross section data on natural and isotopic targets. The potential itself includes six subterms with Woods-Saxon-like radial dependence separated from their energy dependence with a total of forty-six free parameters. Optimization was done using a combination of ``computational steering'' and $\chi^{2}$ minimization, whereby a user manually guided potential parameters until predicted cross sections were visually close to experimental values, then invoked a simulated annealing algorithm to find a parameter-vector optimum. The success of the KD global OMP in reproducing its training data, particularly the elastic-scattering   angular distributions and neutron total cross sections above the resolved-resonance region, as well as its ease of use, have led to widespread adoption as a default OMP in reaction codes such as {\sc talys}~\cite{TALYS1,TALYS2} and Finite Range with Exact Strong COuplings ({\sc fresco}) ~\cite{FRESCO}.

There are limitations of this approach that are discussed in more detail in Sec.~\ref{Sec3C}.

\subsubsection{Dispersive optical potentials}~\
\label{sec:dom}

\vspace{0.2cm}\noindent The Dispersive Optical Model (DOM), first introduced by Mahaux and Sartor~\cite{mangsa86,masa91,masa91rev}, is an optical model which makes use of a dispersion relation~\eqref{eq:full_dispersion}  
that relates the imaginary part of the potential to its real part over all
energies. 
Equation~\eqref{eq:full_dispersion} is a very powerful constraint that provides a variety of advantages over non-dispersive optical models, helping to reduce the number of model parameters (e.g., see Ref.~\cite{Molina:2001}) and to achieve a  better description of neutron induced cross sections for energies below $\sim 5$ MeV \cite{Morillon:2004} compared to traditional OMPs like Koning-Delaroche \cite{koning03}. Dispersion integrals given by Eq.~\eqref{eq:full_dispersion} can be solved numerically \cite{Capote:2001} or analytically for selected imaginary potentials \cite{masa91rev,Quesada:2003a,Quesada:2003b}.  
Several different dispersive OMPs have been derived (starting from Eq.~\eqref{eq:full_dispersion}) for a variety of use-cases. One class of dispersive OMPs was derived to describe deformed nuclei assuming a rigid-rotor structure like Rh, Au, W, Ta, Hf and actinides~\cite{Capote:2005,Soukhovitskii:2005,Capote:2008}. Another was derived for spherical nuclei that are soft relative to vibrations assuming a soft-rotator structure like Fe, Ni, Cr~\cite{SRM-CrNiFe:2013} and Zr~\cite{ripl3}.  Yet another class of dispersive coupled-channel potentials has been used to describe both elastic and inelastic scattering data in a broad energy range up to 200~MeV.
These derived dispersive OMPs~\cite{Soukhovitskii:16,Quesada:2007,Quesada:2013,Zhao:2020} have been shown to be approximately Lane consistent, i.e., the same OMP holds for incident neutrons and protons~\cite{Lane:1962a,Lane:1962b} and the parametrization becomes isospin dependent.  
 The dispersion relation has also been used to provide a consistent description of both bound and scattering data in spherical nuclei~\cite{Dickhoff17,Mahzoon:2014} in some cases allowing ``data-driven'' extrapolations to the drip lines~\cite{charity06,charity07}.

We now briefly discuss efforts that augment the data set constraining the OMP to negative energies by taking in structure information (charge density, energy levels, particle number, etc.), in addition to the
elastic-scattering data corresponding to positive energies. 
As mentioned in the Introduction, the optical potential can be interpreted as the irreducible self-energy  $\Sigma^\star(r,r';E)$, in the Green's function formalism.  Moreover, $\Sigma^\star(r,r';E)$  generalizes the exact
nucleon–nucleus optical potential of Feshbach to include both bound and scattering states (a more detailed discussion can be found in Sec.~\ref{Sec3B2}) . Connecting the optical potential to the Green's
function, along with utilizing the dispersion relation in Eq.~\eqref{eq:full_dispersion}, allows for a complete description of the nucleus over both the positive- and negative-energy domains~\cite{Dickhoff17,Mahzoon:2014}.  Adjusting OMP parameters to describe data using Eq.~\eqref{eq:full_dispersion} guarantees that the irreducible self-energy stays well
defined~\cite{Mahzoon:2017,Atkinson:2018}. Currently, there are DOM fits using this Green's function formalism for spherical targets $^{16,18}$O, $^{40,48}$Ca, $^{58,65}$Ni, $^{112,124}$Sn, and $^{208}$Pb for
$-200 \textrm{ MeV}<E<200 \textrm{ MeV}$~\cite{Mahzoon:2017,Atkinson:2018,Pruitt:2020,Atkinson:2020}.
  In principle, a dispersive optical potential can be applied in the same mass-number and energy range as   a typical non-dispersive potential (such as KD) can. Work is currently underway to implement a global parametrization of a fully-dispersive optical potential for spherical targets~\cite{Pruitt22}. 

Noting that Hartree-Fock potentials are already inherently nonlocal, it was demonstrated
that spatial nonlocality of the self-energy including its imaginary part, must be treated explicitly in order to describe properties 
below the Fermi energy~\cite{Mahzoon:2014}.
To satisfy the dispersion relation in Eq.~\eqref{eq:full_dispersion}, it is at present
assumed that the energy dependence of the imaginary part is the same for all spatial coordinates, which simplifies the numerical effort. Typically, optical potentials approximate the spatial
nonlocality with an energy dependence~\cite{perey62}. However, this added energy dependence does not satisfy the dispersion relation, which would lead to an incorrect description of the negative-energy
observables. Thus, the spatial nonlocality is treated explicitly with the so-called Perey-Buck form~\cite{perey62}
\begin{equation}
   \mathcal{U}(\vec r,\vec r';E) = U\left(\frac{r+r'}{2};E\right) e^{-\frac{(\vec {r}-\vec {r}')^2}{\beta^2}}\pi^{-3/2}\beta^{-3},
   \label{eq:nonlocality}  
\end{equation}
where $\beta$ is a nonlocality parameter which controls how much strength is distributed off the diagonal. It is worth noting that the functional form in Eq.~\eqref{eq:nonlocality} is chosen out of convenience, but is capable to represent essential features of microscopic potentials~\cite{Waldecker:2011,Dussan:2011}. With this treatment of the nonlocality, along with the dispersion relation in its subtracted form, quantities such
as particle numbers, charge densities, and ground-state binding energies are included in the DOM fit. This allows for data-informed predictions of quantities such as the neutron skin of $^{48}$Ca and $^{208}$Pb (see Refs.~\cite{Mahzoon:2017,Pruitt:2020,Atkinson:2020,Pruitt2020PRC} for more details).

\begin{figure}
   \includegraphics[width=\textwidth]{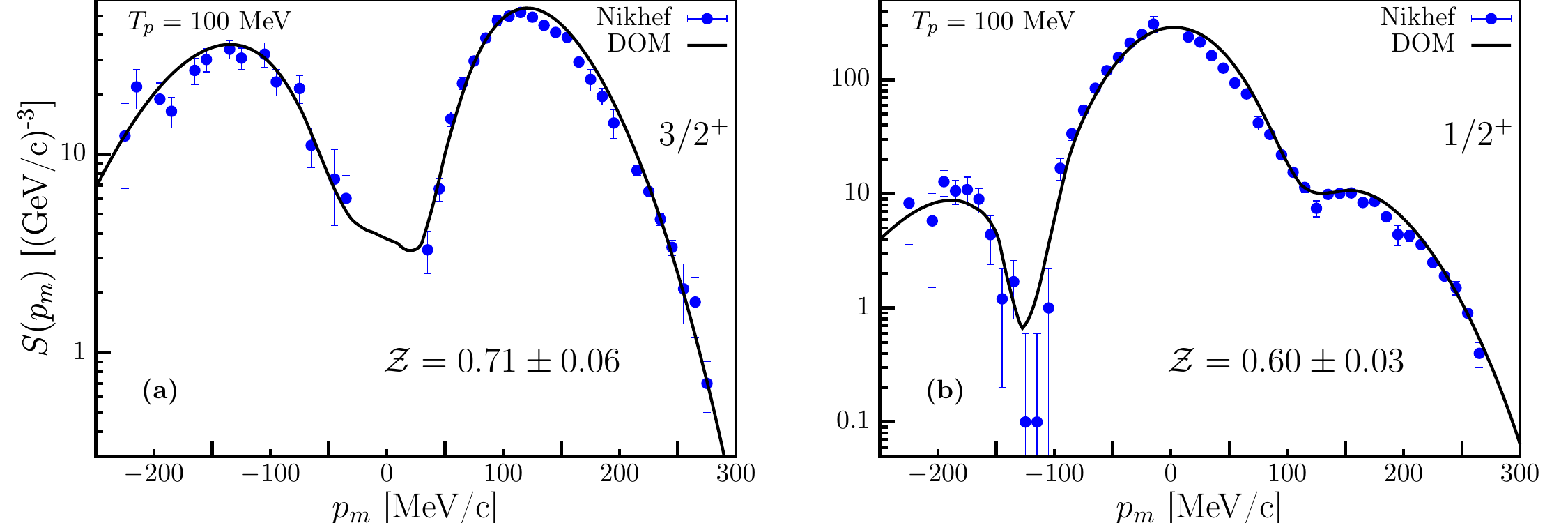}
      \caption{$^{40}$Ca$(e,e'p)$ $^{39}$K spectral functions in parallel kinematics, at an outgoing proton kinetic energy of 100 MeV. The solid line is the calculation using the DOM ingredients, while the points are from the
      experiment detailed in~\cite{Kramer:1989}.
      (a) Distribution for the removal of the $0{d}\frac{3}{2}$ . The curve contains the DWIA for the $3/2^+$ ground state including a spectroscopic factor of 0.71. (b) Distribution for the
      removal of the $1{s}\frac{1}{2}$   proton with a spectroscopic factor of 0.60 for the $1/2^+$ excited state at 2.522 MeV. The figure is adapted
      from Fig. 5 of  Ref.~\cite{Atkinson:2018}.}
   \label{fig:eep100}
\end{figure}

The ability to describe both bound and scattering states of a nucleus is particularly useful in the description of stripping, transfer and knockout reactions needed to fully utilize FRIB~\cite{Petal17,GFK}. As a specific example, we briefly present the DOM
calculation of $^{40}$Ca$(e,e'p)^{39}$K cross sections~\cite{Atkinson:2018}. This reaction, measured at NIKHEF~\cite{Kramer:1989}, can be described using a distorted-wave
impulse approximation (DWIA) which assumes that the virtual photon exchanged by the electron couples to the same proton that is detected and that the final-state interaction can be described using an
optical potential~\cite{Giusti:1988,Boffi:1980}. The ingredients of the DWIA therefore require a distorted wave describing the outgoing proton at the appropriate energy and an overlap function
for the removed proton and its associated spectroscopic factor. The DOM allows for a consistent DWIA analysis in that the bound state wave function, spectroscopic factor, and outgoing proton distorted
wave can all be provided from the same self-energy. The resulting momentum distributions, shown in Fig.~\ref{fig:eep100}, came straight from the DOM self-energy - the $^{40}$Ca$(e,e'p)^{39}$K data was
not used in the DOM fit. The spectroscopic factors coming directly from the DOM self-energy show a good description
of the data - thus updating the previously obtained spectroscopic factors (found by scaling the previous DWIA analysis to match the data points). Furthermore, this analysis was also done for 
$^{48}$Ca$(e,e'p)^{47}$K providing a new perspective on the quenching of spectroscopic factors~\cite{Aumann21,Atkinson:2019}. 
These results demonstrate how, provided that sufficient elastic-scattering and structure data is available to constrain the fit, the DOM potential applied in the appropriate reaction theory is a powerful way to consistently describe knockout reactions and aspects of transfer reactions relevant to FRIB science.

\subsubsection{Uncertainty quantification}~\

\vspace{0.2cm}
\noindent Uncertainty quantification (UQ) for OMPs is an emerging topic, with the majority of publications on the topic dating from the last five years. The earliest systematic attempt at OMP UQ was by the authors of the Chapel Hill global OMP \cite{varner91}, who used a bootstrap method to assess that the variances of parameters in their OMP were very small (on the order of a percent). An attempt to estimate uncertainties of potential parameters was discussed at the IAEA RIPL project, where rough estimates of geometry and potential depth uncertainties were given \cite{ripl3}. Recent analyses using Bayesian techniques\cite{Lovell2020,Koning2015, Pruitt2020PRC,King2019,Phillips_2021} reveal much larger uncertainties (tens of percent for elastic-scattering observables and up to a hundred percent for single-nucleon transfer cross sections) indicating that the statistical assumptions used for training phenomenological potentials can impact predictions as strongly as the data used for training. For a long time,    due to the absence of global, readily employed OMPs, with well-calibrated parametric uncertainties, OMP users resort either to tuning OMP parameters by hand or to performing ad hoc UQ by comparing predictions from multiple OMPs -- neither of which is easily extrapolated to the high-asymmetry regime that will be probed at FRIB. Recently,  Pruitt, Escher and Rahman have developed extension of the global spherical proton and neutron OMPs of KD \cite{koning03}  and CH89 \cite{varner91} with uncertainty quantification of the potential parameters, the so called KDUQ and CHUQ \cite{Pruitt:22_report}. The mass and energy range of validity are the same as the  original KD and CH89, i.e  $24<A<209$ and $0.001 \text{ MeV }<E<200 \text{ MeV}$  for KDUQ and   $40<A<209$ and   $10 \text{ MeV }<E<65 \text{ MeV}$ for CHUQ. This new development will support the quantification of uncertainties in reaction observables.  A natural next step  to improve the constraint of these parametrizations is to enforce dispersion relations, include data on highly asymmetric systems and bound observables.  

\subsection{Microscopic approaches}
\label{Sec3B}
As previously discussed, the phenomenological models have been built using experimental data primarily from stable nuclei. Hence, it is uncertain whether the extrapolations of phenomenological optical potentials to unstable isotopes would be reliable. 
For this purpose, optical potentials based on microscopic or semi-microscopic nuclear structure calculations prove to be vital.
This section contains a description of various recent attempts to link the OMP to the underlying interaction between nucleons in free space. 
All approaches have strengths and weaknesses as well as   limitations to their applicability which are discussed below.

\subsubsection{Constructing Green's function from beyond mean-field approaches: Feshbach formulation, nuclear structure model and optical potentials from  effective Hamiltonians}~\ 
\label{Sec3B1}

\vspace{-0.8cm}
\paragraph{Feshbach formulation.}
One of the approaches used to integrate microscopic nuclear structure information into the construction of optical potentials is by using the Feshbach formulation \cite{feshbach1958unified}, where the optical model potential for a nucleon scattering energy $E$ is given  by Eq.~\eqref{vefff}.  

\begin{figure}
    \centering
    \includegraphics[width=0.9\textwidth]{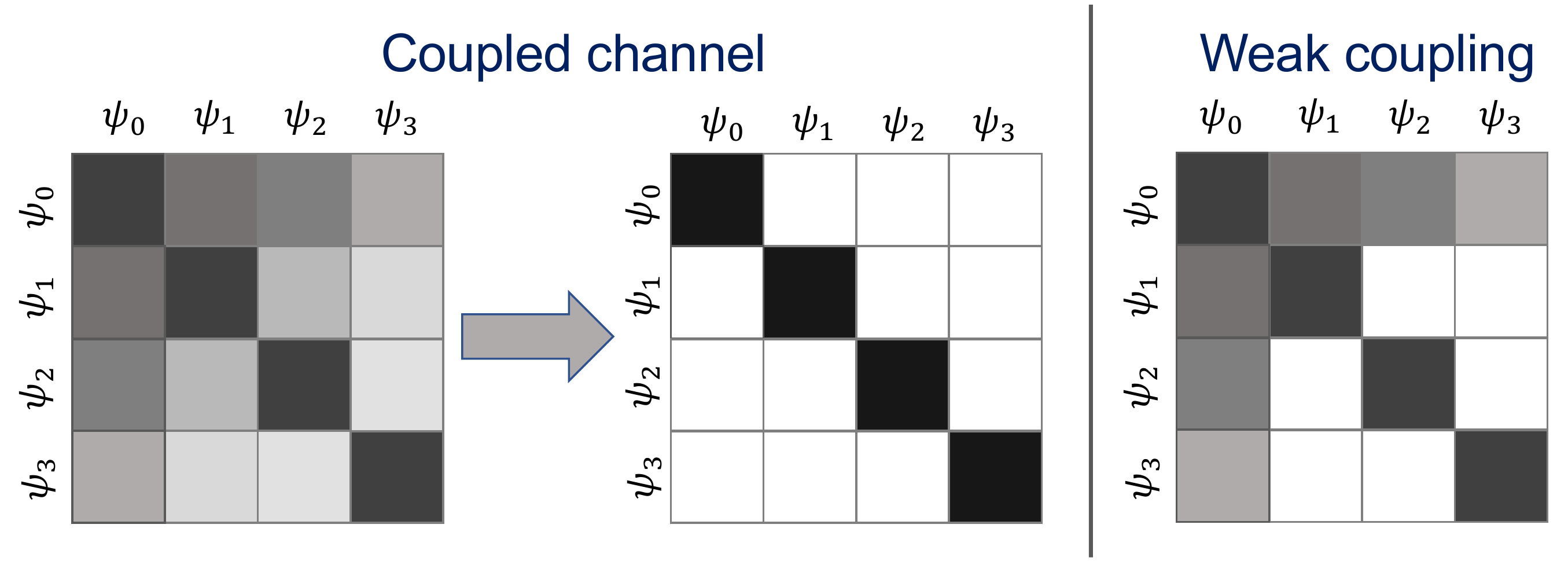}
    \caption{Schematic illustration of the obtention of the ``arrow'' matrix from the weak coupling approximation. $\psi_i$ correspond to the many-body states of the target, coupled together by the incoming nucleon. }
    \label{fig:arrow}
\end{figure}

Equation~(\ref{vefff}) can be obtained from the set of coupled differential equations written in terms of the reaction channels \cite{feshbach1958unified}. 
By solving this set of equations, the Green's function matrix  $G_{jk}$   can be diagonalized in the space of  the excited states $j,k\neq0$, but  one can also use the weak coupling approximation \cite{feshbach1958unified,rao_73, lev_74a,lev_74b,coulter_77}, which neglects the couplings between excited states, i.e. ${\cal V}_{jk}={\cal V}_{0k}\delta_{j0}$ . In this case the coupling potential is expressed in terms of an ``arrow'' matrix (Fig.~\ref{fig:arrow}), and Eq.~\eqref{vefff} becomes  
\begin{equation}
  V_\mathrm{opt}={\cal V}_{00} + \sum_{j \neq 0} V_{0j}G_{jj} {\cal V}_{j0},
  \label{eq:OP1}
\end{equation}
where only the diagonal elements of the Green's function~\eqref{eq:8i} enter.  
The coupling potentials and the Green's functions in the second term of Eq.~\eqref{eq:OP1} can be provided by nuclear structure calculations. 

The method works best for the low-energy region where nuclear structure models can provide reliable and converged calculations (approximately $< 50$ MeV). The versatility of this approach lies in the variety of structure calculations it can accommodate. However, the description of compound nucleus reactions, in which the relevant target-nucleon states are statistical in nature, is an open challenge.

\paragraph{Nuclear Structure Method.}
\indent Another method is to construct the potential from a phenomenological effective NN interaction using the Green's function formalism. It is called the Nuclear Structure Method (NSM). The Green's function formalism allows the hierarchization of correlations and avoids double countings. Antisymmetrization due to the fermionic nature of nucleons is taken into account. The NSM was first proposed for realistic NN interactions by N.~Vinh~Mau in the early 1970s \cite{vinhmau_70}. Then it has been recast in order to be used 
with density functionals such as Skyrme or Gogny \cite{bernard1979microscopic}. \\
\indent When dealing with a spherical target nucleus (without pairing), the NSM potential is made of two contributions: the Hartree-Fock potential and the Random Phase Approximation  potential. The mean-field term is energy-independent. Its exchange term (Fock term) turns out to be nonlocal when using a finite-range interaction. The RPA contribution is a polarization contribution. The absorption results from taking into account the coupling to inelastic channels when the target nucleus is excited. Such excitations are described in the RPA formalism. This term is non-local, energy dependent and complex. The NSM can be interpreted as a Feshbach potential with consistent ingredients as the same functional is used all along 
the calculation. Hence in Eq.~\eqref{eq:OP1}, $V_{00}$ would be the equivalent of the Hartree-Fock potential in the NSM whereas $V_{0i}$'s would be provided by RPA. 

A pioneering application of the method at lowest order with the Skyrme functional has demonstrated the ability of the Hartree-Fock potential to grasp the main features of the real part of the optical potential below 50~MeV incident energy \cite{dover_72}. Follow-up studies have included second-order terms with particle-particle (pp) and particle-hole (ph) correlations. The pp correlations are implicitly contained in the phenomenological NN interaction. Particle-hole correlations are then taken into account explicitly through the Random Phase Approximation (RPA). In the beginning of the 1980s, several groups have worked on the NSM and related methods mostly using Skyrme interactions \cite{bernard1979microscopic,vinhmau_76,bouyssy_81}. More refined versions of the NSM have been proposed including both inelastic excitations and $(n,p)$ charge exchange  
\cite{osterfeld_81,osterfeld_81b,osterfeld_85}. The NSM has been used as well 
to determine $\alpha$-nucleus potentials \cite{lassaut_82,dermawan_82,leeb_85}. 
In the last decade, there has been a renewed interest in the NSM with Skyrme \cite{hao_15,hao_18,hao_20} and Gogny interactions \cite{blanchon_15,blanchon_15b,blanchon_17}. The NSM describes with relative success, for both neutron and proton projectiles, the scattering off target nuclei such as: 
$^{16}$O \cite{hao_20,mizuyama_12c,Hao:15} , $^{40}$Ca \cite{vinhmau_76,osterfeld_81,hao_20,blanchon_15,blanchon_17} , $^{48}$Ca~\cite{hao_20,blanchon_17}  and $^{208}$Pb \cite{hao_20,Hao:15} , for incident energy below 50~MeV.
A calculation fully handling continuum and self-consistency has been proposed with Skyrme interactions \cite{mizuyama_12c}. The method has been applied to 
neutron scattering off $^{16}$O below 30~MeV. This approach is particularly interesting because it allows one to circumvent the pitfall of RPA calculations in a harmonic oscillator (HO)   basis that requires the introduction of \textit{ad hoc} escape and damping widths \cite{blanchon_15}. Methods close to the NSM have also been used to describe proton inelastic scattering \cite{Mizuyama:14}.

The NSM works well for incident energies below 50~MeV. Thus the method is complementary to g-matrix approaches in terms of energy range. In its current versions, it is limited to target nuclei well described within RPA, typically double-closed shell nuclei. However, the extended reach of energy density functional  based on structure calculations (pairing, deformation and odd number of nucleons) \cite{hilaire_07} suggests that further versions of the NSM  will be suitable for a wide range of target nuclei. Some recent attempts have extended the approach to scattering off target nuclei with pairing using Hartree-Fock-Bogolyubov (HFB) formalism \cite{mizuyama_19,mizuyama_21}. The NSM will then be extended to include pairing correlations within the HFB formalism with Quasiparticle-Random-Phase Approximation (QRPA) on top of it. These new developments will eventually allow for the description of nucleon scattering off deformed target nuclei with pairing. 

We mention here also approaches that employ Skyrme \cite{kuprikov_06,kuprikov_09,pilipenko_10,pilipenko_12,xu_13} or Gogny \cite{lopez_21} functionals for infinite nuclear matter (see more extensive discussion of such methods in Sec.~\ref{Sec3B4}). The optical potential 
is then obtained using the local density approximation (LDA) with a 
consistent density. This approach allows a satisfactory description of the elastic scattering observables for energies up to 100~MeV. 
In this context, there have been several attempts to fit new Skyrme 
functionals adding scattering constraints to the more usual structure ones~\cite{pilipenko_10,pilipenko_12,xu_13,PhysRevC.96.024621} .

\paragraph{Optical potentials from effective Hamiltonians.}
While it is alluring to use a single interaction (e.g., Chiral) or pseudo--interaction (e.g., Gogny or Skyrme functionals) to construct the nuclear structure properties for the ground state, excitations, and subsequently the optical potential, it is also possible to combine effective Hamiltonians and interactions without lack of generality. This strategy has the advantage of potentially reducing computational costs or increasing the many--body expansion with respect to a calculation that treats every component on equal footing.

Over the years, several approaches have used different effective interactions and microscopic methods to construct the optical potential. The approach based on Hartree--Fock \cite{Dover:71,bernard1979microscopic} consists in defining the real part of a local potential based on an appropriate Skyrme functional. It is extended with additional couplings and absorption \cite{Hao:15,LeAnh:21}. Eventually, one can consider the optical potential as arising from particle--vibration coupling, even extended in the continuum \cite{Mizuyama:14}. Another prominent example is the nuclear field theory, where single-particle and collective degrees of freedom are combined, eventually employing phenomenological coupling instead of a single consistent Hamiltonian or functional \cite{Bortignon:77}. The nuclear field theory have been lately expanded, both in functional form \cite{Idini:13}, and using an effective coupling between multipolar vibrations and mean field \cite{Idini:12}.
The coupling between degrees of freedom has been recast as the solution to the Dyson equation \cite{idini2011dyson}, closely relating several structure and reaction observables \cite{Potel:13b, Idini:15, Broglia:16, Potel:17}. However, it is difficult for an explicit Dyson procedure to treat cases where symmetry breaking is prominent, e.g.\ deformed nuclei, both in terms of guaranteeing symmetry restored final states and adequate computational costs. 

The generator coordinate method (GCM) can tackle symmetry restoration in both even and odd nuclei. It is also appealing for its analogy with the resonating group method \cite{Friedrich:81}. Developing effective Hamiltonians has the advantage of both formal consistency within the projection procedure that is difficult to maintain using functionals and simplifying the numerical calculations, all with excellent agreement with experimental structure observables \cite{Ljungberg:22}. The GCM has been used to calculate scattering properties of stable and exotic nuclei in several cases \cite{Friedrich:81,Wintgen:83,lukyanov2014using}. There is further work ongoing in connecting the microscopic structure description in the GCM and the reaction observables in the form of the construction of microscopic optical potential for deformed nuclei.

\subsubsection{Computing the self-energy from \textit{ab initio} predictions of nuclei: Self Consistent Green's function and inversion of propagator using \textit{ab initio} wavefunctions}~\ 
\label{Sec3B2}

\vspace{0.2cm}
\noindent As discussed in previous sections, the irreducible self-energy $\Sigma^\star(E)$  generalizes the exact nucleon--nucleus optical potential of Feshbach to include both \emph{bound} and \emph{scattering} states \cite{mangsa86,Escher:02,Dickhoff19}. Diagonalizing the self-energy leads to the one-body Green's function, also known as propagator (see Eq.~\eqref{eq9}). Hence, many-body Green's function theory provides a well grounded connection between structure and  reactions  \cite{Dickhoff08}
and it enables the
direct computations of the self-energy based on the best accurate \textit{ab initio} methods.
We present the state-of-the-art frontiers and challenges of the Green's function approach in the following.

\textit{Ab initio} computations of the self-energy for finite nuclei can be approached in two ways: a) either by direct application of propagator theory to calculate its Feynman diagram expansion, as done in the Self Consistent Green's Function (SCGF)~\cite{Dickhoff04,Barbieri2005op,Barbieri:17}; or b) inverting the propagators computed using an \textit{ab initio} wave function approach,  as done in the coupled-cluster method (CCM)~\cite{Rotureau2020Front}, the No-Core Shell Model (NCSM), and the Symmetry-Adapted No-Core Shell Model (SA-NCSM)~\cite{Burrows21}. 
\textit{Ab initio} methods can construct the optical potential from chiral interactions, or other effective field theory (EFT)   forces, so that they provide a direct link to the underlying symmetry and symmetry-breaking patterns of quantum chromodynamics. They also provide a systematic approach to quantify theoretical uncertainties arising from the nuclear force and the controlled many-body  approximations.  Typical calculations for finite nuclei involve large but truncated model spaces that lead to a discretization of the scattering continuum. In most cases, the greatest challenges relate to dealing with such discretization and to handling the large number of degrees of freedom needed to resolve the dynamics at several scattering energies~\cite{Soma2014GkvII,Hagen_2014}; a very demanding task if compared to typical successful \textit{ab initio} computations of low-energy nuclear structure.

\paragraph{Direct computations of the self-energy.} The SCGF approach involves computing a converged series of Feynman diagrams based on prescriptions that are aimed at preserving conservation laws. Modern nuclear physics applications exploit the Nambu-Gorkov formulation to include pairing in spherical open-shell nuclei~\cite{Soma2011Gkv} and follow the algebraic diagrammatic construction (ADC) technique to devise a systematically improvable hierarchy of many-body truncations~\cite{Barbieri:17}. Up to third order, or ADC(3), accurate ground state observables and low-energy spectroscopy are achieved for several chains of isotopes near the oxygen, calcium, nickel, and tin regions~\cite{Soma20Front,Soma2020prc,Arthuis:20,Soma:21}. The Nambu-Gorkov approach has been applied only at second order for open shells nuclei and it is now being implemented to the previously unavailable ADC(3) level~\cite{Soma:13,Barbieri:22}.

\begin{figure}
    \centering
        {\includegraphics[clip,trim=1cm 1cm 1cm 1cm,width=0.51\textwidth]{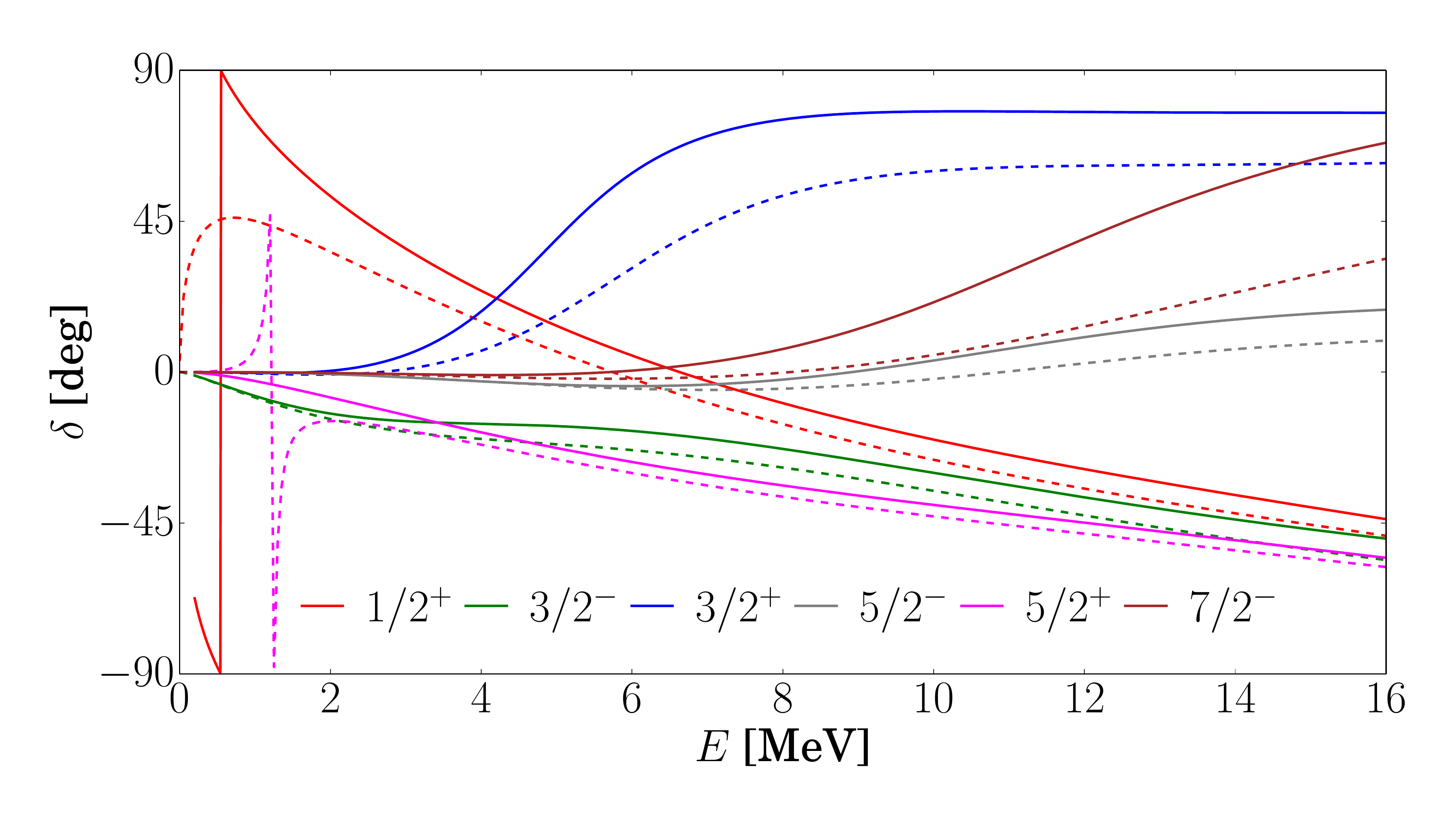}}
    \includegraphics[width=0.43\textwidth]{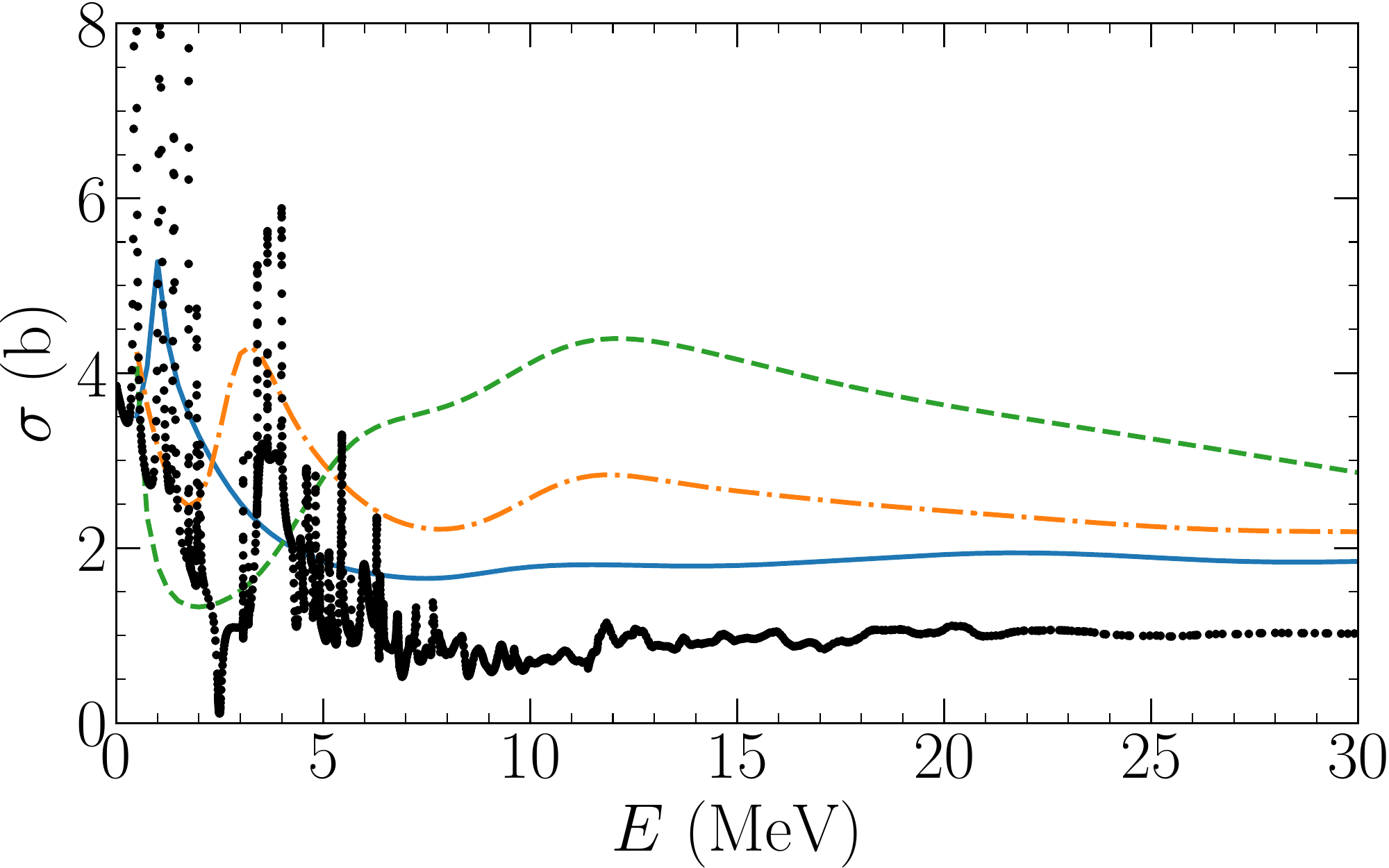}
    \caption{\emph{Left}: A comparison of $n$-$^{\rm 16}$O phase shifts obtained with  SCGF calculations using only  the correlated mean field $\Sigma^\infty$ at $N_{\rm max}=$11 (solid lines) and with the NCSM/RGM (dashed lines), which includes only the ground state of $^{16}$O, at $N_{\rm max}=$11. Both calculations have been made using NNLO$_\textrm{opt}$ nucleon-nucleon interaction and $\hbar \Omega=18$ MeV.
    \emph{Right:} Comparison of the predicted SCGF total cross section for elastic $n$-$^{\rm 16}$O as a function of the energy when including different portions of doorway states: only the correlated mean field $\Sigma^\infty$ (dashed green line), with half of the 2p1h and 2h1p configurations (dotted dashed orange line), and with complete ADC(3) (solid blue line). Computations are  based on the N2LO$_{\rm sat}$ interaction. The experimental data is also shown (black dots). Reprinted from Ref.~\cite{Idini:19}, with permission from the American Physical Society. }  
    \label{fig:B3_GF-SM}
\end{figure}

In SCGF theory the self-energy is naturally split into a mean-field part, denoted as $\Sigma^{(\infty)}$, and a dynamic contribution, $\widetilde\Sigma(E)$, which is energy dependent and accounts for coupling to the virtual inelastic channels that give rise to the dispersion relation~\eqref{eq:full_dispersion}. 
Exploratory SCGF computations of optical potentials are reported in Refs.~\cite{Waldecker:2011,Barbieri2005op}. 
The $\Sigma^{(\infty)}$ from SCGF agrees qualitatively well  with direct scattering computations  with the no core shell model with resonating group method  (NCSM/RGM, see Ref. \cite{navratil16}  for a recent review and references therein) when virtual excitations of the target are suppressed, as seen in Ref.~\cite{Idini:19} and illustrated in
 Fig. \ref{fig:B3_GF-SM} which compares the $n$-$^{\rm 16}$O phase shifts obtained with both methods. 
To reach a more predictive description of  single-particle bound states and resonances, we include virtual states of the target nucleus. By including low-lying excitations of $^{\rm 17}$O, the no-core shell model with continuum (NCSMC) accounts effectively for virtual target excitations and leads to two   bound states, namely $1/2^+$ and $5/2^+$. For SCGF,  virtual  excitations, contained in $\widetilde\Sigma(\omega)$, correct single-particle states and generate a large number of narrow resonances across all scattering energies. Results for low energy states are shown in Ref.~\cite{Idini:19} and there is    qualitative accord with the observation.

In the SCGF approach, the biggest challenge for an \textit{ab initio} theory is calculating $\widetilde\Sigma(\omega)$. The SCGF-ADC(3) construction contains correlations from all two-particles one-hole (2p1h) and one-particle two-holes (1p2h) configurations and has a direct impact on the absorption of the optical model. This is demonstrated for elastic neutron scattering off $^{16}$O by the right hand side of Fig.\ \ref{fig:B3_GF-SM}, where all 2p1h doorway states contribute to the  solid blue line   and are then gradually frozen until only the mean-field $\Sigma^{(\infty)}$ remains (dashed green   line)~\cite{Idini:19}. The 2p1h states become insufficient already at intermediate energies, where more complex configurations must enter into play. Truncations well beyond third order, ADC($n$) with $n \gg$ 3, will most likely   resolve this problem but will require groundbreaking advances in nuclear many-body theory to automatically generate and efficiently sample the exponentially growing number of diagrams.

\paragraph{Inversion of propagators  using  \textit{ab initio} wave functions.}  The indirect approach first evaluates the one-body Green's function in configuration space,
\begin{equation}
	G_{\alpha\beta}(E) = \left\langle \Phi_0\left| a_{\alpha} \frac{1}{E - ({\cal H} - \epsilon_0) + i\eta} a_{\beta}^{\dagger} \right| \Phi_0 \right\rangle + \left\langle \Phi_0 \left| a^{\dagger}_{\beta} \frac{1}{E + ({\cal H} - \epsilon_0) - i\eta} a_{\alpha} \right| \Phi_0 \right\rangle . \label{GFparthole}
\end{equation}
where $\alpha$ and $\beta$ label the single-particle states and with $\eta {\to }  0^+$.  
 One then inverts Eq.~\eqref{eq9} for each scattering energy $E$ to calculate
the optical potential.  To compute $G_{\alpha\beta}(E)$, the completeness of eigenstates
$\ket {\Psi^{A\pm1}}$ can be used, which requires computationally intensive calculations of
many eigenstates, but in   practice, the inverse Hamiltonian operator in Eq.~\eqref{GFparthole} is evaluated using one of a few available Lanczos algorithm methods~\cite{Soma2014GkvII,Marchisio2002} (see also Refs. ~\cite{Barbieri:17,Dagotto:1994_RMP,NND_PRC_2014} and references therein).
Note that if one evaluates Eq.~\eqref{GFparthole}  in the Lehmann representation   using the completeness over 
$\ket {\Psi^{A\pm1}}$ , one is evaluating
the overlap functions $\langle \Psi_n^{A+1} | a_{\alpha}^\dagger | \Phi_0 \rangle$ that are in fact solutions. In this case, the approach is equivalent to computing a discretized   set of scattering waves and then solving an inverse scattering problem. 

The viability of the propagator inversion scheme was demonstrated for oxygen and calcium isotopes using the particle attached and removed CCM, as discussed in Ref.~\cite{Rotureau2020Front} and is being applied within NCSM frameworks~\cite{Burrows21,Burrows22}.
The SA-NCSM provides useful features for nucleon-nucleus scattering such as its suitability describing deformation. {\it Ab initio} descriptions of spherical and deformed nuclei up through the calcium region are now possible in the SA-NCSM \cite{LauneyDD16,DytrychLDRWRBB20,launeymd} without the use of interaction renormalization procedures and effective charges. It has also been  shown that the SA-NCSM can use significantly reduced  model spaces as compared to the corresponding  ultra-large  conventional NCSM model spaces without compromising the accuracy of results for various observables. This allows the SA-NCSM to accommodate larger model spaces  needed for clustering, collective, and continuum degrees of freedom, and to reach heavier nuclei such as $^{20}$Ne \cite{DytrychLDRWRBB20,DreyfussLESBDD20}, $^{21}$Mg \cite{Ruotsalainen19}, $^{22}$Mg \cite{Henderson:2017dqc}, $^{28}$Mg \cite{PhysRevC.100.014322},  as well as $^{32}$Ne and $^{48}$Ti \cite{LauneySOTANCP42018}. 
Moreover,  the construction of self-energies for light nuclei starting from the NCSM/RGM and its extension to the NCSMC are also currently being developed and are of interest as scattering states are  included explicitly in the many-body basis. As the NCSMC reproduces low-energy scattering and bound-state observables \cite{HQN15,DEetal16,PRL11BeNCSMC,HQN19,Ketal20,KQHN20,Ketal22,Hetal22} for light nuclei, the optical potentials derived within this theory, along with the reach of the symmetry-adapted RGM (SA-RGM) to
intermediate-mass nuclei~\cite{launeymd,Mercenne21}, are expected to be accurate in a similar range of energies and masses.

\paragraph{Challenges and opportunities.} 
Computationally, the most demanding task is the accurate evaluation of the self-energy for all relevant scattering energies. Methods that scale polynomially with the mass number, such as SCGF and CCM, are presently limited to simple excitations (e.g., 2p1h and 2h1p) throughout the energy range and converge with respect to the model space up to $\approx$~160~MeV~\cite{Waldecker:2011,IdiniPrep}. However, more complex configurations that are important at intermediate energies are missing (see Fig.~\ref{fig:B3_GF-SM}).  The NCSM family of methods is complementary and it  performs a truncation based on the number of HO excitations, which has two advantages. First, correlated multiple particle-hole configurations are well included at low energies (e.g., SA-NCSM can capture giant resonances, which is important to describe scattering from 1 to 15 MeV per nucleon) but high scattering energies may pose
a challenge. Second, it ensures the exact separation of the motion of the center of   mass.

Indeed, calculations of the one-body Green's function that use
laboratory coordinates may pose issues due to  spurious center-of-mass motions in both the target and the $A\pm1$ systems. Model spaces in typical \textit{ab initio} computations are sufficiently large to decouple the intrinsic and the center of mass  wave functions~\cite{Hagen2019prl}. Moreover, this separation is even exact for NCSM calculations with HO-excitation truncations~\cite{PhysRevC.70.054324}. Nevertheless, the zero point motion of the center of mass is \emph{still present}  and it can be a source of spuriosity. 
Johnson discussed the technical difficulties with expressing a self-energy in a pure laboratory system~\cite{Johnson:17,Johnson:19} and suggested that a proper optical potential theory should be expressed in Jacobi coordinates.  
Center of mass corrections are seen to be sizeable for light nuclei, such as $^{16}$O
but become quickly negligible at larger masses where the self-energy approaches the one in the laboratory frame. We note that the NCSM/RGM and NCSMC methods routinely compute scattering among observables by handling these center-of-mass corrections~\cite{navratil16}. A similar development could also be valuable for reformulating the SCGF self-energy in proper relative projectile-target coordinates.

A related technical question is the discretization of the scattering spectra due to the finite model spaces. Both the Green's function and the optical potential (i.e., the self-energy) develop a real and an imaginary part in the continuum. For practical applications, the finite size of the model space implies a set of discrete poles both in Eq.~\eqref{GFparthole} and in the spectral representation for $\Sigma^\star(E)$.  The correct continuum spectrum is recovered only taking the limit $\eta \to 0$ while \emph{at the same time} letting the density of intermediate states diverge (as per the complete set of configurations in an infinite model space).  Ideally, one would like to use a finite  $\eta$ to impose a width as big as the distance between two neighboring levels, and check that predictions for observables are unaffected by variations of  $\eta$ around such central value. 
Note that the technical issue of handling the $\eta \to 0$ limit  is more compelling for the inversion propagator approaches, since the diverging poles in Eq.~\eqref{GFparthole} can lead to instabilities in the inversion process. This has been studied with the CCM method using Berggren bases with the continuum~\cite{rotureau17,rotureau18}.
For all cases, however, it should be clear that the choice for $\eta$ sets the energy resolution of the optical potential being computed. A higher resolution requires a higher density of intermediate states, posing stronger demand on the \textit{ab initio} method being employed.

To conclude,  constructing the self-energy starting from microscopic computations with a single realistic Hamiltonian allows for a consistent description of the target structure and reaction dynamics, to derive a nucleon-nucleus optical potential and calculate elastic scattering observables. 
 Contrary to phenomenological approaches whose applicability is limited by the reliability of extrapolations, microscopic optical potentials rely only on the knowledge of nuclear interactions and can be built in principle for any nucleus accessible by the theory.  Such nucleon-nucleus optical potentials could be also used for microscopic descriptions of ($d$,$p$) and ($d$,$n$)~\cite{TimoJohn2020PPNP}.  

 Although elegant  and with controlled approximations, \textit{ab initio} methods are computationally intensive, they require suitable approximations and still have to face important challenges. 
In all cases, the quality of   constructed    microscopic potentials  will reflect the current status of high performance computing resources and the accuracy of the many-body approach used.  In particular, the  coupling to possible intermediate states strongly influences the absorption from the elastic channel, i.e., the magnitude of the elastic scattering cross section.   Moreover, the  diffraction pattern, i.e., the position of the minima in the elastic-scattering angular distribution, is determined by the root mean square radius of the target posing important requirements on the quality of the realistic nuclear force used.
The most compelling issue is to advance \textit{ab initio} methods to reach complete and stable description of intermediate configurations. Novel approaches as  those discussed in Refs.~\cite{Arthuis:22,Prokofev2008diagmc} will be  key to reach full predictive power at medium energies.

\subsubsection{Multiple scattering approach}~\ 
\label{Sec3B3}

\vspace{0.2cm}\noindent The theoretical approach to the elastic scattering of a nucleon from a nuclear target pioneered by
Watson~\cite{watson53,Watson1953b}, made familiar by Kerman, McManus, and Thaler (KMT)~\cite{kerman59}, 
and further developed as
spectator expansion of multiple scattering 
theory~\cite{Siciliano:1977zz,Ernst:1977gb,Tandy:1980zz,Crespo:1992zz} is receiving renewed interest as an approach
to the optical potential that can combine advances in nuclear structure with, e.g.,  ~chiral
nucleon-nucleon (NN) interactions. A theoretical motivation for the spectator expansion derives from
our present inability to calculate the full many-body problem when the projectile energy exceeds about {40-50} MeV (see Secs.~\ref{Sec3B1} and~\ref{Sec3B2}). In this case, an expansion is constructed within a multiple scattering theory assuming
that two-body interactions between the projectile and one of the nucleons in the target nucleus play
the dominant role. In the spectator expansion the leading (first) order term involves two-body interactions
between the projectile and one of the target nucleons, the next-to-leading (second) 
order term involves the projectile
interacting with two of the target nucleons, and so forth. 
Hence, this expansion derives its ordering from the number of target nucleons interacting directly with
the projectile, while the residual target nucleus remains `passive'. Due to the many-body nature of the free
propagator for the nucleon-target system, there is an additional aspect to consider in the ordering
of the spectator series.  
The expansion of chiral NN forces not only leads to two-body forces but naturally introduces three-body
forces at next-to-next-to-leading order. The latter will not contribute to the leading order in the
spectator expansion. 
The calculation of an optical potential relies on basic input quantities. For the leading order those
are fully-off-shell NN amplitudes (or t-matrices), representing the current understanding of the
NN force, and fully-off-shell one-body density matrices representing the current understanding of the
ground state of the target nucleus. For any higher order, additional input like 3N amplitudes and
two-body density matrices for the target will be needed.

The standard approach to elastic scattering of a strongly interacting projectile from a target of $A$
particles is the separation of the Lippmann-Schwinger equation for the transition amplitude $T$
\begin{equation}
T = V + V G_0(E) T \label{eq:MS1}
\end{equation}
into two parts, namely an integral equation for $T$
\begin{equation}
T = U + U G_0(E) P T  , \label{eq:MS2}
\end{equation}
where $U$ is the optical potential operator defined by a second
integral equation
\begin{equation}
U = V + V G_0(E) Q U.  \label{eq:MS3}
\end{equation}
In the above equations the operator $V=\sum_{i=1}^A v_{0i}$ consists of
the two-body NN potential $v_{0i}$ acting between the projectile and
the $i$th target nucleon. The free propagator $G_0(E)$ for the
projectile-target system is given by
\begin{equation}
    G_0(E)=\frac{1}{E-H_0+i\eta}
\end{equation}
with $\eta  {\to} 0^+$. 
Though most applications use
targets with $0^+$ ground states, there is no need for this to be the case \cite{Vorabbi:2021kho}. 
In fact, to develop optical potentials valid for exotic nuclei, a variety 
of targets with different ground state spin configurations will need to be considered. 
 
The operators $P$ and $Q$ in Eqs.~(\ref{eq:MS1}) and (\ref{eq:MS2}) are
projection operators with $P+Q=1$, and $P$ being defined such that
Eq.~(\ref{eq:MS2}) becomes a one-body equation. In this case, $P$ is
conventionally taken to project on the elastic channel, such that
 $[G_{0},P]=0$, and is defined as  $ P=|\Phi_0\rangle\langle\Phi_0|/\langle\Phi_0|\Phi_0\rangle$.  
With these definitions, the transition operator for elastic scattering
can be defined as ${T_{el}=PTP}$, in which case Eq.~(\ref{eq:MS2})
can be written as
\begin{equation}
T_{el}=PUP + PUPG_{0}(E)T_{el}.  \label{eq:MS4}
\end{equation}
The choice of the projector $P$ fixes the scattering problem to be considered. A projection
onto the target ground state is the appropriate choice to derive an optical potential describing
the elastic scattering of a nucleon from a target nucleus, but when considering e.g.~inelastic
scattering in a coupled-channel approach, this $P$-space should contain the excited states under
consideration. To our knowledge, this has not been attempted in a multiple scattering approach.

The expression for the optical potential in Eq.~(\ref{eq:MS3}) contains the projection operator $Q$ and
thus, even in the leading order term where $U$ is defined as $U = \sum_{i=1}^{A}\tau_{0i}$, the quantity
$\tau_{0i}$ cannot readily be identified with a NN amplitude derived in free space. Working in momentum space,
it is straightforward to formulate an integral equation for the Watson optical potential~\cite{Chinn:1993zza}
\begin{eqnarray}
\tau_{0i} &=& v_{0i} + v_{0i} G_0(E) Q \tau_{0i} = \hat{\tau}_{0i} - \hat{\tau}_{0i} G_0(E) P \tau_{0i} ,
\label{eq:MS5} 
\end{eqnarray}
where  $\hat {\tau}_{0i}$    is the NN t-matrix given as a solution of a regular two-body Lippmann-Schwinger equation,
in which only the many-body Green's function $G_0(E)$ needs to be considered. The standard impulse
approximation turns this Green's function into a two-body propagator. 
It should be noted that the above equations follow in a straightforward derivation and
correspond to the first-order Watson scattering expansion \cite{watson53,Watson1953b}.
The integration of Eq.~(\ref{eq:MS5}) taking into account
contributions from  the   $Q$ space corresponds to an averaging over inelastic channels and thus
should only be applied for energies higher than $\sim$~30-40~MeV.
Unfortunately a similar formulation as in Eq.~(\ref{eq:MS5}) cannot be made in coordinate space. 
Here the closest to treating
the operator $Q$ is the averaging suggestion made by Kerman, McManus, and Thaler \cite{kerman59} leading to
the KMT factor (A-1)/A in the optical potential.
Ref.~\cite{Chinn:1993zza} showed that the explicit treatment of the operator $Q$ is especially
important for scattering from very light nuclei, where the KMT factor is not close to one. 
The importance of an explicit treatment of $Q$ for nuclei far off the valley of stability needs to be
explored. Studies of reaction cross sections of the helium isotopes at energies below 100~MeV revealed 
that treating $Q$ exactly or via KMT did not lead to major differences: however, not treating $Q$ at all
caused discrepancies of more than 10\% in the reaction cross section.

A further equally important consideration for obtaining the optical potential is to
find a solution of Eq.~(\ref{eq:MS3}), which still has a many-body character due to the propagator
$G_0(E)$. The standard impulse approximation assumes closure, i.e. 
 ignores target excitations .   For projectile energies above $\sim$80~MeV this is generally assumed to be a good approximation 
and errors  have  not been studied yet.
In the impulse approximation and at leading order,  the nucleon-nucleus optical potential for a certain kinetic energy $E$ is given by Eq.~\eqref{eq:MS6}. Let us now discuss a possible extension of this formalism to include effects of the antisymmetrization and to go beyond the impulse approximation.  

The treatment of Pauli antisymmetry effects follows the philosophy growing out of the early work of
Watson~\cite{Takeda:1955zz,GoldbergerWatson} and developed via the spectator expansion
in~\cite{Picklesimer:1981zza}. In the
lowest order the two-body antisymmetry is achieved through the use of
two-body t-matrices which are themselves antisymmetric in the two
``active" variables (corresponding to the weak binding limit in Ref. ~\cite{GoldbergerWatson}).

Going beyond the impulse approximation in the spirit of the spectator expansion means  consider that more nucleons of the target are active. At the next order in the expansion, one needs to consider that two nucleons of the target  $i$ and $j$ are active. The optical potential would depend on  propagators $ {\cal G}_{0j}$ that have the structure of  three-body channel Green's functions. One can describe these propagators within a single particle description as
\begin{equation}
    {\cal G}_{0j}(E) = \frac{1}{E -h_0 - h_j - \sum_{i\neq j}v_{ij} - H^j+i\eta}
\end{equation}
where $\eta  {\to} 0^+$, $H^j$ is the residual target
Hamiltonian involving  $(A-1)$  particles (excluding particles $0$ and $j$), $h_j$ is the kinetic energy
operator for nucleon $j$, and $v_{ij}$ is the interaction between target nucleons $i$ and $j$.  One can then project this Green's function onto a fixed number of eigenstates of  the residual target Hamiltonian $ H^j$.   Due to the presence of
$v_{ij}$ an exact solution will require fully-off-shell two-body density matrices for the target
nucleus as well as three-body dynamics. Two-body density matrices from \textit{ab initio} structure
models are calculable in principle, so there is an opportunity to consistently
estimate the contribution of the next order in the spectator expansion, and thus have a better
understanding of its convergence as function of projectile energy as well as mass number. This will be
a very challenging enterprise.  
A first attempt with nuclear densities derived from HFB mean field calculations for heavier nuclei 
was made in Ref.~\cite{Chinn:1993zz,Chinn:1994xz}
where the interaction $v_{ij}$ was taken as the corresponding nuclear mean field. The result of this
study showed that at projectile energies above  100 MeV the second-order correction is almost negligible, 
while starting to be  evident in the spin observables at energies below 100 MeV. At about 50 MeV, the
second-order correction is quite visible in the differential cross section. 
 
Another approach to take into account the beyond-leading-order effects of three-body forces was
implemented in Ref.~\cite{Vorabbi:2020cgf} by constructing a density-dependent NN interaction that
treats the 3N force in an approximate way \cite{Holt:2009uk,Holt:2009ty}. For energies above 100 MeV, few effects were noticeable in
the differential cross sections, though some effects were observed in the analyzing power and spin
rotation function.

\paragraph{Explicit calculation of the leading order term in the Watson approach.}

For explicit calculations of reaction observables from the leading order term in the spectator
expansion, Eq. (\ref{eq:MS6}), one needs both,
structure information (fully-off-shell one-body density matrices) and reaction information (NN
amplitudes). Current \textit{ab initio} calculations of multiple scattering theory are limited 
in their applicable mass range due to the available \textit{ab initio} structure inputs. To reach target nuclei beyond the $A\sim40$ range, one-body density matrices will need to be calculated
from other structure models, e.g., SCGF, CCM and In-Medium
Similarity Renormalization Group (IM-SRG).

Recent work \cite{burrows20} showed that including the spin of the struck target nucleon has
an effect on the elastic scattering spin observables for neutron-rich systems, which implies consistent
calculations incorporating this term may be necessary to study nuclei off the valley of stability. This
 additional term
requires both, a scalar and spin-dependent one-body density matrix, and guarantees that the scalar
(Wolfenstein A), vector (Wolfenstein C), and tensor (Wolfenstein M, G, H, and D) parts of the NN
interaction are included{\footnote{For the definition and derivation of Wolfenstein amplitudes see Refs.~\cite{wolfenstein-ashkin,wolfenstein-1954}} . For $J=0$ to $J=0$ transitions, it has been shown  that   only A, C, and M contribute
due to parity invariance, though this likely holds for other transitions between the same spin states.
Future work to develop \textit{ab initio} treatments of inelastic scattering in this framework are becoming
possible and will allow for further study of the tensor (Wolfenstein M, G, H, and D) parts of the NN 
interaction.

In addition to the theoretical uncertainties arising from the spectator expansion of the multiple 
scattering theory (e.g.~next-to-leading-order effects for three-body forces, as well as energy-dependencies 
of the NN t-matrix \cite{Elster:1997as}) -- which are expected to be small in the energy regime above 60
MeV -- there 
are additional theoretical uncertainties propagating from the nuclear-structure calculations. This includes both 
model-related uncertainties (e.g.~any residual dependence on the size of the model space), as well as 
uncertainties from the underlying NN interaction (e.g.~uncertainties associated with truncating a chiral 
effective field theory at a certain order, dependence on the fits of the low-energy constants (LECs), and 
others). Different approaches can be taken to address each of these uncertainties, but an accounting of 
them is necessary to extract reliable information about reaction observables. 

\paragraph{Explicit calculation of the leading order term in the $g$-matrix approach.}
This approach starts directly from the general expression for the leading-order term of Eq.~(\ref{eq:MS6}) and realizes that
from quite general considerations~\cite{Arellano:2007np} 
the two-body (NN) amplitude $\hat{\tau}_\alpha$ can be recast 
as
\begin{equation}
\label{wigner}
  \langle{\vec  k}'{\vec  p}'\mid\hat \tau_\alpha(E)\mid{\vec  k}\;{\vec  p}\rangle
  =
\int\;
\frac{d{\vec  z}}{(2\pi)^3}
\;e^{i{\vec  z}\cdot({\vec  W}'-\vec  W)}\;
g_{\vec  z}[\textstyle{\frac12}(\vec  W'+\vec  W); {\vec  b}',{\vec  b}]\;,
\end{equation}
where $g_{\vec  z}$ represents a reduced interaction at the local
coordinate ${\vec  z}$.
In this expression 
${\vec  W}\! =\! {\vec  k} +{\vec  p}$, and
${\vec  b}\! =\! ({\vec  k} -{\vec  p})/2$,
the prior total and relative two-body momenta, respectively.
The same applies to the post momenta, denoted by primed marks.
Additionally,
the ${\vec  z}$ coordinate is given by the average
${\vec  z}\!=\!(
{\vec  r'}\!+\!
{\vec  s'}\!+\!
{\vec  r}\!+\!
{\vec  s})/4$,
the center of gravity of the four coordinates of the two particles,  $\vec r$, $\vec r'$, $\vec s$, $\vec s'$ .

Assuming a density-dependent \textit{NN} effective interaction, 
in Ref.~\cite{Arellano:2007np} it is demonstrated quite generally that
the folding potential in momentum space can be expressed as the
sum of two terms,
\begin{equation}
  U({\vec  k'},{\vec  k};E)=    \sum_{\alpha}    \int d{\vec  P}\;
    \hat\rho_\alpha(\vec  q;\vec  P)\;
   \hat {\tau}_{0} (E\!+\!\epsilon_\alpha) 
    +  U_1({\vec  k'},{\vec  k};E)\;,
\end{equation}
where
  \begin{eqnarray}
    \label{u1}
    U_{1}({\vec  k'},{\vec  k};E) &=&\nonumber \\
    &&\hspace{-4cm}
    -\sum_{\alpha}
    \int_{0}^{\infty}
   \,\,\!\!dz\,\frac{4\pi z^3}{3}
    \int\! \frac{d{\vec  P}}{(2\pi)^3}
    \int\! d{\vec  q'}\,
     \hat{\jmath}_1(z|\vec  q'-\vec  q|)\;
    \rho_\alpha(\vec  q';\vec  P)\,
    {\partial_z}g(\rho_z,E\!+\!\epsilon_\alpha) 
    \;.
  \end{eqnarray}
Here
$\hat{\jmath}_1(x)\!=\!3j_1(x)/x$,   with $j_1(x)$  being the spherical Bessel function of order 1  
and
$\rho_\alpha(\vec  q;\vec  P)\! =\! 
 \varphi_\alpha^\dagger(\vec  P\!+\!\textstyle{\frac12}\vec  q)
 \varphi_\alpha(\vec  P\!-\!\textstyle{\frac12}\vec  q)$,
with $\varphi_\alpha$ the target single-particle wave function 
with energy $\epsilon_\alpha$.  Note that in the sum  only occupied states should be considered.  
While     $\hat {\tau}_{0}$  represents the momentum-space free $t$ matrix (as present in
  the KMT term of the optical potential~\cite{Elster:1989en,Crespo:1990zzb,Arellano:1990xu}),
the fully off-shell $g$ matrix can be modeled with the
infinite nuclear matter
Brueckner-Hartree-Fock $g$ matrix.
Assuming weak isospin asymmetry,
the gradient term 
$\partial_z g\!=\!
[{\partial g}/{\partial\rho}][{\partial\rho}/{\partial z}]$,
is evaluated at a local isoscalar density $\rho_z$.
The resulting nonlocal potential $U({\vec  k'},{\vec  k};E)$
has been applied to nucleon elastic scattering, 
as reported in Refs. \cite{Aguayo:2008hn,Arellano:2011zz}
An interesting interpretation of Eq.~\eqref{u1} for $U_1$ is that
intrinsic medium effects take place
mostly at the surface of the target,
as modulated by $\partial_z\rho$, the gradient
of the density~\cite{Aguayo:2008hn}.

Finally, let us emphasize that there are other similar approaches  which construct in coordinate space nucleon- and nucleus-optical potentials, folding microscopic neutron and proton densities with nucleon-nucleon effective interactions~\cite{DURANT2018668,PhysRevC.102.014622,PhysRevC.105.014606,PhysRevC.66.014610,PEREIRA2009330,PhysRevC.85.044607,PhysRevC.96.059906,PhysRevC.94.034612}.

\subsubsection{Nuclear matter approaches: Whitehead-Lim-Holt potential, G-matrix solutions of the Brueckner-Bethe-Goldstone and the JLM folding model}~\ 
\label{Sec3B4}

\vspace{0.2cm}\noindent Compared to finite nuclei, infinite homogeneous nuclear matter represents a conceptually simpler physical system to study. In particular, calculations of the nucleon optical potential in nuclear matter avoid many of the technical difficulties and practical limitations faced when computing the nucleon optical potential directly in a finite system. When combined with a local density approximation, the nuclear matter approach can also be used to construct nucleon-nucleus optical potentials, provided that the isoscalar and isovector densities of the target nuclei are known. A significant advantage is that the nucleon optical potential in nuclear matter only needs to be computed once over a wide range of densities and proton fractions and then may be applied across large regions of the nuclear chart. Hence, the nuclear matter approach is naturally suited for the construction of global optical potentials, which will be vital for the future of reaction theory for rare isotopes. However, the assumptions of the nuclear matter approach that allow for the ease of constructing global nucleon-nucleus optical potentials also omit phenomena such as surface effects, resonances, and spin-orbit interactions. The nuclear matter approach also tends to produce an overly absorptive imaginary term at high energies. Some of these shortcomings may be straightforwardly remedied while for others the solution remains unclear, for more details see Ref. \cite{Holt22}. Ultimately, the quality of theoretical predictions for reaction cross sections from optical potentials derived within the nuclear matter approach must be assessed by comparisons to experimental data.

The framework for utilizing nuclear matter calculations of the optical potential for finite nuclei was built by Jeukenne, Lejeune and Mahaux in the late 1970s \cite{Jeukenne77,jeukenne76}. They implemented the Local Density Approximation (LDA)
\begin{eqnarray}
U(E;r)_{LDA}&=&V(E;r)_{LDA} + i W(E;r)_{LDA} \nonumber \\
&=& V(E;k_f^p(r),k_f^n(r))_{NM} + i W(E;k_f^p(r),k_f^n(r))_{NM},
\end{eqnarray}
which relates the optical potential at a given position in the nucleus with the optical potential of nuclear matter (denoted by NM)   with the same local density and isospin asymmetry through the neutron $k_f^n(r)$ and proton $k_f^p(r)$ Fermi momenta . A key finding of Ref.\ \cite{Jeukenne77} is that the LDA is insufficient for reproducing elastic-scattering data, which requires a modification called the Improved Local Density Approximation (ILDA) that takes into account the nonzero-range of the nuclear force
\begin{equation}
{U}(E;r)_{ILDA}=\frac{1}{(t\sqrt{\pi})^3}\int U(E;r')_{LDA} e^{\frac{-|\vec{r}-\vec{r}'|^2}{t^2}} d^3r'.
\label{eq:ilda1}
\end{equation}
The ILDA introduces a Gaussian smearing of the optical potential over the range of densities probed across the length scale $t$, typically chosen to be around $t \sim 1.2$\,fm, the effective range of the nuclear force. The most important consequence is that the optical potential in the interior of the nucleus changes little, while the surface diffuseness of the optical potential increases due to finite-range effects.

\paragraph{Whitehead-Lim-Holt global optical potential.} Recent advances \cite{whitehead19,Whitehead20} in the nuclear matter approach to constructing microscopic nucleon-nucleus optical potentials incorporate consistent two-body and three-body forces \cite{sammarruca15} at various orders in the chiral expansion. The nucleon self-energy in nuclear matter is calculated in the framework of many-body perturbation theory (MBPT) \cite{holt13,holt16prc}, which has already been used to produce accurate models of the nuclear equation of state~\cite{bogner05,hebeler11,gezerlis13,coraggio14,drischler16}. In addition to MBPT, there are other many-body frameworks for microscopically calculating the self-energy. One notable example is the work of Rios in SCGF theory~\cite{Rios20}.

In MBPT, the first-order (or Hartree-Fock) contribution to the nucleon self energy in isospin-symmetric nuclear matter is given by
\begin{equation}
\Sigma^{(1)}_{2N}(k)=\sum_{1} \langle \vec{k} \: \vec{h}_1 s s_1 t t_1 | \bar{V}_{2N} | \vec{k} \: \vec{h}_1 s s_1 t t_1 \rangle n_1 ,
\label{eq:2}
\end{equation}
where $\vec k, s, t$ are the momentum, spin, and isospin of the projectile, $n_1$ is the occupation probability $\theta(k_f - h_1)$ for a filled state with momentum $\vec h_1$ below the Fermi surface, and the summation is over intermediate-state momenta $\vec h_1$, spins $s_1$, and isospins $t_1$.

The second-order perturbative contributions to the nucleon self energy in symmetric nuclear matter are expressed as
\begin{equation}
\label{sig2a}
\Sigma^{(2a)}_{2N}(k;E) = \frac{1}{2} \sum_{123} \frac{|\langle \vec{p}_1 \vec{p}_3 s_1 s_3 t_1 t_3 | \bar{V}_{2N} | \vec{k} \vec{h}_2 s s_2 t t_2 \rangle|^2}{E+\epsilon_2-\epsilon_1-\epsilon_3+i\eta} \bar{n}_1 n_2 \bar{n}_3,
\end{equation}
\begin{equation}
\label{sig2b}
\Sigma^{(2b)}_{2N}(k;E) = \frac{1}{2} \sum_{123} \frac{|\langle \vec{h}_1 \vec{h}_3 s_1 s_3 t_1 t_3 | \bar{V}_{2N} | \vec{k} \vec{p}_2 s s_2 t t_2 \rangle|^2}{E+\epsilon_2-\epsilon_1-\epsilon_3-i\eta}   n_1 \bar{n}_2 n_3,
\end{equation}
where the occupation probability for particle states above the Fermi momentum is $\bar n_i = \theta(k_i-k_f)$. The second-order contributions $\Sigma^{(2a)}$ and $\Sigma^{(2b)}$ are energy-dependent and complex.
In Eqs.\ \eqref{sig2a} and \eqref{sig2b}, the single-particle energies $E$ and $\epsilon$ should be computed self-consistently according to 
\begin{equation}
E(k) = \frac{k^2}{2M} + {\rm Re}\,\Sigma(k;E(k)).
\label{dyson1}
\end{equation}

The use of chiral nuclear forces provides several advantages to the phenomenological nuclear forces of the past. By virtue of being an effective field theory, chiral makes a concrete connection to the underlying theory of quantum chromodynamics through its symmetries. Furthermore, chiral nuclear forces are calculated in a perturbative expansion that allows for a direct method of uncertainty quantification by assessing order-by-order convergence \cite{melendez17}.

The wide applicability of the nuclear matter approach and uncertainty quantification capabilities of chiral EFT were employed in Ref.\ \cite{Whitehead21} to construct both the first microscopic global optical potential and the first global optical potential with uncertainty quantification. Both of these advances are central to the development of reaction theory for the rare-isotope beam era, where theoretical predictions for thousands of exotic isotopes will be needed to drive scientific discovery and answer fundamental science questions in nuclear astrophysics. Present microscopic optical potentials have sizable uncertainties, which may be reduced within a Bayesian framework that incorporates experimental nucleon-nucleus scattering and reaction data in Bayesian likelihood functions. In Ref.\ \cite{Whitehead21}, five separate global optical potentials were generated from a set of chiral potentials of different order in the chiral expansion and with varied momentum-space cutoffs. These global optical potentials are expressed in terms of Woods-Saxon functions that are parametrized in terms of energy, target mass, and target isospin asymmetry ($E$, $A$, $\delta$). Assuming the five optical potentials are drawn from a multivariate Gaussian distribution in the space of optical potential parameters, one can then propagate statistical uncertainties to scattering observables. This multivariate Gaussian distribution of nucleon-nucleus optical potentials is referred to as the Whitehead-Lim-Holt (WLH) global optical potential.

In Fig. \ref{fig_potential}, the real (left panel) and imaginary (right panel) terms of the  WLH optical potential for $n$+$^{40}$Ca at $E=5$~MeV and $E=100$~MeV are shown in coordinate space. As the energy increases, the real term decreases in depth while the imaginary term increases in depth. Both terms have larger uncertainties as the energy increases reflecting the enlarging uncertainties of chiral EFT at high energies. These potentials can be easily applied in open source reaction codes to propagate uncertainties to reaction observables. The WLH global optical potential yields good reproductions of experimental elastic scattering cross sections for projectile energies of $E \lesssim 150$ MeV. Total reaction cross sections calculated from WLH are in good agreement with data up to moderate energies, and then overestimate data at larger energies due to an overly absorptive imaginary term. Beyond energies of $E \lesssim 150$ MeV, predictions of the WLH optical potential are expected to have greater discrepancies with data along with larger uncertainties. Predictions of WLH for a wide range of reactions are shown in Sec.~\ref{Sec5}.  

\begin{figure}
\begin{center}
\includegraphics[width=\textwidth]{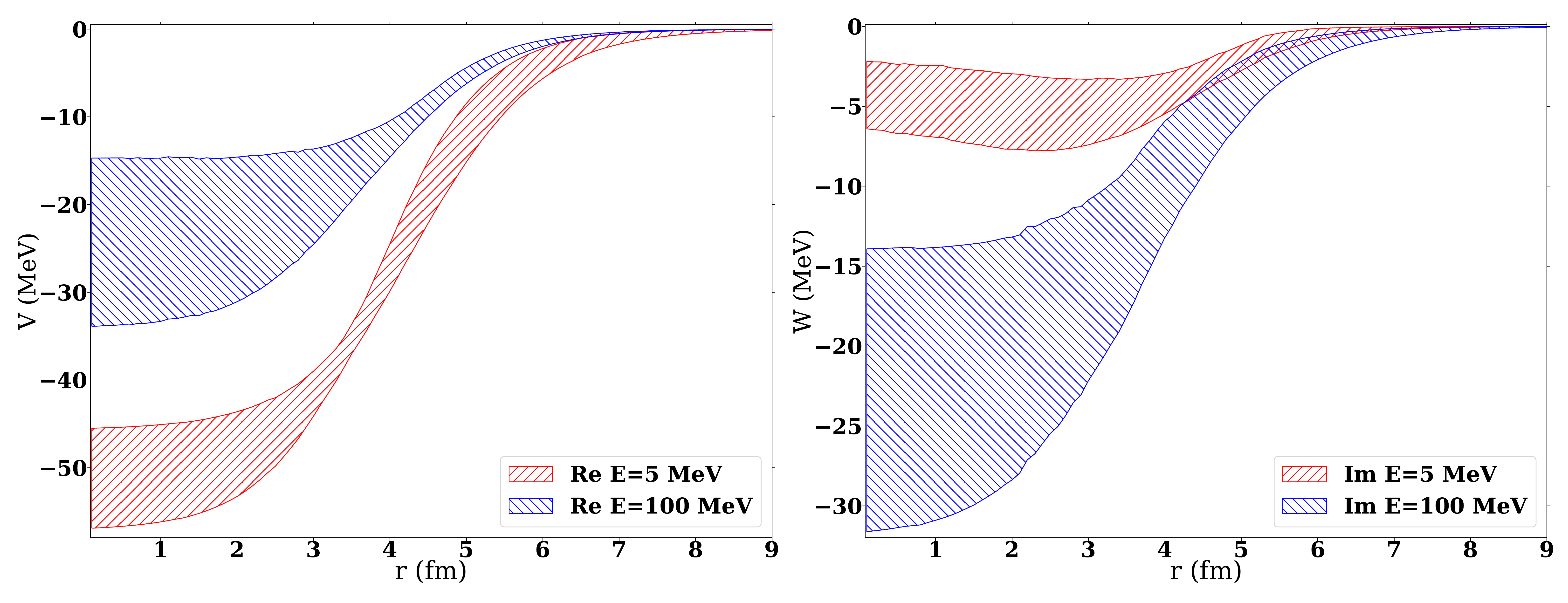}
\caption{The real (left) and imaginary (right) terms of the WLH global optical potential for $n$+$^{40}$Ca at $E=5$\,MeV (red) and $E=100$\,MeV (blue).}
\label{fig_potential}
\end{center}
\end{figure}

\paragraph{G-matrix solutions of the Brueckner-Bethe-Goldstone.}

The nuclear matter approach also gives access to direct inelastic scattering observables. In this case, the effective interaction used to build the microscopic optical potential also serves to build
the transition potentials that enter the definition of the relevant Distorted Wave Born Approximation (DWBA) or coupled-channels equations. 
For instance, nucleon elastic and inelastic scattering were modeled from g-matrix solutions of the Brueckner-Bethe-Goldstone equations in nuclear matter - two well-known examples are the Melbourne~\cite{Amos2000} and the Santiago g-matrices \cite{Arellano2011} - and one-body density matrices through the calculation of non-local optical and transition potentials (an example of the application to inelastic scattering with the Melbourne g-matrix and RPA beyond mean-field approach is given in Refs.\   \cite{Dupuis2008,Dupuis2017}).

\paragraph{JLM folding potential.}
As these approaches have proven less suited at incident energy below 30-50 MeV, one pragmatic solution to cover the missing low-energy range, quite important for energy applications and experimental programs at RIBs, is to still rely on
effective interactions derived from nuclear matter calculations but which are slightly renormalized to account for selected scattering observables. One such approach is the JLM folding model mentioned above, which has been extensively used to describe elastic and inelastic scattering of protons, neutrons, and composite particles within the double folding method, for both spherical and deformed targets. A global Lane-consistent
parametrization of the JLM interaction was given by Bauge \textit{et al.} in 2001~\cite{Baug2001} by adjusting the interaction to reproduce many elastic scattering and charge-exchange observables between 1 keV and 200 MeV. Many reactions were studied  with this parametrization starting with HFB ground state densities and transition densities from the QRPA nuclear structure calculations. Recent examples are the determination of inelastic scattering to discrete states and to the continuum for neutron scattering below 30 MeV off spherical \cite{Dupuis2019}, and axially-deformed targets such as actinides within the coupled-channel framework \cite{Dupuis2015}, as well as the modeling of proton inelastic scattering off unstable targets \cite{Cortes2018}.
The method's ability to provide accurate reaction observables is mostly related to the quality of the nuclear structure input, so it was intensively used to challenge structure theory
with hadron scattering observables. Despite its phenomenological content, the method has displayed good predictive capabilities especially for direct inelastic scattering, as no inelastic observables are used to constrain the interaction.
However, its phenomenological aspect makes the method's precision hard to improve beyond the use of better nuclear structure input, and it relies on simplified nuclear matter calculations with old-fashioned bare interactions. Moreover, resulting potentials from the JLM folding model are local and
non-dispersive, while the optical potential is known to be non-local and to obey dispersion relations. The spin-orbit
component is ad hoc--it does not stem from an underlying nuclear matter calculation--and uses a simple form factor given as
a derivative of the microscopic density. This approach could thus be revisited starting from modern nuclear matter
calculations such as those described above.

One aspect for inelastic scattering that deserves attention is the rearrangement correction in Ref.\ \cite{Cheon1985}, which has a large renormalization effect on inelastic cross sections \cite{Dupuis2016b}. This correction, which stems from the density dependence of the effective interaction used for inelastic scattering, is still now applied in an \textit{ad hoc}    manner when folding models are used. We stress that this correction, which has been known for a long time to induce modifications   as large as the difference between the $t$- and the $g$-matrix \cite{Cheon1985}, should be described from more fundamental principles in order to reach a better description of inelastic scattering within the microscopic framework of folding and full-folding approaches.

\subsection{Synergies between microscopic approaches and phenomenology}
\label{Sec3C}
There are three main limitations of the standard phenomenological approach that was presented in Sec.~\ref{Sec3A}. First, in selecting a potential form and parametrization (such as a Gaussian nonlocality), the practitioner makes simplifying assumptions about the physics at hand, pushing any unknown physics into changes of the potential parameters. As such, any extrapolation away from the region of training data is perilous, especially to weakly bound systems near the drip lines that will be probed with FRIB. Second, training phenomenological potentials requires copious training data, the vast majority of which was collected between 1960 and 2000 in direct kinematics at smaller facilities such as university cyclotrons and tandem accelerators. Without additional high-precision $p$, $n$, $d$, $t$, $^3$He, $\alpha$ scattering data, it is unlikely that traditional phenomenological OMPs can be meaningfully improved (except by including additional physical input such as, e.g., deformation information), nor can microscopic approaches be rigorously tested. If new phenomenological OMPs are to be developed using data from radioactive beams in inverse kinematics, low statistics and large uncertainties in the reaction theory used to  constrain these OMPs with non-elastic cross sections present serious problems. Finally, past optimization approaches for phenomenological OMPs have focused almost exclusively on finding ``best-fit'' parameters but lack meaningful parametric uncertainties. Even the best phenomenological potentials fail to achieve $\chi^{2}/N$ values of $\approx$1 that would indicate reasonable reproduction of the training data, an indication that either important physics are missing from the phenomenological forms, training data uncertainties are underestimated, or both.

While the phenomenological approaches described in Sec.~\ref{Sec3A} offer better accuracy in describing scattering on stable targets, their predictability in unknown regions is weak. Similarly, phenomenological optical potentials that provide an excellent description of one reaction channel, can fail to describe other channels.  In contrast, microscopic approaches, having the correct symmetries, are more promising for extrapolations. However, as detailed in Sec. \ref{Sec3B}, the wide range of microscopic approaches have their own  shortcomings. It is therefore appropriate to develop strategies that marry the best of the two worlds. We next discuss some explicit  ways in which microscopic approaches can benefit from an appropriate phenomenological calibration.

It is well known that for a model to be able to reproduce the scattering diffraction pattern, it is essential that the same model describe the size of the system correctly. This is particularly relevant to \textit{ab initio} methods, since various parametrizations of the chiral potentials used in these models may not reproduce accurately the root mean square radius (see e.g., \cite{radii2022} for selected medium-mass nuclei). Hence, for the description of reactions at low energies, it is important that modern nuclear forces employed in the calculations capture nuclear radii~\cite{ekstrom2015} , while ensuring the proper treatment of dominant correlations, as discussed next. 

Particle threshold energies (or resonance energies) are another important quantity for most reactions and become even more relevant for reactions involving nuclei at the limits of stability. In this regard,  microscopic models cannot provide the level of precision needed for an adequate description of the reaction (of the order of $0.1$ MeV). As such, often microscopic approaches find ways to adjust their calculations 
such that the model reproduces the thresholds exactly~\cite{HQN19,Hetal22}. 
Given that it is unlikely that many-body methods will reach the level of precision needed in the near future, one should better understand how these different adjustments affect the optical potential and propagate to complex reaction observables.

As remarked in Sec.~\ref{Sec3B}, with the exception of the NCSM and derivatives (NCSMC, SA-NCSM, etc), optical potentials derived from \textit{ab initio} methods contain only simple excitations, up to 2p1h or 2h1p. In addition, collective correlations may be suppressed for some methods and chiral potential parametrizations employed~\cite{INT21b}. This leads to an underestimation of the flux removed from the elastic channel, and thus an overestimation of the elastic-scattering cross section\footnote{By construction, NCSM-type approaches include higher orders of complexity in the model and therefore do not, in principle, have the same issue, however they are limited to applications on light to medium-mass nuclei.}. While \textit{ab initio} NCSM-type approaches are applicable to medium mass, currently it is not feasible to extend NCSM methods to heavy systems to include the level of complexity required for a good description of the total absorption occurring in the scattering (for example configurations beyond 2p2h~\cite{Rotureau2020Front,Idini19}).
Consequently,  some groups have devised strategies to incorporate the missing physics by hand as for example the method discussed in Sec. \ref{Sec3B} involving doorway states. 

A microscopically derived optical potential with known uncertainties, that has been well calibrated on the important inputs discussed above, has the potential to perform much better than any phenomenological approach when exploring unknown regions of the nuclear chart. There are a few well identified aspects in which the microscopic optical potential can provide critical information to phenomenology.

First and foremost, since the nucleon-nucleon force contains the correct isospin symmetry,  microscopic approaches should in principle provide important guidance with respect to the isospin dependence of the optical potential. This is particularly relevant to scientific programs in facilities with rare isotope beams. Not only is the optical potential isospin dependence important, but also it is critical to know how it varies with beam energy. Future studies focused on extracting the isospin dependence of the microscopic optical potential from first-principles are encouraged.

Another equally important aspect of the optical potential is the radial dependence of the spin-orbit force. While there is a reason to model the radial dependence of the central force  after the density distribution in the target nucleus, the basis for the radial dependence of the spin-orbit force is not well established and for simplicity is taken to be the derivative of the central term. Microscopic studies focused on determining the radial dependence of the spin-orbit force will be very helpful to reduce ambiguous model dependencies in the global optical potential.

Finally, as stated a number of times throughout this whitepaper, a microscopic optical potential is intrinsically non-local. However, most phenomenological potentials have preferred to make the global potential local to avoid the additional computational cost, with the exception of some DOM potentials (see Sec.~\ref{sec:dom}). This simplification introduces a very strong energy dependence in the parameters. Despite it not being directly probed through elastic scattering, the off-shell effects associated with non-locality do show up in other reaction channels \cite{ross2016,titus2016} and therefore should be considered. In view of the incredible advances in computing capacity, a non-local global optical potential is now feasible and microscopic approaches should provide guidance to the radial form and its range. Current microscopic studies have already shown that the simple Gaussian form for the non-locality factor used by Perey and Buck (introduced in Sec.~\ref{sec:dom}) is not sufficient \cite{rotureau17,arellano_18,arellano_21a}. But one should also assess whether it is appropriate to separate this non-local factor in the first place. Further microscopic studies along these lines will help establish a form for the non-local potential that can then be used in phenomenology.

\subsection{Model uncertainties beyond pairwise effective potentials}
\label{Sec3D}
Complex reactions as discussed in Sec.~\ref{Sec2} are often described within a few-body model and the dynamics are obtained from a Hamiltonian including the relevant degrees of freedom and the corresponding pairwise interactions between the clusters. Thus, even when the full dynamics is calculated, there is still a model uncertainty emerging from the reductions of the  many-body model into the few-body model. Quantifying the uncertainty introduced through this simplification is not trivial. In this section we discuss the first steps taken toward this goal.

As an illustration, we consider reactions involving the deuteron, typically described within a three-body model consisting of a neutron, a proton, and a target that consists of $A$ nucleons, interacting through pairwise phenomenological potentials. An exact solution to this three-body problem is provided by the Faddeev formalism~\cite{Faddeev:1961}. While the Faddeev formalism enables a correct description of the three-body dynamics, its predictive power is limited, in part, by the uncertainties in the effective phenomenological nucleon-nucleus potentials that implicitly include the three-cluster interaction $npA$. Additionally, the formal projection of the many-body problem onto the three-particle space gives rise to an irreducible three-body force (3BF) which cannot be decomposed into a sum of pairwise interactions~\cite{rcjohnson:2014} and thus cannot be constrained using nucleon-nucleus scattering data. Efforts to quantify the effects of the irreducible nucleon-nucleon-nucleus forces have been carried out in Refs.~\cite{nktimofeyuk:2019,mjdinmore:2021} by utilizing multiple scattering theory to estimate the lowest order contributions to the 3BF arising from the excitation of nucleons inside the nucleus $A$. While those works demonstrated that the 3BF corrections to the pairwise potentials had a significant impact on deuteron-induced reaction observables, there is some ambiguity coming from how the phenomenological potentials are defined that does not allow for the disentanglement  of irreducible three-body contributions.          

The effects of the 3BF can be quantified without ambiguities by adopting microscopically computed nucleon-nucleus potentials and grounding all calculations on a single microscopic Hamiltonian using the same $NN$ interactions. Uncertainties arising from the omission of the irreducible 3BF in three-body model calculations were performed for deuterium-$^4$He scattering and the $^6$Li ground state~\cite{hlophe:2022}. This system has the advantage that it can be well-described using microscopic reaction theory. First, the NCSM/RGM~\cite{Quaglioni:2009mn, Navratil:2011ay}  was used to compute effective $n/p-\alpha$ potentials. Then the three-body Faddeev equations \cite{AGS} are used to compute the $^6$Li ground state as well as $d+\alpha$ scattering observables. In parallel, the same scattering observables are obtained directly from NCSM/RGM. The comparison between the two approaches discloses the effects of the 3BF, arising from the antisymmetrization. 
The study finds that the irreducible three-body force has a sizable effect on observables. Specifically, the Faddeev approach yields a $^6$Li ground state that is approximately $600$~keV shallower than the one obtained  with the NCSM/RGM. Additionally, the $d$-$\alpha$ three-body calculations yield a $3^+$ resonance that is located approximately $400$~keV higher in energy compared to  the NCSM/RGM result (see Fig.~\ref{fig1}). The shape of the $d$-$\alpha$ angular distributions computed using the two approaches also differ, owing to the difference position of the $3^+$ resonance. 
\begin{figure}
\begin{center}
\includegraphics[width=0.7\linewidth]{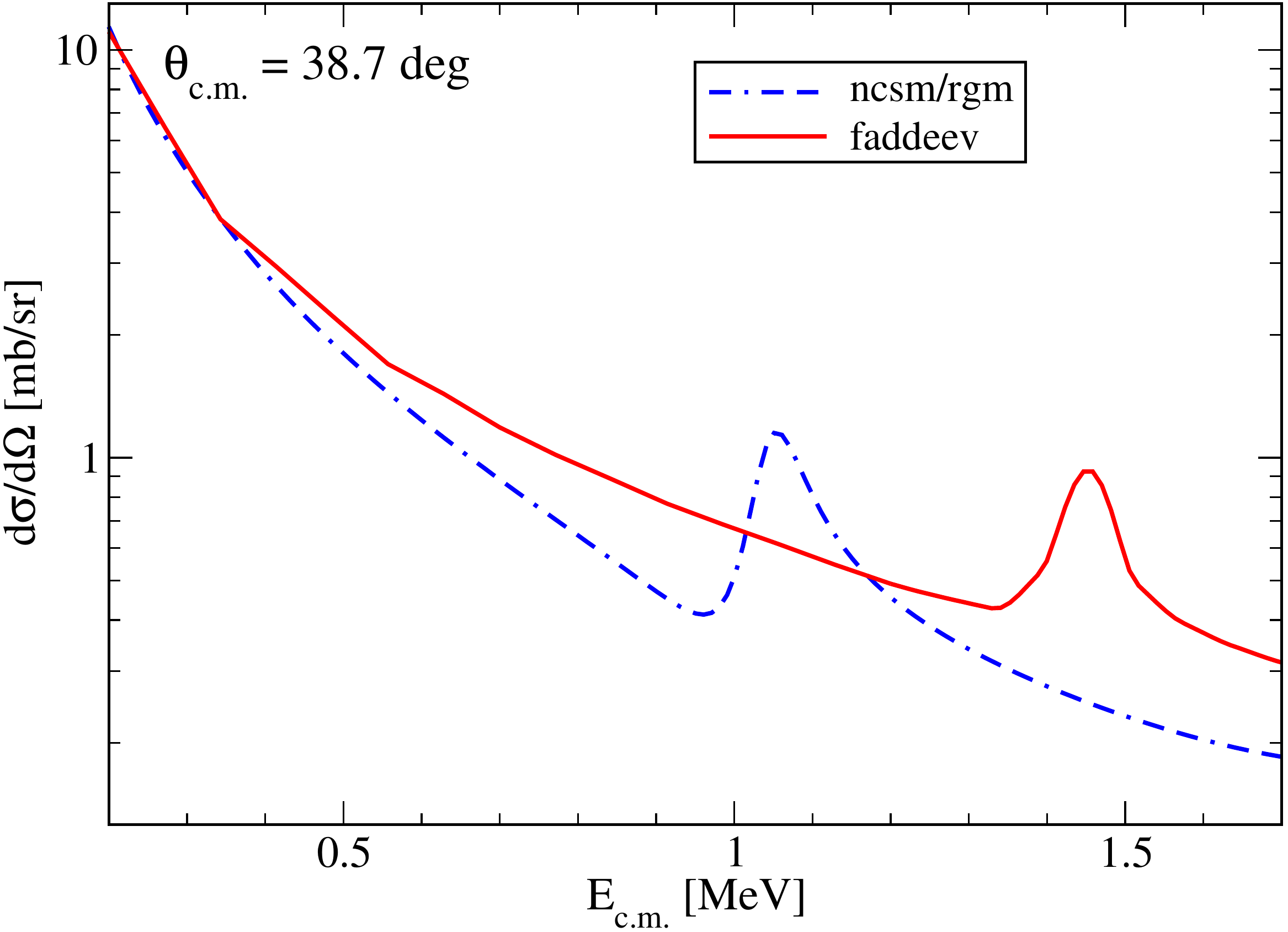}
\caption{The differential cross section  for elastic $d+\alpha$ scattering as a function of the center-of-mass energy $E_{c.m.}$ at the scattering angle $\theta_{c.m.}=38.7$~deg. The solid lines shows phase shifts computed using the NCSM/RGM while the Faddeev results are depicted by dashed lines. The model space for the Faddeev calculation is restricted to a total two-body angular momentum of $J_{np}\le3$ and $J_{n/p-\alpha}\le9/2$ for the $np$ and $n/p$-$\alpha$ subsystems~\cite{hlophe:2022}.   
}
\label{fig1}
\end{center}
\end{figure}
While the utilization of the NCSM/RGM allows for the determination of the contributions  to the 3BF stemming from Pauli exclusion effects, a similar study based on the no-core shell model with continuum~\cite{Baroni:2012su,Baroni:2013fe} (NCSMC) is necessary for the quantification of additional components arising from excitations of the nucleons in $^4$He. Lastly, a similar study that encompasses several nuclei and a broader energy range can shed light on the the mass and energy dependence of the 3BF. Such work would inform the parametrization of the latter and thus lead to improved three-body calculations for reactions.

\section{Tools and resources}\label{S4}

As discussed in Sec.~\ref{Sec3}, there are multiple efforts to build reliable optical potentials which are constructed from different approaches and have therefore various forms, e.g., local/non-local and parametrized/numerical.
To facilitate the development of accurate optical potentials and their use in applications, it is essential to efficiently share our codes and newly-developed optical potentials. Having access to well-documented publicly-available reaction codes is a real asset to developers of potentials. In this way, the potentials for various observables, such as direct and compound
nuclear reaction cross sections (see Sec. \ref{Sec2}), can be easily tested and their accuracy can be quantified. 
From the user side, because parametrizations of optical potentials often involve  many functionals, integrating new  optical potentials in a reaction code can be cumbersome.

To streamline the use of these potentials and to compare efficiently their accuracies for reaction observables, it is timely to list available resources in one platform. To fulfill these needs, in the context of the workshop ``Optical potentials in nuclear physics" held in March 2022 at FRIB, a \href{https://sites.google.com/view/opticalpotentials/optical-potentials-in-nuclear-physics}{\textcolor{blue}{website}}~\cite{webpageoptpot}   gathering reaction codes and optical potential parametrizations was created. We briefly present  here the different resources that this website contains (more details can be found in this website and in the references therein). Let us emphasize that the list presented is non-exhaustive, this is a selected overview of tools available to the community. In the future, the website will be updated with any resources that developers want to share.

\paragraph{Reaction codes.} Because optical potentials approximate the absorption from the elastic channel, their accuracy is often evaluated looking at elastic scattering and polarization data. Different codes, Scattering WAves off NonLocal Optical Potentials in the presence of Coulomb interaction ({\sc {\sc swanlop}}) \cite{SWANLOP1,SWANLOP2}, Schrödinger Integro-Differential Equation Solver ({\sc sides}) \cite{SIDES}, Equations Couplées en Itérations Séquentielles ({\sc ecis}) ~\cite{ECIS}, Optical Model with Nonaxiality ({\sc optman})   \cite{OPTMAN1,OPTMAN2,OPTMAN3} and {\sc fresco} \cite{FRESCO}, provide these observables  for any numerical potential given as input. These  solvers have complementary advantages, we emphasize here  some of their capabilities.
{\sc swanlop} and {\sc sides}, developed  by  Arellano, Blanchon \textit{et al.},  can handle non-local optical potentials exhibiting a Gaussian non-locality, such as the one proposed in the early work of Perey and Buck  \cite{perey62}. Moreover, {\sc swanlop} can read optical potentials expressed in both coordinate and momentum spaces. 
  {\sc ecis} and {\sc optman}, developed respectively by  Raynal and  Soukhovitski, are connected to a comprehensive database of parameters of local optical potentials as part of the IAEA RIPL project~\cite{ripl3},  allowing to calculate consistently scattering cross sections for many targets in a broad energy range.
{\sc fresco}  \cite{FRESCO}, developed by Thompson, also contains   a wrapper code {\sc sfresco}, 
that can be used to fit the optical potentials parameters to experimental data.

For more complex direct reactions, such as transfer and breakup, the codes {\sc fresco}  and NonLocal Adiabatic Transfer ({\sc nlat})~\cite{NLAT}  are the tools of choice.  {\sc Fresco}   calculates virtually any direct or multi-step nuclear reaction which can be expressed in terms of countable coupled-channels. In particular, {\sc fresco}  provides  various cross sections for breakup and transfer, obtained within the Continuum Discretized Coupled-Channel method~\cite{CDCC1,CDCC2,CDCC3,CDCC4,CDCC5,CDCC6} (CDCC), Coupled Reaction Channels~\cite{CRC1,CRC2} (CRC) or the DWBA. R-matrix and Lagrange methods allow
non-local potentials to be included non-iteratively. {\sc Nlat}, developed by Titus \textit{et al.}, calculates  transfer cross sections for single-nucleon transfer reactions, $(d,p)$, $(d,n)$, $(n,d)$ or $(p,d)$, including nonlocal nucleon-target interactions, within the finite-range adiabatic distorted wave approximation \cite{Johnson74} (ADWA) and DWBA.    This code is suitable for deuteron induced reactions in the range of $E_d\sim 10-70$ MeV.

To compute  compound reactions (see Sec. \ref{Sec2B}),  {\sc talys} \cite{TALYS1,TALYS2}, Yet Another Hauser Feshbach Code ({\sc yahfc}) \cite{YAHFC}    and {\sc empire} \cite{EMPIRE}, relying on the Hauser-Feshbach formalism~\cite{HauserFeshbach:52}, are available.  These three codes, which have been widely used by the community, provide predictions of nuclear reactions, including direct, pre-equilibrium and compound nucleus reactions, through multiple different methods and inputs. In particular, they can treat various optical models, spherical or  deformed, through coupled-channels methods. These codes are also connected to different optical potential libraries, {\sc empire}  is connected to RIPL database \cite{ripl3}, and both  {\sc yahfc} and {\sc talys} have some popular global parametrizations integrated in the code. There are some ongoing efforts to update {\sc yahfc} to integrate the extended global spherical proton and neutron optical potentials of CH89 \cite{varner91}  and KD \cite{koning03}  with uncertainty quantification of the potential parameters, respectively CHUQ and KDUQ \cite{Pruitt:22_report} (see Sec. \ref{Sec3A}) .

To solve coupled-channels problems in nuclear physics,  the subroutine {\sc rmatrix} \cite{rmatrix}, developed by Descouvemont, is also a  practical tool as it can be easily integrated in any  code.   This routine takes in input local or non-local coupling potentials at different nucleus–nucleus distances. It also includes an efficient way to deal with long-range potentials  with  propagation techniques, which significantly speeds up the calculations.

\paragraph{Available optical potential parametrizations.}

A comprehensive database of parameters for local optical potentials for many targets in a broad energy range was developed during the IAEA RIPL project~\cite{ripl3}. A retrieval code ({\sc omget}) is available from the {RIPL3} webpage~\cite{riplwebpage}   at tab ‘‘OPTICAL". This is a FORTRAN code that can prepare inputs for the optical solvers  {\sc ecis} and {\sc optman} using the RIPL OMP library. 

As emphasized in Sec. \ref{Sec3A}, there are a multitude of phenomenological optical potentials that have been developed for the last fifty years. Some of these parametrizations for neutron-, proton-, deuteron- and $\alpha$-target systems have been collected  by Kay in a excel spreadsheet. Having such a compilation of phenomenological potentials makes it easier for the user who wants to  compare observables obtained with various optical potentials. This spreadsheet is a work in progress and any suggestion is welcome.

Another recent effort has been made by Pruitt, Escher and Rahman~to quantify the uncertainties of the potential parameters in  the global spherical proton and neutron optical potentials of KD \cite{koning03} and CH89 \cite{varner91} with uncertainty quantification, the so called KDUQ and CHUQ \cite{Pruitt:22_report}.  The mass and energy range of validity are the same as the  original KD and CH89, i.e  $24<A<209$ and $0.001 \text{ MeV }<E<200 \text{ MeV}$   for KDUQ and   $40<A<209$ and   $10 \text{ MeV }<E<65 \text{ MeV}$ for CHUQ.
The optical potential parameters  and tools for sampling are available in the supplemental material of Ref.~\cite{Pruitt:22_report}.

The recently developed microscopic global WLH nucleon-nucleus potential  with quantified uncertainties \cite{whitehead19} (see Sec. \ref{Sec3B4}) has been parametrized to be easily integrated into modern reaction codes, using a local Woods-Saxon form  with parameters that vary smoothly in energy, mass, and isospin asymmetry. This global potential is valid for targets with mass $12 \leq A \leq 242$ and energies $0\leq E \leq 150$ MeV. 
A python script sampling the  WLH global parametrization can be downloaded on the website~\cite{WLHwebpage}.

\paragraph{Recommendations.} Historically,  optical potentials have been parametrized using a local Woods-Saxon radial form with parameters depending smoothly on the beam energy and mass of the target.  Because these global potentials  have  simple expressions, they are easily shared and often used for applications.  In general,  the newly developed microscopic optical potentials are  non-local and do not have an analytical parametrization.  To facilitate collaboration between theorists and experimentalists, 
 we propose recommendations  for the  non-local optical potentials to be shared as easily as possible between  makers and  users.

Consider the case where the non-local potentials do not couple different partial waves $lj$. The radial forms $U_{lj}(r,r';E)$,
defined on a two-dimensional $(r,r')$ radial grid, may then be represented as a matrix and their eigen-expansions determined.
At the specific incident energies $E$ where these potentials are targeted, some of its  eigenvectors will have much higher overlaps with the scattering wave functions than the others. 
It would therefore be more efficient   if just the eigensolutions with the largest product of overlaps and eigenvalues could be retained without significant loss of accuracy. 
These eigenvectors would be the one-dimensional form factors, 
such as $f_{i,lj,E} (r)$ corresponding to eigenvalues 
$\lambda_{i,ljE}$, in the expansion
      \begin{equation} \label{eq:nonloc-sep}
        U_{lj}(r,r';E)=U_{ljE}(r)\ \delta(r{-}r') 
         +\sum_{i=1}^n \lambda_{i,ljE} \ f_{i,lj,E} (r) \ f_{i,lj,E} (r')\ . 
    \end{equation}
The evaluator would choose expansion size $n$ just sufficient to describe the important physical effects that are not described by the local potential. 

As well as saving space in publications, these separable 
expansions would allow the fast solution of the scattering equations without needing R-matrix or Lagrange mesh bases that require solving a set of linear equations defined by the radial grid. That is because the scattering
equation with a potential like Eq. ~\eqref{eq:nonloc-sep} can be solved by the method of Eq. (5) of Ref. ~\cite{Frantz-PhysRevLett.1.340}, where a linear combination of inhomogeneous solutions (from each $f_{i,lj,E} (r)$ as a driving term) is added to the regular solution to reproduce the effects of the potential (\ref{eq:nonloc-sep}). Groups of only $n{+}1$ linear equations now need to be solved.

Further efficiency would follow if the form factors $f_{i,lj,E} (r)$ could have analytic forms in their radial and energy dependence. In general the principal eigenvectors will not have analytic shapes, but if the potential makers could fit some parametrized analytic forms, this would make the interchange of non-local potentials even easier.

Despite the recent efforts to treat  non-local optical potentials, there are still many methods and codes that have not been generalized to non-local interactions.  To move forward as a community,  we suggest that reactions codes need to be extended to deal with  both local and non-local potentials.

\section{Comparing approaches}
\label{rm}

\label{Sec5}

In this section, we provide a critical assessment and comparison of the different approaches presented in this work, both in order to  illustrate the content of the previous sections, and to set the stage for the next one. We show in Figs. \ref{fig:11}-\ref{fig:14} a systematic comparison of predictions for a variety of observables. We consider two broad categories: phenomenological (solid lines) and microscopic or semi-microscopic (dashed lines) models. The models for which an uncertainty quantification (UQ) study has been performed are represented by their 95\% uncertainty band. The acronyms used in this section,  referring to the optical potentials discussed throughout this paper, are listed in Table \ref{tab2}. Features of each optical potential, e.g., applicability in mass and in energy,  are summarized in Table \ref{tab1}.

\begin{table}
	\centering
	\begin{tabular}{|c|l|}
		\hline
		\multicolumn{2}{|c|}{List of abbreviations}\\
		\hline		
		KD		& Koning-Delaroche\\
		\hline
		KDUQ       & Koning-Delaroche with Uncertainty Quantification\\
		\hline
		
		DOM (STL)		& Dispersive Optical Model (Saint Louis)\\
		\hline
		MR		& Morillon-Romain\\
		\hline
		MBR		& Morillon-Blanchon-Romain\\
		\hline
		NSM		&  Nuclear Structure Model\\
		\hline        
		SCGF		&Self-Consistent Green's Function\\
		\hline
		MST-B		&  Multiple Scattering Theory - Burrows\\
		\hline
		MST-V		& Multiple Scattering Theory - Vorabbi\\
		\hline                        
		WLH       & Whitehead-Lim-Holt\\
		\hline                        
		JLMB		& Bruyères Jeukenne-Lejeune-Mahaux\\
		\hline
	\end{tabular}
	\caption{List of abbreviations used to denote the OMPs discussed in the text.}\label{tab2}
\end{table}

\begin{table}

\begin{tabular}{|c|c|c|c|c|c|c|c|}
				\hline
              & Mass & Energy & D. & Mic. & UQ & Bib & Sec. \\
              \hline
              KD		& $24\leq A\leq 209$  & 1 keV $\leq E\leq 200$ MeV & \xmark & \xmark & \xmark & \cite{koning03} & \ref{S3.1}\\
                        \hline
              KDUQ       & $24\leq A\leq 209$ & 1 keV $\leq E\leq 200$ MeV & \xmark & \xmark & \cmark & \cite{Pruitt:22_report} & \ref{S3.1} \\
                        \hline
              
              \multirow{1}{*}{DOM }		& C, O, Ca, Ni,& \multirow{2}{*}{$-\infty < E < 200$ MeV} & \multirow{2}{*}{\cmark} & \multirow{2}{*}{\xmark} & \multirow{2}{*}{\cmark} & \multirow{1}{*}{\cite{Atkinson:2020}} & \multirow{2}{*}{\ref{sec:dom}}\\
              (STL)&Sn, Pb isotopes&&&&&\cite{Pruitt2020PRC}&\\
                        \hline
              MR		& $12<Z<83$ & $E<200$ MeV & \cmark & \xmark & \xmark & \cite{Morillon:07} & \ref{sec:dom}\\
                        \hline
              MBR		& $12<Z<83$ & $E<200$ MeV & \cmark & \xmark & \xmark & & \ref{sec:dom}\\
                        \hline
              NSM		& $^{40}$Ca, $^{48}$Ca, $^{208}$Pb  & $E<$ 40 MeV & \cmark & \cmark & \xmark & \cite{blanchon_15} & \ref{Sec3B1}\\
                     \hline                        
              SCGF		& O, Ca, Ni isotopes & $E<100$ MeV & \cmark & \cmark & \xmark & \cite{Idini:19} & \ref{Sec3B2}\\
                        \hline
              MST-B		& $A\leq20$ &  $ E\gtrsim$ 70 MeV & \xmark & \cmark & \xmark & \cite{burrows20} & \ref{Sec3B3}\\
                        \hline
              \multirow{2}{*}{MST-V}		& \multirow{2}{*}{$4\leq A \leq 16$} & \multirow{2}{*}{$E\gtrsim$ 60 MeV} & \multirow{2}{*}{\xmark} & \multirow{2}{*}{\cmark} & \multirow{2}{*}{\xmark} & \cite{Vorabbi:2020cgf} & \multirow{2}{*}{\ref{Sec3B3}}\\
              &&&&&&\cite{Vorabbi:2021kho}&\\
                        \hline                        
              WLH       & $12\leq A\leq 242$ & $0\leq E\leq 150$ MeV & \xmark & \cmark & \cmark & \cite{Whitehead21} & \ref{Sec3B4} \\ \hline                        
              \multirow{2}{*}{JLMB}		& \multirow{2}{*}{$A>30$} & \multirow{2}{*}{1 keV $<E<$ 340 MeV} & \multirow{2}{*}{\xmark} & \multirow{2}{*}{\cmark}  & \multirow{2}{*}{\xmark} & \cite{Bauge:98} & \multirow{2}{*}{\ref{Sec3B4}}\\
              &&&&&&\cite{Baug2001}&\\ \hline
			\end{tabular}
			\caption{Summary of the Optical Potential models discussed in the text, identified with the acronyms detailed in Table~\ref{tab2} and used in Figs. \ref{fig:11}, \ref{fig:12}, \ref{fig:13}, and \ref{fig:14}. In the second and third columns, we indicate the applicability ranges in terms of mass and bombarding energy, respectively. The fourth column (D.) identifies dispersive potentials, while the fifth one differentiates between microscopic (Mic.), i.e., based on structure calculations, and phenomenological potentials. 
   The sixth column indicates whether an uncertainty quantification (UQ) analysis has been performed. The seventh column  (Bib.) points to the relevant references, and the eighth column  (Sec.) to the section in which the model is discussed.}\label{tab1}

\end{table}

In Fig. \ref{fig:11} we show angular differential cross sections for elastic scattering of neutrons and protons on $^{40,48}$Ca, $^{16}$O, and $^{12}$C at several beam energies, indicated by each line on the figure around zero degree. The overall reproduction of the data is encouraging:
the agreement of all phenomenological models (KD, DOM-STL, MR, MBR, KDUQ), semi-microscopic (JLMB) and most microscopic models (MST-B, MST-V, NSM, WLH)  is excellent at small angles, while both the consistency between approaches and the agreement with the data deteriorates at larger angles. 
This is to be expected since large scattering angles receive  contributions from other non-elastic channels, e.g., inelastic excitation,  transfer and breakup, tracing back to the imaginary part of the optical potentials. 
Another positive feature, the diffraction pattern of maxima and minima in the angular distribution agrees well with the data, suggesting that the bulk properties of the matter density distribution (such as the mean squared radius) are reproduced. 

    \begin{figure}
      \centering
      \includegraphics[width=\textwidth]{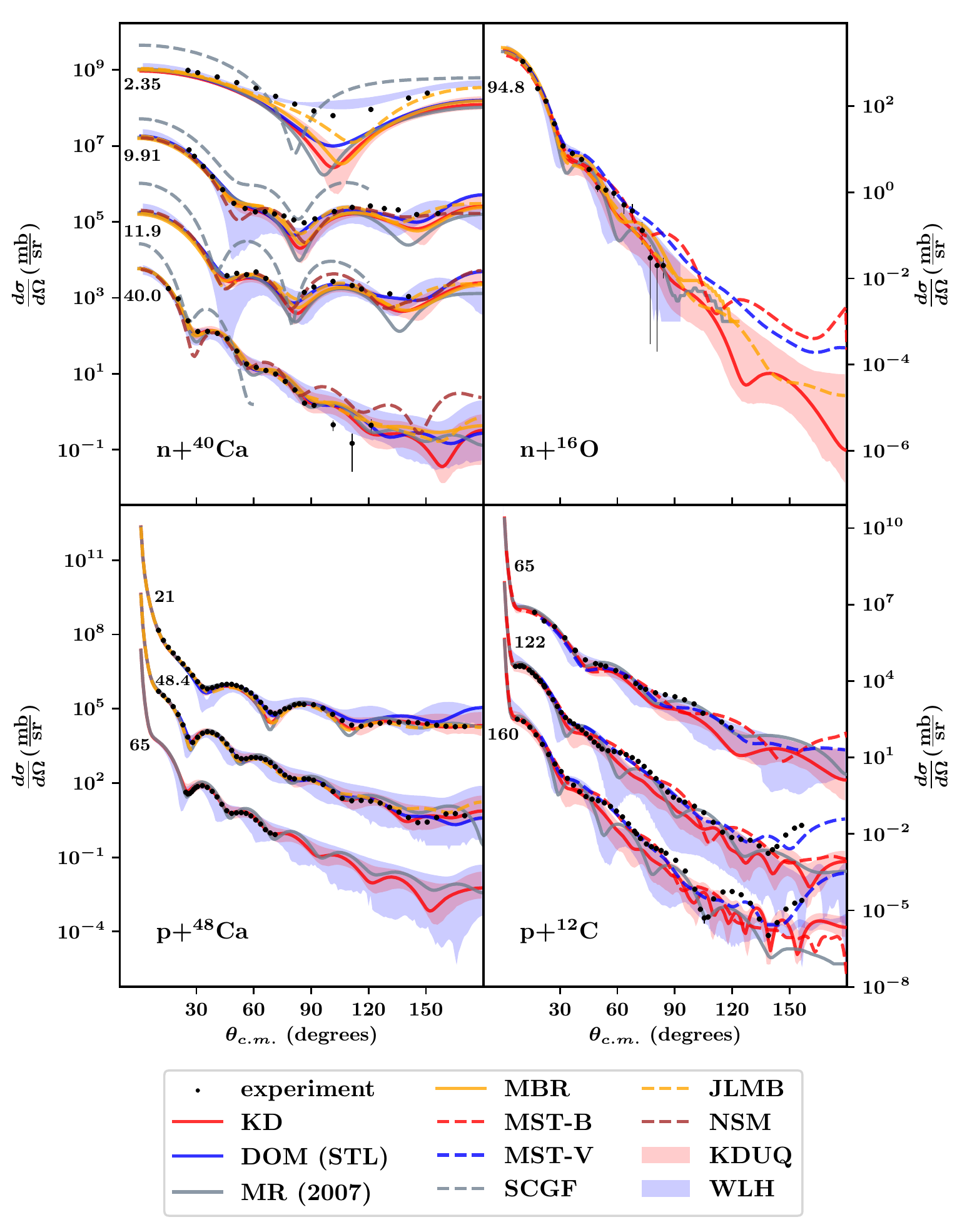}
      \caption{Elastic-scattering angular differential cross sections for neutrons on $^{40}$Ca and $^{16}$O and protons on $^{48}$Ca and  $^{12}$C, computed making use of the optical models indicated in the legend. The numbers by the curves around zero degree correspond to the nucleon bombarding energies in MeV . The shaded areas correspond to  95\% confidence intervals. }
      \label{fig:11}
  \end{figure}

In addition to these general remarks, there are a number of specific aspects that emerge from these comparisons:

\begin{itemize}
\item 
In contrast with the other  \textit{ab initio} models,  the  SCGF optical potential, featured in the $n+^{40}$Ca calculation, shows a consistent over-prediction of the elastic cross section, resulting from a lack of absorption.  This is likely a consequence of the fact that the level density at energies relevant for scattering phenomena is typically underpredicted by current  \textit{ab initio} calculations, leading to a small imaginary component in the optical potential.   Note that the  predictions presented here have been obtained with the NNLO-sat chiral interactions, using other two- and three-nucleon forces will influence the diffraction pattern and the absorption. 
One should also emphasize that the \textit{ab initio}-derived SCGF potential is the only one shown in Figs.~\ref{fig:11}--\ref{fig:12} that does not  rely on static densities or mean field approximations. This characteristic is shared by all methods discussed Sec.~\ref{Sec3B2}.
\item For the $n+^{40}$Ca at 2.35 MeV, only WLH is able to reproduce the shape of the angular distribution and absolute value of the cross section. That can be easily understood by noting that WLH is the only calculation including the contribution of compound elastic processes, which are important for this case.  Note that if  these contributions were included in the other calculations (KD, DOM-STL, MR, MBR,  JLMB and KDUQ), they may   also reproduce the data.

\item The calculations based on multiple scattering theory (MST-B, MST-V) are only applied to the $n+^{16}$O and $p+^{12}$C elastic scattering at high energies, consistent with their range of validity. These two models exhibit a similar level of agreement when compared to the data. Although they are similar for smaller angles,  there are significant differences at larger angles. This may result from the different treatment of spin used in both approaches as discussed in section \ref{Sec3B3}.

\item The approach based on an RPA description of the collective low-energy nuclear spectrum (NSM) are designed to be used at lower energies, where they indeed perform well. 

\item The  semi-microscopic model JLMB is accurate over the whole energy range, as its parameters were adjusted to reproduce neutron and proton elastic scattering observables in the 1 keV - 200 MeV energy range for $A\geq 40$.   One can see here that although the JLMB was not fitted on lighter nuclei, it agrees fairly well  with experimental data on $^{16}$O target. This indicates that the JLMB  remains a good starting point for $A<40$ nuclei. 
\item WLH performs well for all energies considered, however the uncertainty intervals from WLH are larger than those from KDUQ. This is particularly evident for  $p+^{48}$Ca scattering at large angles. While KDUQ was  fitted to data, WLH results from a microscopic calculation of the nucleon-nucleon interaction in nuclear matter and therefore there is no reason why these uncertainties should be of the same magnitude.
\end{itemize}

We now turn to energy distributions (total cross sections in Fig. \ref{fig:12} and reaction cross sections in Fig. \ref{fig:13}). As for the angular distribution, here the phenomenological potentials agree well with the $n+^{40,48}$Ca data at all energies. This is expected for the DOM parametrization which was fitted to reproduce both $^{40,48}$Ca data. However, since the MR, MBR, KD and KDUQ global optical potentials were only fitted to $^{40}$Ca data, the good agreement with the $^{48}$Ca data suggest that the mass and isospin dependencies of these parametrizations are accurate.     The WLH also agrees well with the data within its uncertainty, although the uncertainty interval is very large. For this observable too, the current imperfections of SCGF \textit{ab initio} calculations are apparent. Even if the available data are more scarce for the reaction cross sections than for the total cross sections, the picture drawn by the comparisons in Fig. \ref{fig:13} is  similar to that from Fig. \ref{fig:12}.  The reaction cross section  is largely associated with the imaginary part of the optical potential, reflecting the role of the open reaction channels in removing flux from the elastic one.  

While the phenomenological approaches perform well, the NSM approach underestimates the reaction cross section, which might point to the fact that  it fails to account for important reaction processes at lower energies. Since  the NSM takes explicitly into account direct excitation of collective states in the low-lying spectrum, this result suggests that other reaction channels, such as compound nucleus formation, charge exchange, and  transfer, need to be included in order to account for the total absorption.  Note that the sharp variation in the NSM results compared to the other potentials can be explained by the discrete energy positions of the collective states predicted by RPA calculations, causing peaks in the reaction cross section and hence the total cross section.    Overall, Figs. \ref{fig:12} and \ref{fig:13} reflect the difficulty encountered by microscopic theories in describing the variety  of relevant reaction channels. However, the good behaviour of the WLH potential (despite its wide uncertainty interval), based on microscopic calculations of the nucleon-nucleon interaction in nuclear matter, is noteworthy.

  \begin{figure}
      \centering
      \includegraphics[width=0.92\linewidth]{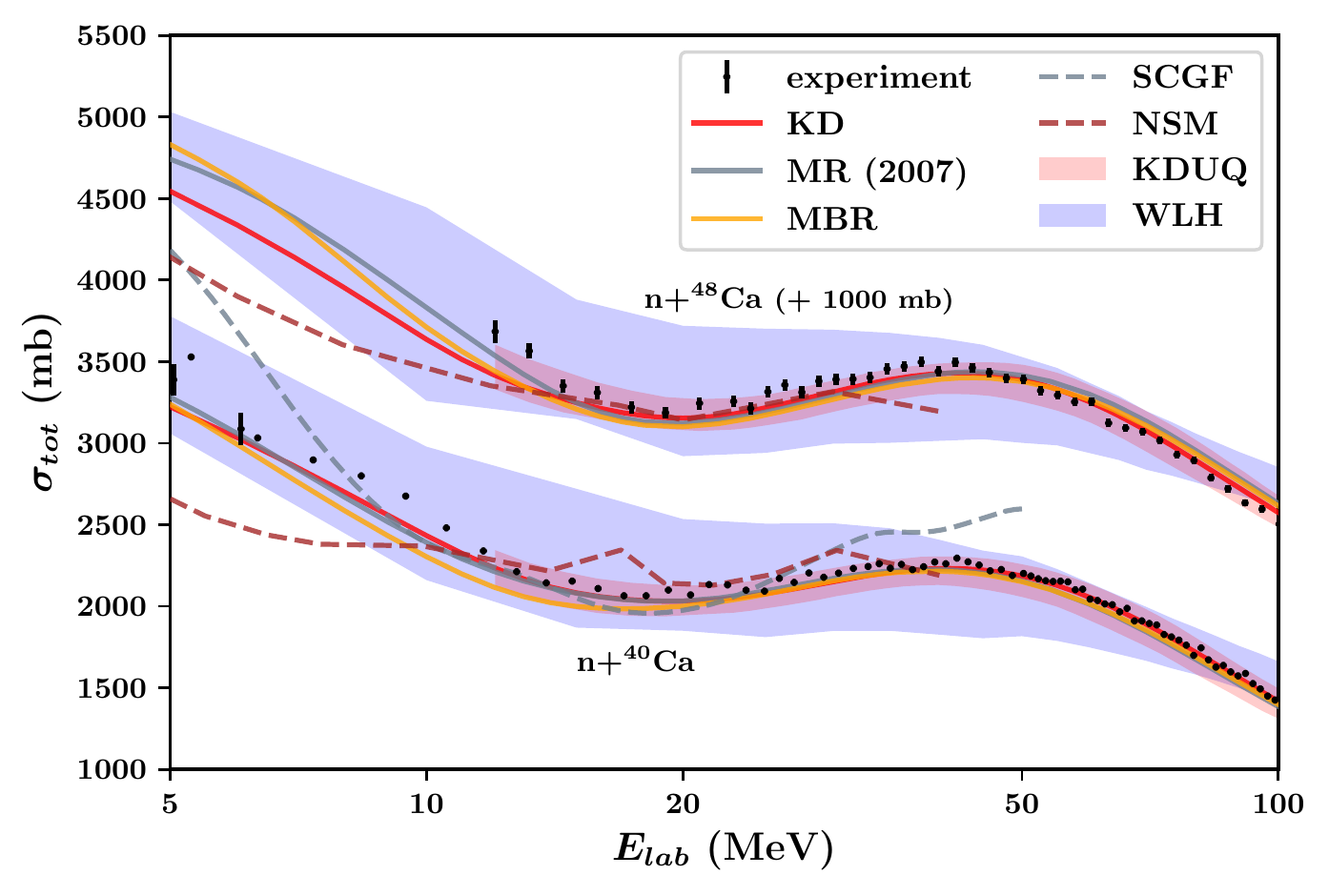}
      \caption{Total cross section as a function of neutron bombarding energy for $^{40,48}$Ca. The shaded areas correspond to  95\% confidence intervals.}
      \label{fig:12}
  \end{figure} 
    \begin{figure}
      \centering
      \includegraphics[width=0.92\linewidth]{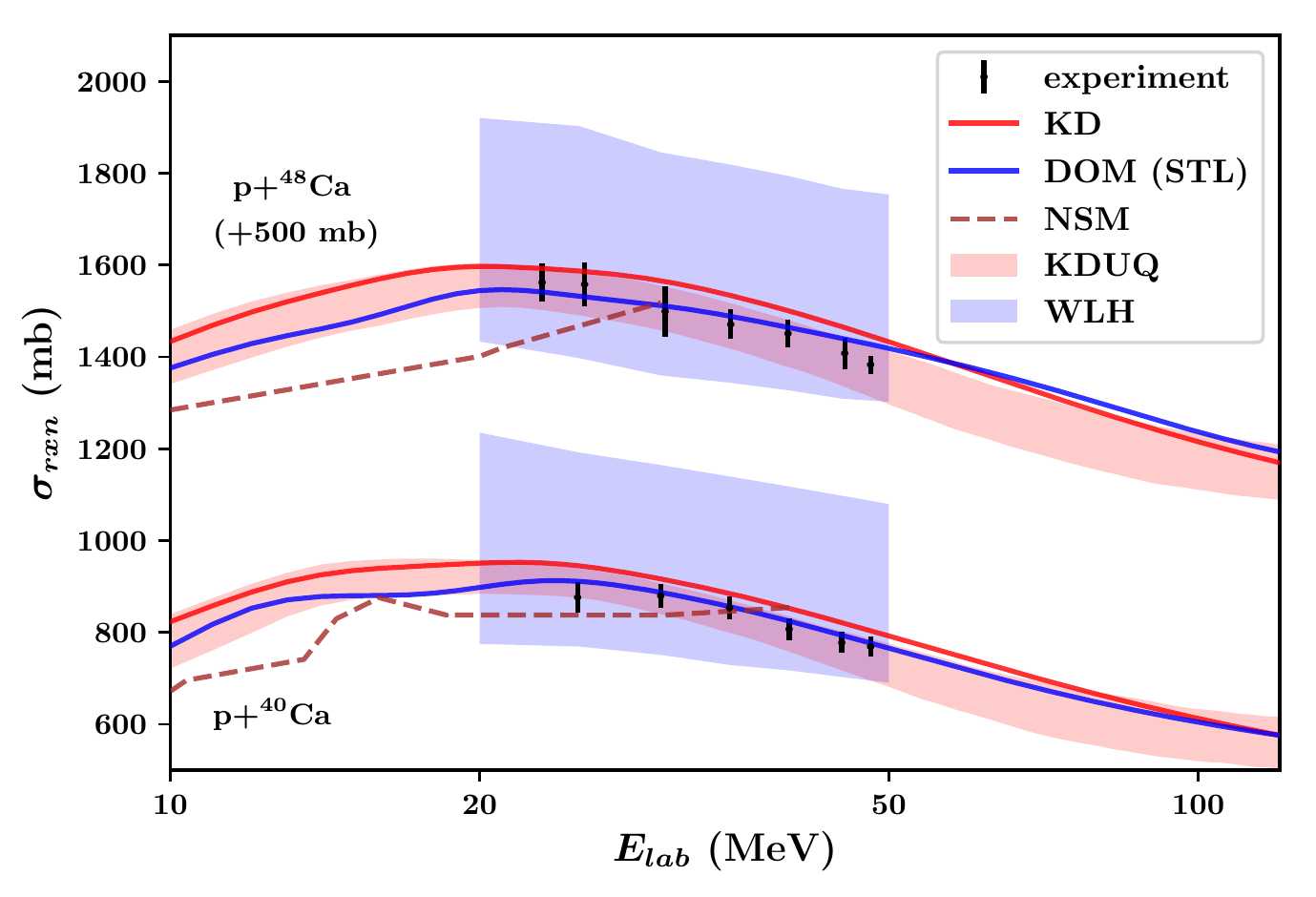}
      \caption{Reaction cross section as a function of proton bombarding energy for $^{40,48}$Ca.  The shaded areas correspond to  95\% confidence intervals.}
      \label{fig:13}
  \end{figure}

Finally we consider the asymmetry of the total cross section  between $^{40}$Ca and $^{48}$Ca as shown in Fig. \ref{fig:14}. This observable is very sensitive to the difference between neutron and proton densities, and the fact that  all the models reproduce the trend reasonably well suggests that they can account for the isospin dependence of the cross section around stability. However, the WLH potential somewhat underpredicts the energy of the first dip at around 30 MeV. 
  The uncertainty bands associated with KDUQ and WLH capture the data at the 95\% confidence level, and contrary to  the previous observables,  they both have a  similar width. A likely explanation is  that uncertainties associated with ratios of observables obtained consistently within the same theory tend to cancel. In the WLH, all  nuclei are derived consistently within the same framework, and uncertainties can be traced back to its specific approximations (included here are the truncation level of the chiral EFT). It is unclear how systematic uncertainties evolve for different nuclei in the case of phenomenological approaches. The phenomenological dispersive optical potentials MR and MBR perform remarkably well below 30 MeV, while KD is somehow worse. This situation is reversed above 40~MeV. While this ratio of observable offers, in principle, an excellent constraint  for the isospin dependence of the optical potential, we must note the large experimental uncertainties associated with the data in Fig. \ref{fig:14}. Dedicated experimental efforts  devoted to new measurements with extended range and improved error bars would be necessary for it to be useful in the extraction of the isospin dependence of the optical potential. 
Since exotic systems are characterized by an extreme neutron-to-proton ratio, we expect the cross section asymmetry to be a strong indicator of the reliability of potentials away from stability. 

\begin{figure}
      \centering
      \includegraphics[width=0.92\linewidth]{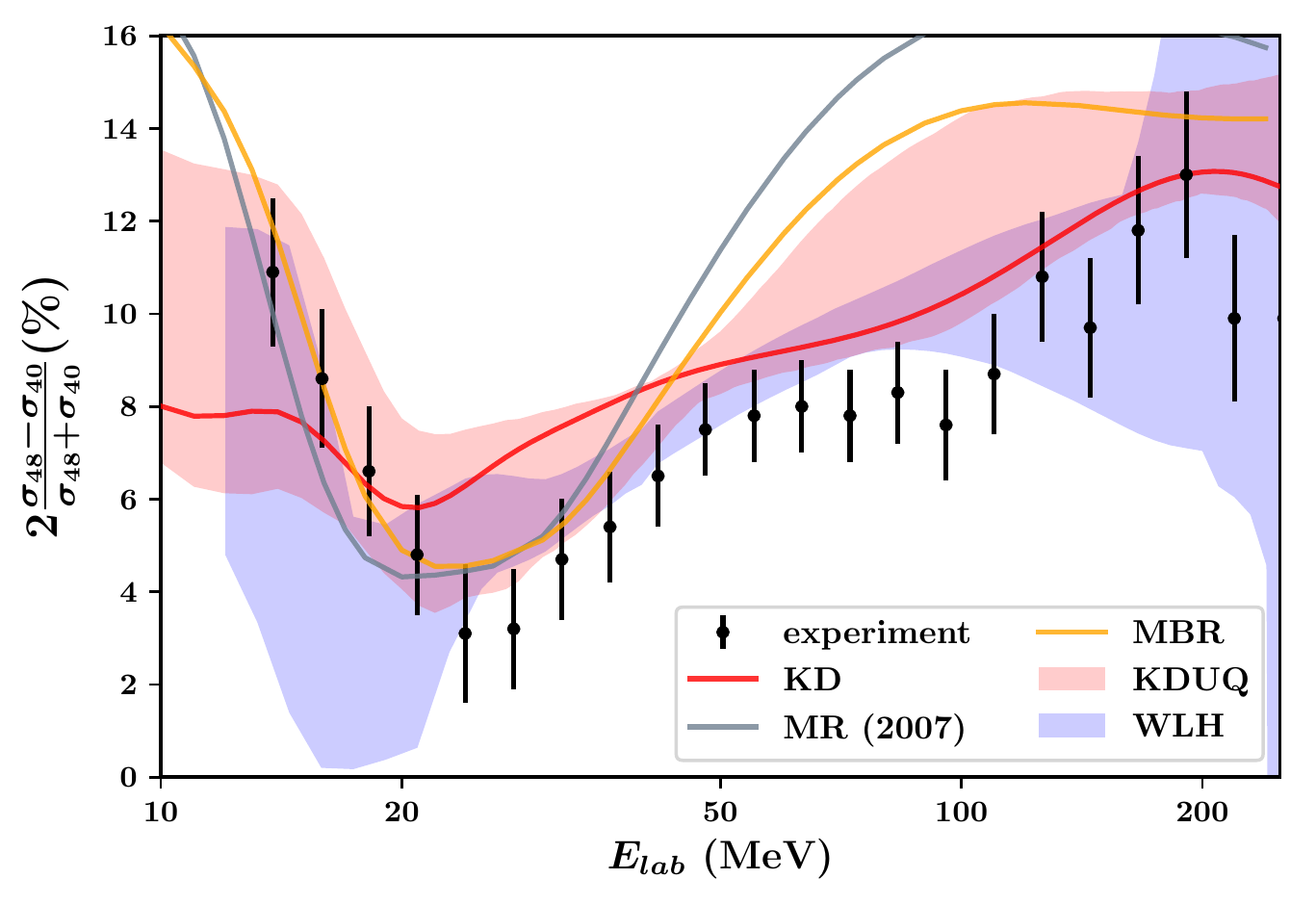}
      \caption{Asymmetry of the total cross section between $^{40}$Ca and $^{48}$Ca, defined as twice the ratio between the difference and the sum of the total cross sections, expressed in percentage.  The shaded areas correspond to  95\% confidence intervals.}
      \label{fig:14}
  \end{figure}

Overall, phenomenological optical potentials  with quantified uncertainties perform  well for stable nuclei even for observables and energies they have not been fitted to, as testified by their ability to reproduce the asymmetry data shown in Fig. \ref{fig:14}.  A rigorous estimation of the associated uncertainties, such as implemented in the KDUQ and WLH models, is a very desirable feature. We also want to stress the importance of the implementation of dispersivity in the DOM-STL, MR, and MBR phenomenological potentials. 
Some microscopic approaches provide a good reproduction of elastic scattering for their ranges of validity: low energy for  NSM, and high energy for the multiple scattering  potentials.   
However, advances in  fully \textit{ab initio} potentials, represented here by the SCGF and multiple scattering theory models, are still needed before they can be reliably used far from stability.

  \section{Outlook and recommendations}

In conclusion, optical potentials are ubiquitous in nuclear physics. In this white paper we discuss multiple ongoing efforts in the theory community aimed at improving their reliability and at quantifying associated uncertainties. Overall, results for nucleon elastic scattering on stable targets show that the various methods do  capture most of the physics, although as expected discrepancies amongst the methods and with the data increase at backward angles. The situation for nuclei away from stability is much more dire and requires dedicated future programs. In this section, we summarize the key points that should be kept in mind moving forward.

First and foremost, improving the determination of the optical potential for rare isotopes requires a close collaboration between theory and experiment. Experiments specifically targeted on constraining the optical potential are an imperative. While  recent \textit{ab initio} advances on the optical potential are impressive, it is clear that to obtain predictability away from stability, a careful validation of the current methods for systems with large  isospin asymmetry is essential. This implies working closely with the experimental community, such that theorists engage in both ends of the experimental endeavor, namely in providing input to experimental design and helping in the interpretation of the measurements. We emphasize the benefit of systematic studies, experimental setups that can measure multiple channels simultaneously or cover a range of beam energies. To obtain constraints on the isospin dependence of the optical potential, it is important to be able to extract observables that are particularly sensitive to isospin asymmetry. As such, experiments that span multiple isobars or measure a long isotopic chain are invaluable.

Secondly, future theoretical studies should strive to include uncertainty quantification in deriving the optical potential. The field is ripe for merging the knowledge obtained from microscopic approaches with experimental data within a Bayesian analysis. Such a statistical framework provides a natural avenue, not just for uncertainty quantification, but also for interpolating and extrapolating the optical potential,  assessing the information content of various observables, and for quantitatively discriminating between models. Theorists must be better informed on the experimental data used, and a concerted effort needs to be made such that experimental error bars incorporated in the optical model constraints include not just  statistical but also systematic errors.

Concerning the progress in theory, there are many thrusts that need to be pursued and the field can greatly benefit from stronger collaborations between theorists with differing expertise. In general, \textit{ab initio} methods need to be expanded beyond current truncations, so they can include additional correlations and ameliorate the lack of absorption resulting from the unphysically low density of states. As more \textit{ab initio} methods extend into the continuum, reaction data needs to be part of the standard protocols of validating \textit{ab initio} theories (currently theorists use mostly bound state spectra and root mean square radii to determine the quality of their model). It is important to address nuclear collectivity in microscopic optical model approaches at a global level. Although going beyond two-particle two-hole contributions is very challenging for some many-body frameworks, the current status demonstrates that including higher-order correlations is unavoidable. There are a variety of structure quantities calculated within the microscopic approaches that serve as inputs to the construction of optical potentials  (e.g., one-body densities and two-body densities). It is important that structure theorists calculate these quantities, test them for convergence and make them widely available. 

Several recent studies have demonstrated the added benefit of including the dispersion relation, enforcing causality as a constraint on the optical potential. Especially when considering a global optical potential spanning several energy regimes, it is desirable to correct the optical potential such that the real and imaginary parts of the extracted interaction are  related in the appropriate manner.

 It is understood that non-locality does affect reactions beyond the elastic channel and therefore its impact should be carefully considered. While in principle the optical potential is non-local, following the earlier work by Perey and Buck, a global non-local nucleon optical potential remains to be implemented. Studies have shown that the Gaussian shape assumed in the Perey and Buck parametrization is likely too simplistic. Since the optical model non-locality cannot be measured directly, this information must come from theory. Non-locality should be inspected when extracting the optical potential from microscopic theories, particularly to understand its full off-shell behavior and the dependencies on  model-space truncations. For most of the existing codes, local potentials are computationally much more efficient. However, methods to include non-local potentials can be very  efficient  when the optical potentials are expressed in separable form.
 
 In addition to non-locality, we identified two other features of the optical potential that must rely mostly on theory. As one moves away from stability, details on the isospin dependence become ever more important. Even with the new facilities, experiment will not be able to cover the whole nuclear landscape and such extrapolations in isospin will be reliant on theory. Testing this aspect of the microscopically derived optical potential is of paramount importance. Another important term in the optical potential is the spin-orbit force. The interplay of the spin-orbit force and the central term has been shown to be very important for loosely-bound systems. Elastic scattering is not strongly sensitive to this term and therefore, again, theory must provide guidance. 
 
Despite the discussion in this white paper being mostly focused on the nucleon optical potential, we must underline the necessity of optical potentials for complex probes (beyond the neutron and the proton). Many experimental programs at rare isotope facilities require the use of complex probes and progress on modern formulations for the nucleus-nucleus optical potential has been slow. Future theoretical studies should include global deuteron, triton, alpha and heavy-ion microscopic optical potentials, valid for nuclei away from stability.

Ultimately, progress in the theory for optical potentials does not immediately ensure integration into the many applications in our field. It is critical that tools be updated so that the whole community can benefit from their improvements. An integral part of this work was the creation of a \href{https://sites.google.com/view/opticalpotentials/optical-potentials-in-nuclear-physics}{\textcolor{blue}{website}} for the purpose of concentrating in one place the existing relevant codes. These codes need to be regularly updated by their authors to incorporate the latest optical potential developments. Only then will the field fully benefit from the theoretical advances.

Finally, it must be noted that while there has been  increasing interest by the many-body nuclear structure community in investigating the connection to the optical potential, the gap between the existing effort and the needs is still very large. Workforce development in this area is still critical and involves a particular skill set, including many-body nuclear formalisms, few-body reaction theory and modern statistical methods.

With the RIB factory in RIKEN in full force, FRIB having started operations earlier this year and numerous other facilities around the world, we expect a plethora of rare isotope data, directly relevant for the optical potential, in the coming years. Therefore, we anticipate this topic will need to be revisited in the next 5-6 years.

\section*{Acknowledgements}
The authors  thank H. Arellano for giving information on the the SCL-Bruyères g-matrix approach for the optical potential and K. Launey for critical feedback and insights in the SA-NCSM subsection.  The authors also thank S. Nikas and P. Gastis for sharing the calculations plotted in Figs. 4 and 5.  
This work is supported by the U.S. Department of Energy, Office of Science, Office of Nuclear Physics, under the FRIB Theory Alliance award no.\ DE-SC0013617. 
The work at Lawrence Livermore National Laboratory (LLNL) is performed under the auspices of the U.S. Department of Energy under Contract DE-AC52-07NA27344 and was supported in part by the LLNL-LDRD Program under Project No. 21-ERD-006.
The work at Brookhaven National Laboratory is sponsored by the Office of Nuclear Physics, Office of Science of the U.S. Department of Energy under Contract No. DE-AC02-98CH10886 with Brookhaven Science Associates, LLC.
The work at Ohio University is supported by the U.S.\ Department of Energy Office of Science under Grants {DE}-{FG02}-93ER40756. 
This work was also supported by  the U.S. Department of Energy Office of Science under grants  DE-SC0021422, {DE}-{SC}0019209, DE-SC001920, DE-SC0019521, DE-AC02-06CH11357 and DE-NA0003841. 
This work is supported by the National Science Foundation under Grant no.\ PHY-1913728, PHY-2209060, PHY1652199, PHY1912643, and PHY2207756.
Computing support for the NCSM/RGM and Faddeev results presented in Figs.~\ref{fig:B3_GF-SM} and~\ref{fig1} came from the LLNL Institutional Computing Grand Challenge program.
Computing support for the SCGF results presented in Figs.~\ref{fig:B3_GF-SM} come from the DiRAC DiAL system at the University of Leicester, UK, (funded  by  the  UK  BEIS via  STFC  Capital  Grants No.~ST/K000373/1 and No.~ST/R002363/1 and STFC DiRAC  Operations  Grant  No.~ST/R001014/1) and from the National Energy Research Scientific Computing Center (DOE Contract No.~DE-AC02-05CH11231) using NERSC award NP-ERCAP0020946.

\section*{References}
\addcontentsline{toc}{section}{References}

\bibliography{BiblioFull}

\end{document}